\documentclass[12pt]{article}
\usepackage{caption,amsmath,graphicx,url,amssymb,latexsym,epsfig,tabularx,setspace,multirow,threeparttable,longtable,pdflscape,tabu,appendix,comment,todonotes,subfigure} 
\usepackage{algpseudocode}
\usepackage{algorithm}  

\usepackage[margin=1in]{geometry}
\usepackage{xspace}

\begin{document}

\def\algname{\textsc{GPUTraj\-DistSearch}\xspace}
\def\periodic{\textsc{Periodic}\xspace}
\def\periodicbest{\textsc{PeriodicBest}\xspace}
\def\periodicgood{\textsc{PeriodicGood}\xspace}
\def\setsplit{\textsc{SetSplit}\xspace}
\def\setsplitF{\textsc{SetSplit-Fixed}\xspace} %SetSplit-A, SetSplitA
\def\setsplitMax{\textsc{SetSplit-Max}\xspace} %SetSplit-B, SetSplitB
\def\setsplitMM{\textsc{SetSplit-MinMax}\xspace} %SetSplit-C, SetSplitC
\def\greedysetsplit{\textsc{GreedySetSplit}\xspace}
\def\greedysetsplitMin{\textsc{GreedySetSplit-Min}\xspace} %GreedySetSplit-A, greedysetsplitA 
\def\greedysetsplitMax{\textsc{GreedySetSplit-Max}\xspace} %GreedySetSplit-B, greedysetsplitB
\def\galaxy{\textsc{Galaxy}\xspace}
\def\random{\textsc{RandWalk}\xspace}
\def\randomuniform{\textsc{RandWalk-Uniform}\xspace}
\def\randomnormal{\textsc{RandWalk-Normal}\xspace}
\def\randomnormalfive{\textsc{RandWalk-Normal5}\xspace}
\def\randomexp{\textsc{RandWalk-Exp}\xspace}

\begin{centering}
\textbf{Parallel Distance Threshold Query Processing for Spatiotemporal Trajectory Databases on the GPU\\}
\vspace{0.3cm}
Michael Gowanlock\\ 
Department of Information and Computer Sciences and NASA Astrobiology Institute\\ University of Hawai`i, Honolulu, HI, U.S.A.\\
Email: gowanloc@hawaii.edu\\
\vspace{0.3cm}
Henri Casanova\\
Department of Information and Computer Sciences\\ University of Hawai`i, Honolulu, HI, U.S.A.\\
Email: henric@hawaii.edu

\end{centering}

\begin{abstract}
Processing moving object trajectories arises in many application domains
and has been addressed by practitioners in the spatiotemporal database and
Geographical Information System communities.  In this work, we focus on a
trajectory similarity search, the distance threshold query, which finds all
trajectories within a given distance $d$ of a search trajectory over a time
interval.  We demonstrate the performance of a multithreaded implementation
which features the use of an R-tree index and which has high parallel
efficiency (78\%-90\%).  We introduce a GPGPU implementation which avoids
the use of index-trees, and instead features a GPU-friendly indexing
method.  We compare the performance of the multithreaded and GPU
implementations, and show that a speedup can be obtained using the latter.
We propose two classes of algorithms, \setsplit and \greedysetsplit, to
create efficient query batches that reduce memory pressure and
computational cost on the GPU.  However, we find that using fixed-size
batches is sufficiently efficient in practice.  We develop an empirical
performance model for our GPGPU implementation that can be used to predict
the response time of the distance threshold query. This model can be used
to pick a good query batch size.
\end{abstract}

\section{Introduction}
%outline the query

Applications in a wide range of domains process datasets that contain
trajectories of moving objects.  Trajectory data can be generated from the
motions of people and objects captured in the form of traces from Global
Positioning System (GPS) devices. It can also be generated from scientific
applications, such as moving animals in ecological simulations, vehicles in
traffic simulations, and applications that utilize Geographical Information
Systems (GIS). Regardless of the manner in which trajectory data is
generated, these applications process spatiotemporal trajectory datasets to
gain insight into their target domains.  In this work, we focus on
historical continuous trajectories \cite{Forlizzi2000}, where a database of
trajectories is given as input and supports searches over these
trajectories. More specifically, we study the following \emph{distance
threshold search}: Find all trajectories within a distance $d$ of a given
query trajectory over a time interval [$t_0$,$t_1$].  An example of this
search in the context of an ecological simulation would be to find all
prey within 500 m of a query predator over the course of a day.

Numerous methods to index and process moving object trajectories
efficiently have been developed by researchers in the field of spatial and
spatiotemporal databases.  Most works focus on out-of-core implementations
where only part of the database resides in memory while its majority resides on
disk.  As a result, indexing techniques are designed to optimize data
layouts so as to reduce disk accesses. Furthermore, many of these works
focus on sequential query processing.  However, current architectures and
data storage capacities offer attractive alternatives. In particular,
many-core GPUs (Graphic Processing Units) have become mainstream, are
programmable for general purpose computing, and should be well-suited to
the large number of moving distance calculations required for processing
queries on spatiotemporal databases.  In addition, current off-the-shelf
compute nodes and GPU devices have relatively large memories.  Given that a
spatiotemporal database can be easily partitioned (e.g., temporally) and
queried across multiple compute nodes, query processing can be performed in
parallel and entirely in-memory.

In this work, we focus on in-memory distance threshold search processing using the GPU, which
to the best of our knowledge has not been studied previously.  As typical
out-of-core indexing techniques are no longer effective in this setting,
new indexing approaches must be developed. In this context we make the
following contributions:

\begin{itemize}
\item We develop an indexing technique that is suitable for efficient distance threshold searches on the GPU.

\item We implement a GPU kernel to perform the distance threshold search,
minimizing branch instructions to achieve good parallel efficiency on the
GPU.

\item We compare our GPU implementation to a previously developed CPU-only
      implementation that uses an in-memory index tree, and show that
      using the GPU can afford significant speedup.

\item Efficient searches are predicated on grouping query trajectories in
batches, and we propose two classes of algorithms to create such batches.
We find that creating same-size batches is sufficient to achieve good
performance in practice.

\item We develop a performance model of the search response time which considers
the underlying spatiotemporal properties of the datasets, and the expected
CPU and GPU execution times.  We demonstrate that this model can predict
response times within a reasonable margin, thus making is possible to
select a good query batch size.
\end{itemize}

%outline of the paper
The rest of this paper is organized as follows: In
Section~\ref{sec:related_work}, we outline the motivation for this work,
present background material and review related work.  In Section~\ref{sec:problem_outline}, we
formally define our in-memory, on-GPU, distance threshold search problem.
Section~\ref{sec:traj_indexing} describes our indexing
technique and Section~\ref{sec:search_alg} details our implementation of a
GPU kernel that processes queries using this indexing technique.
Section~\ref{sec:set_algs} proposes several algorithms that group queries
into batches.  The performance of the search when using these algorithms is
evaluated in Section~\ref{sec:exp_eval}.  In
Section~\ref{sec:performance_model}, we propose our response time performance model.
Section~\ref{sec:conclusions} concludes with a
summary of our findings and perspectives on future work.

\section{Motivation and Related Work}\label{sec:related_work}
\subsection{Motivating Example}\label{sec:project_motivation}

One motivation application for this work is in the area of
astrophysics~\cite{2011AsBio..11..855G}. The study of the habitability of
the Earth suggests that life can exist in a multitude of environments.  The
past decade of exoplanet searches implies that the Milky Way, and hence the
universe, hosts many rocky, low mass planets that may be capable of
supporting complex life.   The Galactic Habitable Zone is thought to be the
region(s) of the Galaxy that may favor the development of complex life.
With regards to long-term habitability, some of these regions may be
inhospitable due to transient radiation events, such as supernovae
explosions or close encounters with flyby stars that can gravitationally
perturb planetary systems. Studying habitability thus entails solving the
following two types of \emph{distance threshold searches} on the
trajectories of (possibly billions of) stars orbiting the Milky Way:
(i)~Find all stars within a distance $d$ of a supernova explosion, i.e., a
non-moving point over a time interval; and  (ii)~Find the stars, and
corresponding time periods, that host a habitable planet and are within a
distance $d$ of all other stellar trajectories.  Our work aims to have a
direct impact on such applications.

\subsection{Background and Related Work}

%What a trajectory is
A trajectory is defined by a set of positions that describe the motion of a
moving object in Euclidean space over a time interval.  The continuous
nature of trajectories requires that each point traversed by the trajectory
be approximated by a polyline, where points are connected via line
segments.  The goal of trajectory similarity searches is to find
trajectories with similar attributes. Various types of trajectory
similarity searches have been studied in a number of domains, such as convoys
\cite{Jeung:2008:DCT:1453856.1453971}, flocks
\cite{Vieira:2009:ODF:1653771.1653812}, and swarms
\cite{Li:2010:MMM:1807167.1807319}.  The most studied similarity search
is the $k$ Nearest Neighbors ($k$NN)
search~\cite{Frentzos2005,Frentzos2007,Gao2007,Guting2010}.

%Trajectories in the context of spatiotemporal databases
The field of spatiotemporal databases provides a number of methods and
perspectives on processing spatiotemporal data. The typical approach is
search-and-refine, by which an index is searched and yields a preliminary
result set, which is then refined to produce a final result set. Thus, many
data structures have been advanced to efficiently index trajectory data,
which have been based on the success of the R-tree~\cite{Guttman-R_tree},
such as TB-trees~\cite{Pfoser2000}, STR-trees~\cite{Pfoser2000},
3DR-trees~\cite{Theodoridis-1996} and SETI~\cite{Chakka2003}.
Additionally, systems have been designed to process and analyze
trajectories, such as TrajStore~\cite{Cudre-Mauroux2010} and
SECONDO~\cite{Guting2010}.  The common approach is to map nodes in an
index-tree to pages stored on disk, and the performance of applications is
largely a function of index-tree node accesses, where fewer node accesses
can improve response time by avoiding costly data transfers between memory
and disk. Index-trees have been used extensively for performing $k$NN
searches.

% Distance Threshold queries
In this work, we study distance threshold queries, which can be viewed
as $k$NN searches with an unknown value of $k$ and thus unknown 
result set size. As a result, the typical search-and-refine strategy
is not necessarily well-suited to these searches. Furthermore, several of the
aforementioned index-trees, while efficient for $k$NN queries, are 
not efficient for distance threshold queries.  These queries have not
received a lot of attention in the literature.  Our previous work
in~\cite{Gowanlock2014} studies \emph{in-memory} sequential distance threshold
searches, using an R-tree to index trajectories inside hyperrectangular
minimum bounding boxes (MBBs).  The main contribution is an indexing method
that achieves a desirable trade-off between the index overlap, the number
of entries in the index, and the overhead of processing candidate
trajectory segments.  An interesting finding is that a good empirical
metric that can be used to achieve such a trade-off is cache reuse, showing
in particular that minimizing MBB volume is not as important as having an
upper bound on the fraction of a trajectory stored in each MBB for
in-memory trajectory databases.  The work in~\cite{Arumugam2006} solves a
similar problem, i.e., finding trajectories in a database that are within a
query distance $d$ of a search trajectory, but the algorithm does not
return the time intervals in which this occurs.  The authors compare four
query processing strategies, one which utilizes an R-tree and three that
use a plane-sweep approach.  They find that an adaptive plane-sweep
approach yields the best performance. A key difference with the work
in~\cite{Gowanlock2014} is that they consider an out-of-core scenario, with
part of the database residing on disk.  Also, and
unlike~\cite{Gowanlock2014}, their R-tree implementation does not attempt
to find a good compromise between index overlap and index size.

% Queries on the GPU
Methods have been recently proposed for indexing spatial and spatiotemporal
databases on
GPUs~\cite{Zhang2014,Zhang:2012:USH:2390226.2390229,You:2013:PSQ:2534921.2534949,Luo2012},
some of which employ more straightforward data structures in comparison to
the aforementioned index-trees. The efficient execution of $k$NN
searches has been investigated on the
GPU~\cite{Pan:2011:FGL:2093973.2094002,CPE:CPE1718} and on hybrid CPU-GPU
environments~\cite{Krulis2012}, although not in the context of
spatiotemporal databases.  In this work, we target the GPU, but we focus on
distance threshold searches.
Although related to the $k$NN search, the distance threshold
search has significant differences that make it difficult to reuse $k$NN
search techniques~\cite{Gowanlock2014}.

\section{Problem Definition}\label{sec:problem_outline}

Let $D$ be a spatiotemporal database that contains
$n$ 4-dimensional (3 spatial dimensions + 1 temporal dimension) line segments.
A line segment $l_i$, $i=1,\ldots,n$, is defined by a spatiotemporal starting point
($x_i^{start}$, $y_i^{start}$, $z_i^{start}$, $t_i^{start}$), a
spatiotemporal ending point ($x_i^{end}$, $y_i^{end}$, $z_i^{end}$,
$t_i^{end}$), a segment id and a trajectory id.  Segments belonging to the
same trajectory have the same trajectory id and are ordered temporally by
their segment ids. We call $t_i^{end}-t_i^{start}$ the \emph{temporal extent}
of $l_i$ and the line segments in $D$, \emph{entry segments}.

We consider \emph{historical continuous searches} that search for entry
segments within a distance $d$ of a query $Q$, where $Q$ is a
set of line segments that belong to a moving object's trajectory. We call the line segments in $Q$ \emph{query segments}. The
search is continuous, such that an entry segment may be within the distance
threshold $d$ of particular query segment for only a subinterval of that
segment's temporal extent. The result set thus contains a set of entry segments, and for each segment, a time interval.  For example, for a query segment with temporal extent [0,1], the search may return ($l_1$,[0.1,0.3]) and ($l_2$,[0.6,0.9]).

We consider a platform that consists of a host, with RAM and CPUs, and a GPU
device with its own RAM (global, local, constant) and compute units.
We consider an \emph{in-memory database}, meaning that $D$ is stored once
and for all in global memory on the GPU.  We 
focus on an \emph{online} scenario where the objective is to minimize the
response time for an arbitrary set of queries. This is the typical
objective considered in other spatiotemporal database works such as the
ones reviewed in Section~\ref{sec:related_work}. 
We consider the case in which $D$ and
$Q$ cannot fit together on the GPU, with a twofold rationale. First, the
memory on the GPU is limited and in practice a single database is subjected
to a large number of queries. Second, memory for the
result set must be allocated statically since dynamic memory
allocation is not permitted on the GPU.  However, the result set size is
non-deterministic and depends on the spatiotemporal nature of the data. As a
result, memory allocation for the result set must be conservative and
overestimate the amount of memory required. This overestimated size grows
linearly with $|Q|$, thereby creating even more
memory pressure on the GPU.  For these two reasons we partition $Q$ in batches
that are processed in sequence.  Note that such incremental query
processing is also useful when multiple users query the database
simultaneously, and would thus compete for memory space on the GPU.

\section{Trajectory Indexing}\label{sec:traj_indexing}
%\todo[inline]{HC: Make sure I am not totally off toward the end of this
%paragraph when talking
%about~\cite{Zhang2014,Zhang:2012:USH:2390226.2390229}. The original text
%said that that work was great for ``online batch'' processing... I think
%it's great for offline, right?. \\MG: Wrong, at least by the wording in
%their work.  I guess it's a question of semantics: The distinction between
%online and offline is not the same as the distinction between batch
%processing and not batch processing. They say that batch processing, is an
%online approach and not an offline approach.  For example, in some of their
%papers (the Zhang, You et al ones), they say they are doing online work, in
%the context of Online Analytical Processing (OLAP).  Its confusing to me as
%well, but no where in the 3 papers we cite do they say they are doing
%offline implementations.  For example, in
%\cite{Zhang:2012:USH:2390226.2390229}, where they use grid files, they say
%``based on these observations, our goal is to design a prototype system
%that can utilize commodity hardware capacities, including parallel
%computing power and large memory capacity, to boost the performance of OLAP
%type queries in a batch mode for U2S trajectory data''. I modified the end
%of the paragraph accordingly.  I guess the difference between our work and
%their work is what was written in the previous section: that we consider
%the search in the context of memory constraints, but both are online
%implementations.}

Many efficient indexing methods have been proposed in the spatiotemporal
database literature assuming that processing takes place on a CPU.  The GPU
architecture is markedly different from that of the CPU.  The GPU uses the
Single Instruction Multiple Data (SIMD) execution model, requiring that
work-items (GPU threads) that take different execution paths be executed
sequentially~\cite{Han:2011:RBD:1964179.1964184}.  Therefore, limiting the
amount of conditional branching in GPU implementations is important to achieve good
performance.  As a result, efficient CPU implementation approaches (which
can, e.g., benefit from branch prediction techniques) are likely to be
vastly inefficient when applied directly to the GPU.  

Our previous work uses an in-memory R-tree index  for processing distance
threshold searches on the CPU~\cite{Gowanlock2014}.  For a given query, the
search phase of the computation finds candidate segments as stored in
MBBs in the leaf nodes of the
R-tree, and the refine phase reduces these candidates to find those that
should be part of the result set. The majority of the computation is spent
in the search phase, which has many branch instructions to follow R-tree
node pointers from the root to the leaves, which should be avoided on the
GPU.  Similar observations have been made in the literature when indexing
spatial and spatiotemporal databases on the
GPU~\cite{Zhang2014,Zhang:2012:USH:2390226.2390229}.  The authors
in~\cite{Zhang2014} note that it is not clear whether index-trees should be
used at all. The work in~\cite{Zhang2014,Zhang:2012:USH:2390226.2390229}
utilizes grid files, or ``flatly structured grids,'' data structures in
which polylines are converted to MBBs and are assigned to cells on a grid
to spatially partition and index the data as an alternative to using index
trees.  In this work, we design a GPU-friendly indexing method for scenarios
in which large query sets must be partitioned into batches that are
processed iteratively.

In light of the above, we propose the following approach to index the
database.  We first sort the entry segments by non-decreasing $t_{start}$
values. Without loss of generality we assume that the entry segments are
numbered in that order (i.e., $t^{start}_{i} \leq t^{start}_{i+1}$ ).  The
full temporal extent of database $D$ is [$t_0,t_{max}$] where $t_0 =
\min_{l_i \in D} t_i^{min}$ and $t_{max} = \max_{l_i \in D} t_i^{max}$. We
divide this temporal extent logically into  $m$ bins of fixed length $b =
(t_{max} - t_0) / m$. We say that an entry segment $l_i$, $i=1,\ldots,n$,
belongs to bin $B_j$, $j=1,\ldots,m$, if $\lfloor t_i^{start} / b \rfloor =
j$. For bin $B_j$ we can then define $B_{j}^{start} = j\times b$ and $B_j^{end} = \max_{l_i \in B_j} t_i^{end}$.
We call [$B_{j}^{start},B_{j}^{end}$] the temporal extent of bin
$B_j$. We then define $B_j^{first} = \arg\min_{i | l_i \in
B_j} t_i^{start}$ and $B_j^{last} = \arg\max_{i | l_i \in B_j}
t_i^{start}$.  [$B_{j}^{first},B_{j}^{last}$]  is thus the index range of
the entry segments in $B_j$.
Bin $B_j$ is thus fully described as
($B_{j}^{start}$,$B_{j}^{end}$, $B_{j}^{first},B_{j}^{last}$).  The set of bins
is the ``index'' of the database.

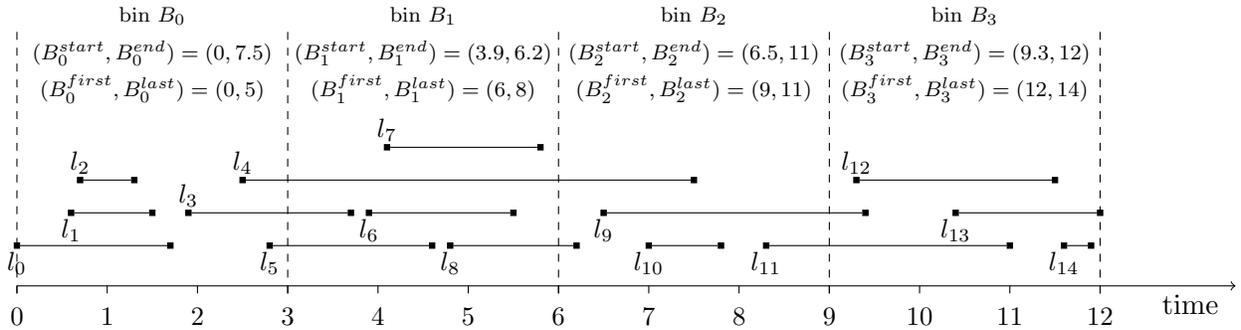
\begin{figure}[t]
\centering
  \begin{tikzpicture}

  \def\xscale{1.2}

  % draw the time axis
  \draw [->] (0,0) -- (\xscale*13.5,0);
  \node (timelabel) at (\xscale*13.0,-0.25) {\small{time}};
  \foreach \x in {0,...,4} {
  	\draw [dashed] (\xscale*3*\x,-0.1) -- (\xscale*3*\x,3.48);
  }
  \foreach \x in {0,...,12} {
	\draw (\xscale*\x,0) -- (\xscale*\x,-0.1);
	\node [label=south:\footnotesize{\x}] at (\xscale*\x , -0.0) {};
  }

 % draw bin information
\def\bins{
	0/0/7.5/0/5,
	1/3.9/6.2/6/8,
	2/6.5/11/9/11,
	3/9.3/12/12/14
	}
  \foreach \x/\a/\b/\c/\d in \bins {
	\def\height{4}
	\def\increment{0.45}
	\node [label=south:\scriptsize{bin $B_{\x}$}] at (\xscale*1.5+\xscale*3*\x , \height) {};
	\node [label=south:\scriptsize{($B_{\x}^{start},B_{\x}^{end})=(\a,\b)$}] at (\xscale*1.5+\xscale*3*\x , \height-\increment) {};
	\node [label=south:\scriptsize{($B_{\x}^{first},B_{\x}^{last})=(\c,\d)$}] at (\xscale*1.5+\xscale*3*\x , \height-2*\increment) {};
  }

  % Draw the segments
  \def\positions{
	0.0/1.7/0.2/0/south,
 	0.6/1.5/0.4/1/south,
	0.7/1.3/0.6/2/north,
	1.9/3.7/0.4/3/north,
 	2.5/7.5/0.6/4/north,
	2.8/4.6/0.2/5/south,
	3.9/5.5/0.4/6/south,
 	4.1/5.8/0.8/7/north,
	4.8/6.2/0.2/8/south,
	6.5/9.4/0.4/9/south,
	7.0/7.8/0.2/10/south,
	8.3/11/0.2/11/south,
	9.3/11.5/0.6/12/north,
	10.4/12/0.4/13/south,
	11.6/11.9/0.2/14/south}

  \def\yscale{2.18}
  \def\yoffset{0.10}
  \tikzstyle{every node} = [draw,fill=black,inner sep=1.0pt]
  \foreach \a/\b/\y/\z/\p in \positions {
	\ifthenelse{\equal{\p}{south}}{
  	\node [label=south:\footnotesize{$l_{\z}$}] (start) at (\xscale*\a,\y*\yscale+\yoffset) {}; 
        }{
  	\node [label=north:\footnotesize{$l_{\z}$}] (start) at (\xscale*\a,\y*\yscale+\yoffset) {}; 
        }
        \node (end) at (\xscale*\b,\y*\yscale+\yoffset) {}; 
        \draw (start) -- (end);
  }

\end{tikzpicture}  
    \caption{Example indexing of line segments into bins.}
   \label{fig:example_index}
\end{figure}

Figure~\ref{fig:example_index} shows and example for a database with 14
entry segments along the time axis with a total temporal extent of 12 
(segments are simply shown as non-overlapping horizontal lines as we do not
depict their spatial dimensions or orientations).  The temporal extent of
the database is logically divided into 4 bins, and for each bin we indicate
the $B_{}^{start}$,$B_{}^{end}$, $B_{}^{first}$ and $B_{}^{last}$ values.
For instance, bin $B_1$ contains the three entry segments with $t^{start}$
in the [$3,6)$ interval, i.e., $l_6$,$l_7$ and $l_8$.  Therefore, $B_{1}^{first} =
6$ and $B_{1}^{last} = 8$. Among the three entry segments in bin $B_1$,
$l_8$ has the highest $t^{end}$ value at 6.2. Therefore, $B_{1}^{start} = 3$ and $B_{1}^{end} = 
6.2$.

%\todo[inline]{MG: In practice, I indexed the bins in an R-tree because I
%couldn't figure out how to binary search in the context of a range.  The
%R-tree search just does one search with one query 1-D MBB. I figured this
%was sufficient, since even the stupid search I did which scanned all of the
%bins took a negligible amount of time.}

Given the database and set of bins, we consider a query set $Q$.
We first sort the query segments by non-decreasing $t_{start}$
values in $O(|Q| \log |Q|)$ time, which gives the temporal extent of the query
(the combined temporal extent of the query segments). We then determine the
set of (contiguous) bins that temporally overlap the temporal extent of the
query. We do this determination in $O(\log m)$ time by using an index-tree
in which we store the bins' temporal extents.  Given this set of bins,
$\mathcal{B}$, we compute $first = \min_{B\in \mathcal{B}} B_{j}^{first}$
and $last = max_{B\in \mathcal{B}} B_{j}^{last}$ in $O(1)$ time.
 We thus obtain $E_Q = \{l_i \in D | first \leq i
\leq last\}$, the set of the candidate entry segments that may be part of
the result set. Each query segment must then be compared to
each candidate segment in $E_Q$. We term each such a comparison an
\emph{interaction}, and we have a total of $|Q| \times |E_Q|$ interactions.

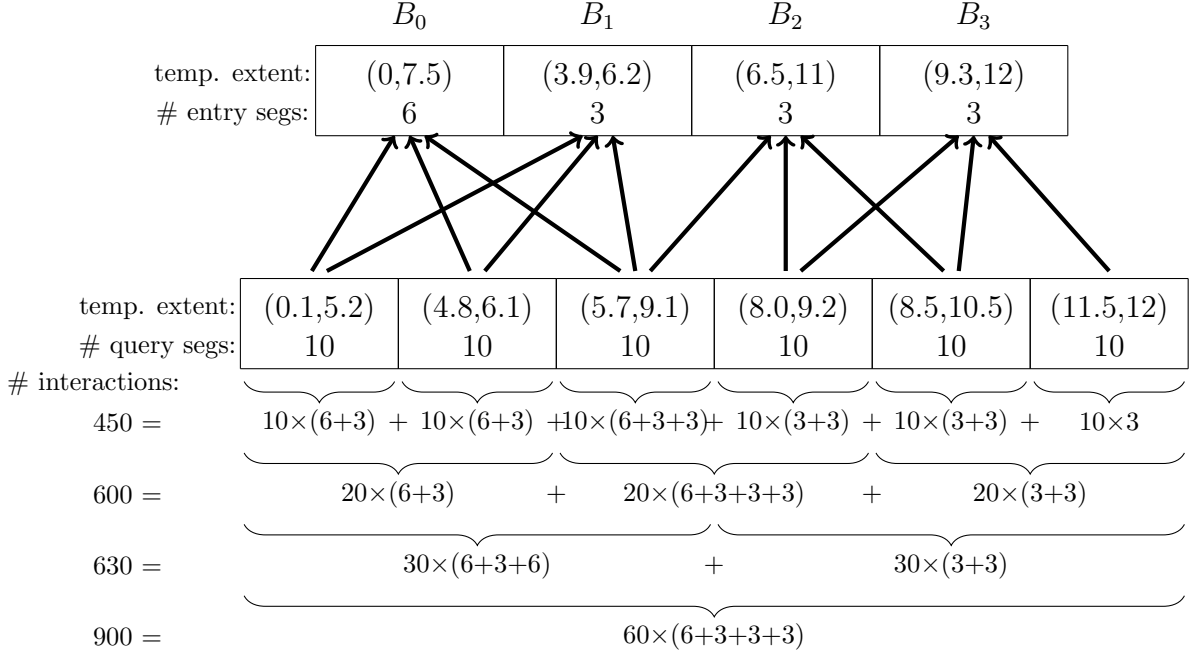
\begin{figure}[t]
\centering
  \begin{tikzpicture}
\usetikzlibrary{arrows}
\usetikzlibrary{decorations.pathreplacing}

\def\bins{
	0/0/7.5/0/5,
	1/3.9/6.2/6/8,
	2/6.5/11/9/11,
	3/9.3/12/12/14
	}

\def\binwidth{2.5}
\def\height{1.2}
\def\voffset{0.8}
\def\hoffset{1.0}
\def\topbin{3}
\foreach \x/\a/\b/\c/\d in \bins {
  	\draw (\hoffset+\binwidth*\x,\topbin) -- (\hoffset+\binwidth*\x+\binwidth,\topbin);
  	\draw (\hoffset+\binwidth*\x,\topbin) -- (\hoffset+\binwidth*\x,\topbin+\height);
  	\draw (\hoffset+\binwidth*\x+\binwidth,\topbin) -- (\hoffset+\binwidth*\x+\binwidth,\topbin+\height);
  	\draw (\hoffset+\binwidth*\x,\topbin+\height) -- (\hoffset+\binwidth*\x+\binwidth,\topbin+\height);
	\node at (\hoffset+\binwidth*\x + \binwidth/2,\topbin+\height+\voffset-0.4) {$B_\x$};
	\node at (\hoffset+\binwidth*\x + \binwidth/2,\topbin+\height-0.4) {(\a,\b)};
	\node at (\hoffset+\binwidth*\x + \binwidth/2,\topbin+\height-\voffset-0.1) {\pgfmathparse{int(round(\d-\c+1))}\pgfmathresult};
}
\node at (\hoffset-1.1,\topbin+\height-0.4) {\footnotesize{temp. extent:}};
\node at (\hoffset-1.1,\topbin+\height-\voffset-0.1) {\footnotesize{\# entry segs:}};

\def\batches{ % num queries, %t_start, %t_end$}
	0/10/0.1/5.2,
	1/10/4.8/6.1,
	2/10/5.7/9.1,
	3/10/8.0/9.2,
	4/10/8.5/10.5,
	5/10/11.5/12}

\def\topb{-0.1}
\def\widthb{2.1}
\def\heightb{1.2}
\def\voffsetb{0.8}
\foreach \x/\a/\b/\c in \batches {
  	\draw (\widthb*\x,\topb) -- (\widthb*\x+\widthb,\topb);
  	\draw (\widthb*\x,\topb) -- (\widthb*\x,\topb+\heightb);
  	\draw (\widthb*\x+\widthb,\topb) -- (\widthb*\x+\widthb,\topb+\heightb);
  	\draw (\widthb*\x,\topb+\heightb) -- (\widthb*\x+\widthb,\topb+\heightb);
	\node at (\widthb*\x + \widthb/2,\topb+\heightb-0.4) {(\b,\c)};
	\node at (\widthb*\x + \widthb/2,\topb+\heightb-\voffsetb-0.1) {\a};
}
\node at (-1.1,\topb+\heightb-0.4) {\footnotesize{temp. extent:}};
\node at (-1.1,\topb+\heightb-\voffsetb-0.1) {\footnotesize{\# query segs:}};

\draw [->,ultra thick] (0.5*\widthb-0.1,\topb+\heightb+0.1) -- (\hoffset+0.5*\binwidth-0.2,3);
\draw [->,ultra thick] (0.5*\widthb+0.1,\topb+\heightb+0.1) -- (\hoffset+1.5*\binwidth-0.2,3);

\draw [->,ultra thick] (1.5*\widthb-0.1,\topb+\heightb+0.1) -- (\hoffset+0.5*\binwidth,3);
\draw [->,ultra thick] (1.5*\widthb+0.1,\topb+\heightb+0.1) -- (\hoffset+1.5*\binwidth,3);

\draw [->,ultra thick] (2.5*\widthb-0.2,\topb+\heightb+0.1) -- (\hoffset+0.5*\binwidth+0.2,3);
\draw [->,ultra thick] (2.5*\widthb,\topb+\heightb+0.1) -- (\hoffset+1.5*\binwidth+0.2,3);
\draw [->,ultra thick] (2.5*\widthb+0.2,\topb+\heightb+0.1) -- (\hoffset+2.5*\binwidth-0.2,3);

\draw [->,ultra thick] (3.5*\widthb-0.1,\topb+\heightb+0.1) -- (\hoffset+2.5*\binwidth,3);
\draw [->,ultra thick] (3.5*\widthb+0.1,\topb+\heightb+0.1) -- (\hoffset+3.5*\binwidth-0.2,3);

\draw [->,ultra thick] (4.5*\widthb-0.1,\topb+\heightb+0.1) -- (\hoffset+2.5*\binwidth+0.2,3);
\draw [->,ultra thick] (4.5*\widthb+0.1,\topb+\heightb+0.1) -- (\hoffset+3.5*\binwidth,3);

\draw [->,ultra thick] (5.5*\widthb,\topb+\heightb+0.1) -- (\hoffset+3.5*\binwidth+0.2,3);

%%%%%%%%%%
\def\eps{0.05}
\def\vspacing{0.95}
\foreach \x/\a/\b/\c in \batches {
	\draw [decorate,decoration={brace,amplitude=8pt,mirror},xshift=0pt,yshift=-2pt] (\x*\widthb+\eps,\topb-0.1) -- (\x*\widthb+\widthb-\eps,\topb-0.1);
 	\pgfmathparse{Mod(\x,2)==0?1:0}
    	\ifnum\pgfmathresult>0
		\draw [decorate,decoration={brace,amplitude=8pt,mirror},xshift=0pt,yshift=-2pt] (\x*\widthb+\eps,\topb-0.1-\vspacing) -- (\x*\widthb+2*\widthb-\eps,\topb-0.1-\vspacing);
	\fi
 	\pgfmathparse{Mod(\x,3)==0?1:0}
    	\ifnum\pgfmathresult>0
		\draw [decorate,decoration={brace,amplitude=8pt,mirror},xshift=0pt,yshift=-2pt] (\x*\widthb+\eps,\topb-0.1-2*\vspacing) -- (\x*\widthb+3*\widthb-\eps,\topb-0.1-2*\vspacing);
	\fi
 	\pgfmathparse{Mod(\x,6)==0?1:0}
    	\ifnum\pgfmathresult>0
		\draw [decorate,decoration={brace,amplitude=8pt,mirror},xshift=0pt,yshift=-2pt] (\x*\widthb+\eps,\topb-0.1-3*\vspacing) -- (\x*\widthb+6*\widthb-\eps,\topb-0.1-3*\vspacing);
	\fi
}

\node at (-1.95,\topb-.2) {\footnotesize{\# interactions}:};
\node at (-1.5,\topb-.1-0.65*\vspacing) {\footnotesize{450 =}};
\node at (-1.5,\topb-.1-1.65*\vspacing) {\footnotesize{600 =}};
\node at (-1.5,\topb-.1-2.65*\vspacing) {\footnotesize{630 =}};
\node at (-1.5,\topb-.1-3.65*\vspacing) {\footnotesize{900 =}};

\node at (0*\widthb + .5*\widthb,\topb-.1-0.65*\vspacing) {\footnotesize{10$\times$(6+3)}};
\node at (0*\widthb +  1*\widthb,\topb-.1-0.65*\vspacing) {\footnotesize{+}};
\node at (1*\widthb + .5*\widthb,\topb-.1-0.65*\vspacing) {\footnotesize{10$\times$(6+3)}};
\node at (1*\widthb +  1*\widthb,\topb-.1-0.65*\vspacing) {\footnotesize{+}};
\node at (2*\widthb + .5*\widthb,\topb-.1-0.65*\vspacing) {\footnotesize{10$\times$(6+3+3)}};
\node at (2*\widthb +  1*\widthb,\topb-.1-0.65*\vspacing) {\footnotesize{+}};
\node at (3*\widthb + .5*\widthb,\topb-.1-0.65*\vspacing) {\footnotesize{10$\times$(3+3)}};
\node at (3*\widthb +  1*\widthb,\topb-.1-0.65*\vspacing) {\footnotesize{+}};
\node at (4*\widthb + .5*\widthb,\topb-.1-0.65*\vspacing) {\footnotesize{10$\times$(3+3)}};
\node at (4*\widthb +  1*\widthb,\topb-.1-0.65*\vspacing) {\footnotesize{+}};
\node at (5*\widthb + .5*\widthb,\topb-.1-0.65*\vspacing) {\footnotesize{10$\times$3}};

\node at (0*\widthb + 2*.5*\widthb,\topb-.1-1.65*\vspacing) {\footnotesize{20$\times$(6+3)}};
\node at (0*\widthb + 2*\widthb,\topb-.1-1.65*\vspacing) {\footnotesize{+}};
\node at (2*\widthb + 2*.5*\widthb,\topb-.1-1.65*\vspacing) {\footnotesize{20$\times$(6+3+3+3)}};
\node at (2*\widthb + 2*\widthb,\topb-.1-1.65*\vspacing) {\footnotesize{+}};
\node at (4*\widthb + 2*.5*\widthb,\topb-.1-1.65*\vspacing) {\footnotesize{20$\times$(3+3)}};

\node at (0*\widthb + 3*.5*\widthb,\topb-.1-2.65*\vspacing) {\footnotesize{30$\times$(6+3+6)}};
\node at (0*\widthb + 3*\widthb,\topb-.1-2.65*\vspacing) {\footnotesize{+}};
\node at (3*\widthb + 3*.5*\widthb,\topb-.1-2.65*\vspacing) {\footnotesize{30$\times$(3+3)}};

\node at (3*\widthb,\topb-.1-3.65*\vspacing) {\footnotesize{60$\times$(6+3+3+3)}};

\end{tikzpicture}  
    \caption{Example matching between query batches and entry bins.}
   \label{fig:index_lookup}
\end{figure}

Some of the computed interactions are certain to not add
the candidate to the result set.  For instance, in the context of the
example in Figure~\ref{fig:index_lookup}, consider a query with a single
query segment with temporal extent [8,10]. The query segment overlaps the
temporal extents of bin $B_2$ and bin $B_3$, meaning that it will be
compared to $l_9,\ldots,l_{14}$. And yet, $l_{10}$, $l_{13}$, and $l_{14}$
cannot overlap the query segment's temporal extent.  More generally, the
larger $|Q|$, the larger the number of interactions,
and thus the larger the number of ``wasteful'' interactions. 
This observation provides a motivation for processing query segments in
relatively small batches (in addition to the fact that using batches is
necessary because memory for the result set must
be allocated statically--see Section~\ref{sec:problem_outline}).

Figure~\ref{fig:index_lookup} shows an example of how using batches decreases the number of interactions. 
The top of
the figure shows the same set of bins as in Figure~\ref{fig:example_index},
without showing the entry segments but indicating temporal extents and
numbers of entry segments. The bottom of the figure shows a set of 60 query
segments partitioned into 6 batches. For each query batch we indicate its
temporal extent and its number of segments. An arrow is drawn between a
query batch and an entry bin if the query segments in the batch must be
compared to the entry segments in the bin. Below the batches we show the
number of interactions necessary to process the
query.  For instance, batch 2 has a temporal extent (5.7,9,1), which
overlaps with the temporal extents of bins $B_0$, $B_1$, and $B_2$, which
contain 6, 3, and 3 entry segments, respectively. Therefore, the processing
of batch 2 entails 10$\times$(6+3+3)=120 interactions. Using 10-segment
query batches results in a total of 450 interactions. The figure also shows the
number of interactions using larger batches.  For instance, while processing
10-segment batch 2 requires 120 interactions and processing 10-segment batch
3 requires 60 interactions, processing the aggregate 20-segment batch leads
to 20$\times$(6+3+3+3) = 300 $>$ 180 interactions.  In this example, processing all query segments as
a single 60-segment batch would lead to 900 interactions,
twice the number of interactions when using 10-segment batches.
Processing each query segment individually (batch size of 1)
minimizes the number of segment interactions. However, processing each batch
incurs the non-negligible overhead of sending data from the host to the GPU
and of invoking a GPU kernel. Consequently, one of the questions we
investigate in this work is that of choosing batch sizes that minimize
query response time.

Note that more advanced indexing methods could be envisioned that inform
each individual entry what queries temporally overlap to avoid computing
wasteful interactions.  However, these methods lead to more data transfer
overhead between the host and the GPU.  Preliminary results show that in
practice this overhead leads to significant increases in total response time
in spite of reducing the number of wasteful interactions.

%\begin{figure}[t]
%\centering
  %\includegraphics[width=1.0\textwidth]{index_lookup.pdf}  
    %\caption{index lookup}
   %\label{fig:index_lookup}
%\end{figure}

Given the above, we propose the following general approach for implementing
query distance threshold searches on the GPU. The entry segments in $D$,
sorted by non-decreasing $t^{start}$ values are stored contiguously in the
global memory of the GPU. The database index, i.e., the description of the
bins, and the query segments in $Q$ (sorted by non-decreasing $t^{start}$
values) are stored in RAM on the host.  The query segments are partitioned
in batches (not necessarily of identical sizes).  For each batch, the index
range of the candidate entry segments is calculated using the bins. The
query segments in a batch and the index range, which is encoded as two
integers, are sent from the host to the GPU.  The candidate entry segments
are then compared to the query segments, generating a result set that is
returned to the CPU.  Our indexing method guarantees that these candidate
entry segments are stored contiguously in memory, which allows for
efficient memory transfers between global, local and private memory spaces
on the GPU, and which reduces the use of branches.  The search is complete
when all batches have been processed in this manner. In
Section~\ref{sec:search_alg} we describe our GPU kernel for performing the
search, while in Section~\ref{sec:set_algs} we describe approaches for
picking good batch sizes.

\section{Search Algorithm}\label{sec:search_alg}
%\todo[inline]{Add in the response times of other kernel designs.}

In this section we describe an algorithm, \algname, that performs distance
threshold searches using the indexing and search techniques outlined in
Section~\ref{sec:traj_indexing}. This algorithm is implemented as a GPU
kernel using OpenCL, and optimized to use as few branch instructions as
possible.  To take advantage of the high number of hardware threads on the
GPU and of its fast context-switching we simply use one GPU thread for each
candidate entry segment.  Each thread then compares its candidate entry
segment to all query segments in the batch. Using $Q_{batch}$ to denote
a query batch, which is a subset of $Q$, each thread then computes
$|Q_{batch}|$ interactions. 
Another option would be to use one thread per query segment, but it runs
the risk of not fully utilizing all available hardware threads since $|Q_{batch}|$,
unlike $|D|$, is not large.  More specifically, the kernel
takes as input: (i)~$Q_{batch}$, an array of query trajectory segments sorted by
$t^{start}$ values; (ii)~\emph{firstCandidate}, the index in $D$ of the
first candidate entry segment (recall that the entire database $D$ is
stored on the GPU once and for all); (iii)~\emph{numCandidates}, the number of
candidate entry segments; (iv)~$d$, the threshold distance; and
(v)~\emph{setID},  a global index that keeps track of the location in memory where the next result set item should be written. $Q_{batch}$, \emph{firstCandidate}, \emph{numCandidates} are computed on the host before executing the kernel and transferred to the GPU 
along with $d$ and \emph{setID}.  The kernel returns
a set of time intervals annotated by trajectory ids.

The pseudo-code of the kernel is shown in
Algorithm~\ref{alg:GPU_trajdistsearch}.  The threads in OpenCL are numbered
using a global id (\emph{gid}$\geq 0$). As we use only \emph{numCandidates}
threads, all threads with \emph{gid} larger than \emph{numCandidates} do
not participate in the computation
(lines~\ref{algline.pre_start}-\ref{algline.pre_end}).  Once the result set
is initialized to the empty set (line~\ref{algline.init}), the relevant
candidate segment is copied into the thread's private memory (variable
\emph{entrySegment}) line~\ref{algline.entry_into_private_memory}. The
algorithm then loops over all query segments to compute interactions
between the candidate segment and the query segments
(line~\ref{algline.query_loop}). Given the candidate segment and the
current query segment, function \emph{temporalIntersection()} generates new
candidate and query segments that span the same time interval
(line~\ref{algline.interpolate}).  The algorithm then computes the interval
of time during which these two segments are within a distance $d$ of each
other (line~\ref{algline.time}), which involves computing the coefficients
of and solving a degree two polynomial~\cite{Guting2010}.  If this interval
is non-empty, then \emph{setID} is incremented atomically
(line~\ref{algline.atomicinc}).  The interval is annotated with the
trajectory id and added to the result set (line~\ref{algline.found}). The
full result set is returned once all interactions have been computed.

\begin{algorithm}
\caption{Pseudo-code for the \algname kernel algorithm.}
\label{alg:GPU_trajdistsearch}
\begin{algorithmic}[1]

\begin{small}
\Procedure{\algname}{$Q_{batch}$, {firstCandidate}, {numCandidates}, $d$, {setID}}

\State {gid} $\leftarrow$ getGlobalId()\label{algline.pre_start}
\If {{gid}$\geq${numCandidates}}
\State return
\EndIf\label{algline.pre_end}

\State {resultSet} $\leftarrow$ $\emptyset$\label{algline.init}

%\ForAll{querySegmentMBB in Q.MBBSet}\label{algline.outerloop}
%\State CandidateSet $\leftarrow$ T.Search(querySegmentMBB, d)\label{algline.search}

\State {entrySegment} $\leftarrow$ $D$[{firstCandidate} + {gid}]\label{algline.entry_into_private_memory}

\ForAll{{querySegment} $\in$ $Q_{batch}$}\label{algline.innerloop}\label{algline.query_loop}

\State ({entrySegment}, {querySegment})  $\leftarrow$ temporalIntersection(\label{algline.interpolate}
\Statex ~~~~~~~~~~~~~~ ~~~~~~~{entrySegment},{querySegment})

\State {timeInterval}  $\leftarrow$ calcTimeInterval(\label{algline.time}
\Statex ~~~~~~~~~~~~~~~~~~~~~{entrySegment},{querySegment},$d$)
\If {{timeInterval} $\neq$ $\emptyset$}
\State {resultID} $\leftarrow$ atomic\_inc({setID})\label{algline.atomicinc}
\State {resultSet}[{resultID}]  $\leftarrow$ {resultsSet}[{resultID}] $\cup$ {timeInterval}\label{algline.found}
\EndIf
\EndFor
%\EndFor

\State return {resultSet}[0:setID]\label{algline.return}
\EndProcedure

\end{small}
\end{algorithmic}
\end{algorithm}

%\todo[inline]{HC: I am not sure I follow/undersatnd the 3rd sentence. It
%should say how much memory we do allocate for the result set, and what we
%do if we don't have enough. \\MG: The 3rd sentence was confusing, as it had
%a double negative.  I included the amount of memory we allocate now. W.r.t.
%what we do if we don't have enough memory: in practice, if the result set
%is large, that means we have a large batch size, which is inefficient.  So,
%if I'm running an experiment which demands a large number of queries per
%batch (depending on how the sets are made), and one needs more memory than
%I have allocated, it just crashes and I don't consider the result, because
%I know that trial was far from the best response time.  For example,
%consider using a batch size of 1000.  You can see from the results that
%this is going to be far from the best batch size.  Some of the
%deterministic setsplit (\setsplitA) set sizes can be quite large and cause
%it to crash as well.}

The size of the result set for a kernel invocation could be as high as the
number of interactions, $|Q_{batch}|\times$\emph{numCandidates}. However,
in practice, only a small fraction of the interactions add to the result
set.  Since memory for the result set must be allocated statically, in our
experiments we conservatively allocate enough memory for a result set with
as many items as there are entries in the dataset. In practice, one could
allocate much less memory, and in the rare cases in which more memory is
needed one would simply re-attempt the kernel execution with more allocated
memory.

\section{Generation of Query Batches}\label{sec:set_algs}

\begin{figure}[t]
\centering
  \includegraphics[width=0.5\textwidth]{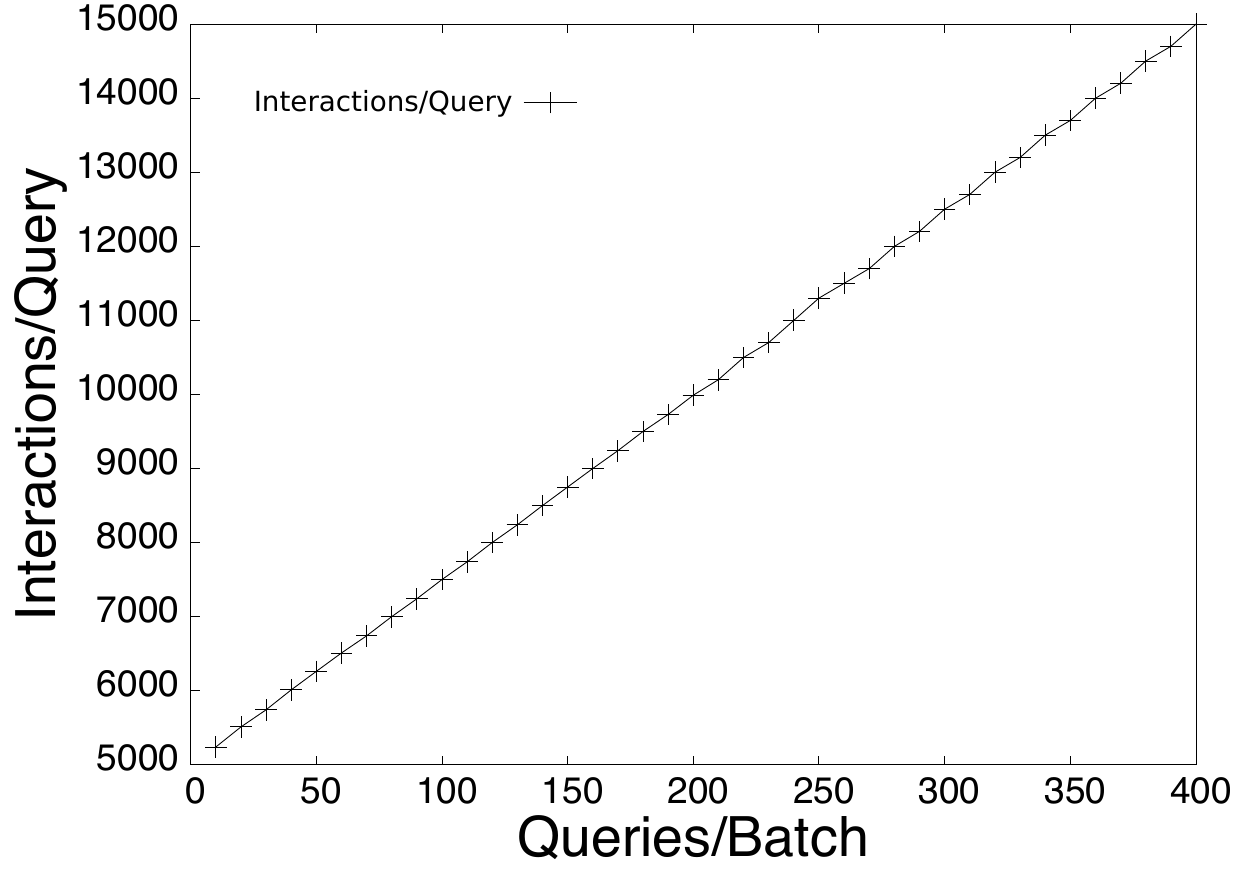}  
    \caption{The number of interactions per query vs. batch size, for the \galaxy dataset ($10^6$ entry trajectory segments), with 40,000 query trajectory segments.}
   \label{fig:interactions_vs_batch_size}
\end{figure}

As explained in Section~\ref{sec:traj_indexing}, an important question is
that of choosing appropriate, perhaps optimal, query batch sizes.  Using
small batches increases the total number of kernel invocations, and each
such invocation has a non-negligible overhead.  Conversely, using large
batches increases the number of wasteful interactions. This increase was
demonstrated in Figure~\ref{fig:index_lookup} as an example.
Figure~\ref{fig:interactions_vs_batch_size} shows the actual number of
interactions per query segment vs. the number of queries per batch for a
total of 40,000 query segments over the \galaxy dataset with $10^6$ entry
segments (see Section~\ref{sec:datasets} for details on the datasets and queries
used for experimental evaluations). As expected, the number of computed
interactions, and thus the number of wasteful interaction computations,
grows almost perfectly linearly with the batch size.

Beyond the above trade-off between between high overhead and high numbers
of wasteful interactions, the temporal properties of the dataset should
guide how one groups the query segments into batches.  For instance,
consider the example shown in Figure~\ref{fig:index_lookup}.  The first and
second sets of 10 query segments both overlap with the same set of entry
segments (entry bins $B_0$ and $B_1)$.  Therefore, it is likely a good idea
to group the first 20 query segments in a batch, since no extra wasteful
interactions will be generated by this grouping (a total of 180
interactions).  Consider now grouping together the third set of 10 query
segments (which requires 10$\times$(6+3+3)=120 interactions) and the fourth
set of 10 query segments (which requires 10$\times$(3+3)=60 interactions).
This grouping leads to 20$\times$(6+3+3+2)=280 interactions, for
280-120-60=100 extra wasteful interactions. As seen in this example, while
picking a good batch size is important, it is also important to group
together query segments that together do not overlap too many entry bins.
In light of these considerations in what follows we propose several
algorithms to group query segments into batches.

%\item If too large a number of queries are sent per kernel invocation, then
%there will be many wasteful interactions, as the fraction of $Q$ that
%overlap temporally with $E$ will decrease.  For example, consider two
%scenarios: a) $x$ queries are sent per kernel invocation and overlap $n/e$
%entries, where $e$ is the total number of entries and b) $2x$ queries are
%sent, and overlap $r/2e$ entries.  However, $r<2n$, and this effect
%increases with increasing $Q$.  For an increasing set size of $Q$, which we
%denote $P$, there are fewer relevant interactions, and the GPU performs
%more wasteful computations.
%
%\item The temporal properties of the datasets will also influence the best
%number of queries to send per batch.
%
%\end{enumerate}

\subsection{Periodic}

A simple
approach to define query batches is to pick a single batch size, $s$, as in
Figure~\ref{fig:index_lookup}. Each consecutive subsets of $s$ queries in
$Q$ are then grouped together in a batch, for a total of $b = |Q|/s$
batches and thus $b$ kernel invocations.  We call this approach \periodic.

\subsection{SetSplit}

We propose a class of $O(|Q|^2)$ algorithms, called \setsplit,  that attempt
to group query segments together in a way that reduces wasteful
interactions while yielding batches that are not too small.

The first algorithm, \setsplitF (Algorithm~\ref{alg:setsplitA}), produces a
specified number of batches. More specifically, \setsplitF takes as input a
set of query segments, $Q$, and the number of batches to generate,
$numBatches$, and outputs a set of batches. The first step 
is to create a list of batches, $B$, in which each element is a single
query segment (line~\ref{algsetsplitA:init}). While the number of batches
is larger than $numBatches$, the algorithm iteratively merges two adjacent
batches into a single batch (loop at line~\ref{algsetsplitA:loop}).  For
each possible such merge (loop at line~\ref{algsetsplitA:innerloop}), we
compute the sum of the numbers of interactions of two adjacent batches
(line~\ref{algsetsplitA:unmerged}) and the number of interactions of the
merge of these two batches (line~\ref{algsetsplitA:merged}). We determine
the potential merge operation that would lead to the smallest increase in
number of interactions (line~\ref{algsetsplitA:test}), keeping track of the
index of the first batch in that merge, $bestMerge$. We then replace batch
$bestMerge$ by a batch obtained by merging batch $bestMerge$ and batch
$bestMerge+1$, and remove batch $bestMerge+1$ (lines~\ref{algsetsplitA:do1}
and~\ref{algsetsplitA:do2}).
 The algorithm returns an array built from list $B$.

%\begin{algorithm}
%\caption{Pseudo-code for the \setsplitA algorithm.}
%\label{alg:setsplitA}
%\begin{algorithmic}[1]
%
%\begin{small}
%\Procedure{\setsplit}{R, int M, int P}
%\State Merges $\leftarrow$ M-P \label{algsetsplitA:merges}
%\For {Merges}
%\ForAll {M} 
%\State id1 $\leftarrow$ FindCandidateMerge() \label{algsetsplitA:candidatemerge1}
%\State id2 $\leftarrow$ FindCandidateMerge() \label{algsetsplitA:candidatemerge2}
%\State IntUnmerged $\leftarrow$ CalcInts(id1) + CalcInts(id2) \label{algsetsplitA:unmerged}
%\State IntMerged $\leftarrow$ calcMerged(id1,id2) \label{algsetsplitA:merged}
%\State $\epsilon$ $\leftarrow$ IntMerged-IntUnmerged \label{algsetsplitA:epsilon}
%
%\If{$\epsilon < \epsilon_{min}$} \label{algsetsplitA:epsilon_if}
%\State $\epsilon \leftarrow \epsilon_{min}$
%\State Mid1 $\leftarrow$ id1
%\State Mid2 $\leftarrow$ id2
%\EndIf
%\EndFor
%\State Merge(Mid1,Mid2) \label{algsetsplitA:merge_fn}
%\EndFor
%
%
%\EndProcedure
%
%\end{small}
%\end{algorithmic}
%\end{algorithm}

\begin{algorithm}
\caption{Pseudo-code for the \setsplitF algorithm.}
\label{alg:setsplitA}
\begin{algorithmic}[1]

\begin{small}
\Procedure{\setsplitF}{$Q$, $numBatches$}\label{algsetsplitA:init}
\State $B \leftarrow$ list($Q$)
\While {$|B| > numBatches$}\label{algsetsplitA:loop}
\State minDiff $\leftarrow$ $+\infty$
\For {$i = 0, \ldots, |B|-2$}\label{algsetsplitA:innerloop}
	\State numIntsUnmerged $\leftarrow$ numInts($B[i]$) + numInts($B[i+1]$)\label{algsetsplitA:unmerged}
	\State numIntsMerged $\leftarrow$ numInts(merge($B[i]$,$B[i+1]$))\label{algsetsplitA:merged}
	\If{numIntsMerged - numIntsUnmerged $<$ minDiff}\label{algsetsplitA:test}
		\State minDiff $\leftarrow$ numIntsMerged - numIntsUnmerged
		\State bestMerge $\leftarrow$ $i$
	\EndIf
\EndFor
\State $B[$bestMerge$]$ $\leftarrow$ merge($B[$bestMerge$]$,$B[$bestMerge+1$]$)\label{algsetsplitA:do1}
\State $B$.removeElementAt(bestMerge+1)\label{algsetsplitA:do2}
\EndWhile
\State \Return array($B$)
\EndProcedure

\end{small}
\end{algorithmic}
\end{algorithm}

A drawback of \setsplitF is that it can produce many small batches, and in
fact many batches that contain a single query segment, and thus lead to
high overhead.  Using a lower $numBatches$ value leads to more merge
operations and thus lower overhead. However, it is unclear how to pick the
best value for this parameter since it depends on the temporal properties
of the datasets.  To address these shortcomings, we propose another
algorithm, \setsplitMM, that generates batches while imposing constraints on
minimum and maximum batch sizes.

The pseudo-code of \setsplitMM is shown in Algorithm~\ref{alg:setsplitBC}.
\setsplitMM takes as input a set of query segments, $Q$, a lower bound on
the batch size, $min$, and an upper bound on the batch size, $max$. It
outputs a set of batches. The first phase of the algorithm 
(lines~\ref{algsetsplitBC:phase1_begin}-\ref{algsetsplitBC:phase1_end})
is similar to Algorithm~\ref{alg:setsplitA} but for the fact that
merges that would lead to a batch with more than $max$ query segments
are ignored (line~\ref{algsetsplitBC:ignore_too_big}).  The second phase of
the algorithm
(lines~\ref{algsetsplitBC:phase2_begin}-\ref{algsetsplitBC:phase2_end})
loops until no batch remains that contains fewer than $min$ query segments. For each
such batch, the algorithm considers the merge with the predecessor batch if any
(lines~\ref{algsetsplitBC:phase2_test1_begin}-\ref{algsetsplitBC:phase2_test1_end}),
and with the successor batch if any
(lines~\ref{algsetsplitBC:phase2_test2_begin}-\ref{algsetsplitBC:phase2_test2_end}).
The algorithm then performs the merge that leads to the smallest
increase in number of interactions (lines~\ref{algsetsplitBC:phase2_select_begin}-\ref{algsetsplitBC:phase2_select_end}). 
The algorithm returns an array built from list $B$.

We also consider an algorithm, \setsplitMax, that is a special case of
\setsplitMM with $min = 1$, i.e., with no constraint imposed on the minimum
batch size.

\newcommand{\algorithmicbreak}{\textbf{break}}
\newcommand{\Break}{\State \algorithmicbreak}
\newcommand{\algorithmiccontinue}{\textbf{continue}}
\newcommand{\Continue}{\State \algorithmiccontinue}

\begin{algorithm}
\caption{Pseudo-code for the \setsplitMM algorithm.}
\label{alg:setsplitBC}
\begin{algorithmic}[1]

\begin{small}
\Procedure{\setsplitMM}{$Q$, $min$, $max$}
\State $B \leftarrow$ list($Q$)\label{algsetsplitBC:phase1_begin}
\While {true}\label{algsetsplitBC:loop}
\State minDiff $\leftarrow$ $+\infty$
\For {$i = 0, \ldots, |B|-2$}
	\If{numSegments(merge($B[i]$,$B[i+1]$)) $>$ $max$}\label{algsetsplitBC:ignore_too_big}
		\Continue
	\EndIf
	\State numIntsUnmerged $\leftarrow$ numInts($B[i]$) + numInts($B[i+1]$)
	\State numIntsMerged $\leftarrow$ numInts(merge($B[i]$,$B[i+1]$))
	\If{numIntsMerged - numIntsUnmerged $<$ minDiff}
		\State minDiff $\leftarrow$ numIntsMerged - numIntsUnmerged
		\State bestMerge $\leftarrow$ $i$
	\EndIf
\EndFor
\If {minDiff = $+\infty$}
	\Break
\EndIf
\State $B[$bestMerge$]$ $\leftarrow$ merge($B[$bestMerge$]$,$B[$bestMerge+1$]$)
\State $B$.removeElementAt(bestMerge+1)
\EndWhile\label{algsetsplitBC:phase1_end}
\While {there exists $B[i]$ such that numSegments($b$) $<$ $min$}\label{algsetsplitBC:phase2_begin}
	\If {($i > 0$}\label{algsetsplitBC:phase2_test1_begin}
		\State numIntsLeft = numInts(merge($B[i-1]$,$B[i]$))
	\Else
		\State numIntsLeft = $\infty$
	\EndIf\label{algsetsplitBC:phase2_test1_end}
	\If {($i < |B|-1$}\label{algsetsplitBC:phase2_test2_begin}
		\State numIntsRight = numInts(merge($B[i]$,$B[i+1]$))
	\Else
		\State numIntsRight = $\infty$
	\EndIf\label{algsetsplitBC:phase2_test2_end}
	\If {numIntsLeft $<$ numIntsRight}\label{algsetsplitBC:phase2_select_begin}
		\State $B[i]$ $\leftarrow$ merge($B[i-1]$,$B[i]$)
		\State $B$.removeElementAt($i-1$)
	\Else
		\State $B[i]$ $\leftarrow$ merge($B[i]$,$B[i+1]$)
		\State $B$.removeElementAt($i+1$)
	\EndIf\label{algsetsplitBC:phase2_select_end}
\EndWhile\label{algsetsplitBC:phase2_end}
\State \Return array($B$)
\EndProcedure
\end{small}
\end{algorithmic}
\end{algorithm}

\subsection{GreedySplit}

In this section we present a class of $O(|Q|)$ algorithms, called
\greedysetsplit. Like \setsplit, \greedysetsplit also attempts to avoid
small batches and to reduce wasteful interactions, but with lower complexity.
The main idea behind \greedysetsplit is to first do all the ``free'' merges, i.e.,
those merges that do not increase the number of interactions, and then to
merge contiguous batches using a single pass through the set of batches.
The \greedysetsplitMin algorithm imposes a lower bound on the minimum batch size,
while the \greedysetsplitMax algorithm imposes an upper bound on the maximum batch
size. We consider a single constraint (either minimum or maximum batch sizes), as designing a \greedysetsplit algorithm that would
impose both constraints and terminates is difficult
(a batch that is too large may need to be broken into batches that may then be
too small). 
%\todo[inline]{HC: We should explain why we don't consider an algorithm that
%does both min and max\\MG: the reason is that if you first go through with
%say, a minimum criteria, and make sets, and then have a maximum criteria,
%then you have to go about breaking query sets up to satisfy the maximum
%criteria, which make smaller sets, potentially below the minimum, so both
%criteria can't be satisfied.  Ex: if you have a min batch size of 40, and a
%max of 50, and you end up with a batch with 70 elements, then 20 elements
%need to be removed to satisfy the maximum criteria, but then this breaks
%the minimum criteria}

\greedysetsplitMin takes as input 
a set of query segments, $Q$, and a lower bound on
the batch size, $bound$, and it outputs
 a set of batches. Its pseudo-code is shown
in Algorithm~\ref{alg:greedysetsplit}. In the first
phase of the algorithm
(lines~\ref{alggreedysplitA:phase1_begin}-\ref{alggreedysplitA:phase1_end}),
the algorithm traverses the set of batches, $B$, and merges two adjacent batches
if this merge does lead to an increase in number of interactions. 
In a second phase 
(lines~\ref{alggreedysplitA:phase2_begin}-\ref{alggreedysplitA:phase2_end}),
the algorithm iteratively merges a batch with its successor if the batch
contains fewer than $min$ query segments (line~\ref{alggreedysplitA:phase2_test}). In the \greedysetsplitMax, 
line~\ref{alggreedysplitA:phase2_test} is 
replaced by a ``\emph{numSegments(}$B[i]$)$>$\emph{bound}'' test and the if and else clauses are swapped.
The algorithm returns an array built from list $B$.

\begin{algorithm}
\caption{Pseudo-code for the \greedysetsplitMin algorithm.}
\label{alg:greedysetsplit} 
\begin{algorithmic}[1]
\begin{small}
\Procedure{\greedysetsplit}{$Q$, $bound$}
\State $B \leftarrow$ list($Q$)
\State $i$ $\leftarrow$ 0\label{alggreedysplitA:phase1_begin}
\While {$i < |B|-1$}
	\If {numInts(merge($B[i]$,$B[i+1]$)) = numInts($B[i]$) + numInts($B[i+1]$)}
		\State $B[i]$ $\leftarrow$ merge($B[i]$,$B[i+1]$)
		\State $B$.removeElementAt($i+1$)
	\Else
		\State $i$ $\leftarrow$ $i+1$
	\EndIf
\EndWhile\label{alggreedysplitA:phase1_end}

\State $i$ $\leftarrow$ 0\label{alggreedysplitA:phase2_begin}
\While {$i < |B|-1$}
	\If {numSegments($B[i]$) $<$ $bound$}\label{alggreedysplitA:phase2_test}
		\State $B[i]$ $\leftarrow$ merge($B[i]$,$B[i+1]$)
		\State $B$.removeElementAt($i+1$)
	\Else
		\State $i$ $\leftarrow$ $i+1$
\EndIf
\EndWhile\label{alggreedysplitA:phase2_end}
\State \Return array($B$)
\EndProcedure

\end{small}
\end{algorithmic}
\end{algorithm}

\section{Experimental Evaluation}\label{sec:exp_eval}

\subsection{Datasets}
\label{sec:datasets}

We evaluate our query processing scheme using several datasets, all of
which are 4-dimensional (3 spatial dimensions, 1 temporal dimension). Our
first dataset, called \galaxy, contains trajectories of stars moving in the
Milky Way's gravitational field (as generated by the astronomy application
described in Section~\ref{sec:project_motivation}).  More specifically,
this dataset contains $10^6$ trajectory segments, corresponding to 2,500
trajectories of 400 timesteps each.  Since each trajectory has the same
number of timesteps and about the same temporal extent, the temporal
profile of active trajectories is roughly uniform. However, since our
approach relies on temporal data partitioning, we also generate synthetic
datasets with various temporal profiles of the number of active
trajectories. Such profiles occur, for instance, in datasets of vehicular
traffic trajectories with nighttime, daytime, and rush hour patterns.  

\begin{figure}[!htp]
\centering

        \subfigure[\randomuniform]{
            \includegraphics[width=0.40\textwidth]{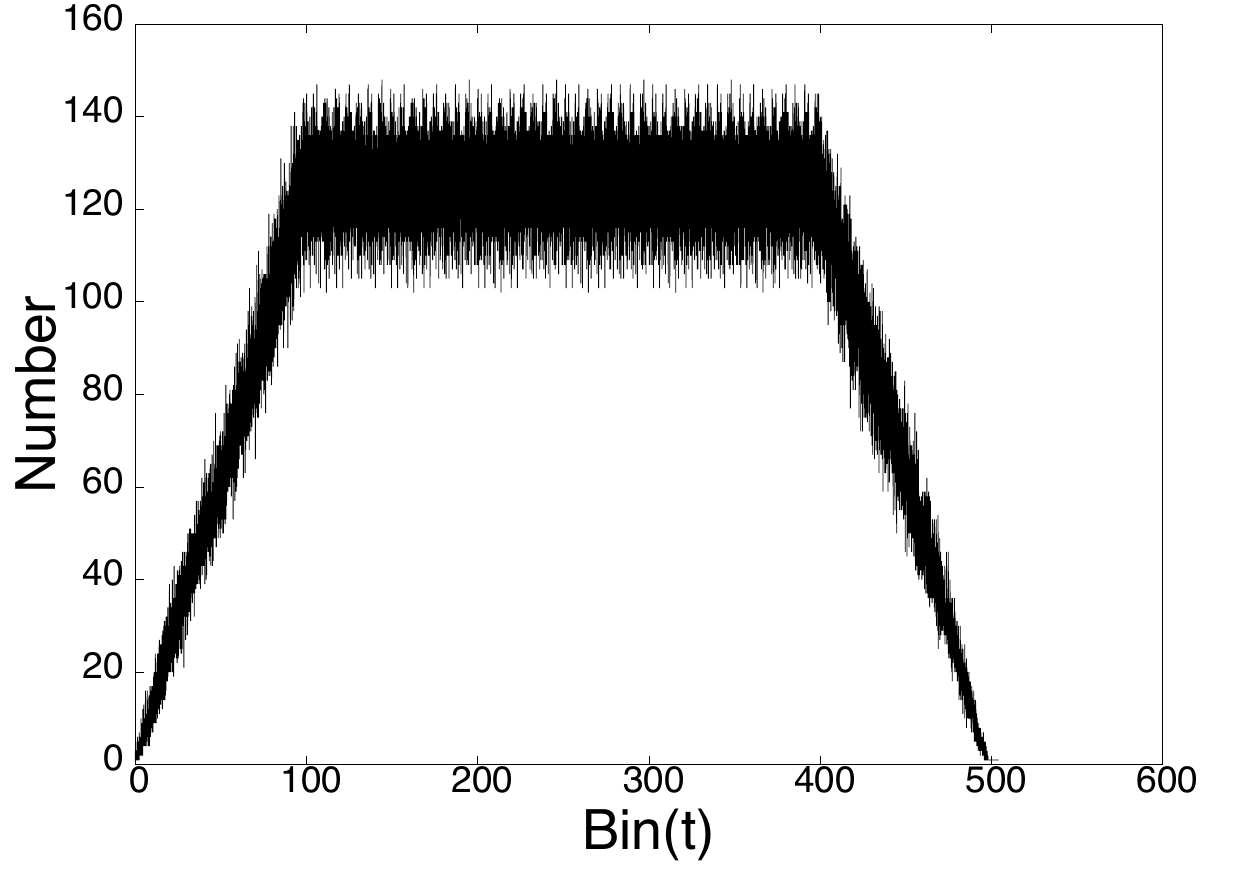}
        }
        \subfigure[\randomnormal]{
            \includegraphics[width=0.40\textwidth]{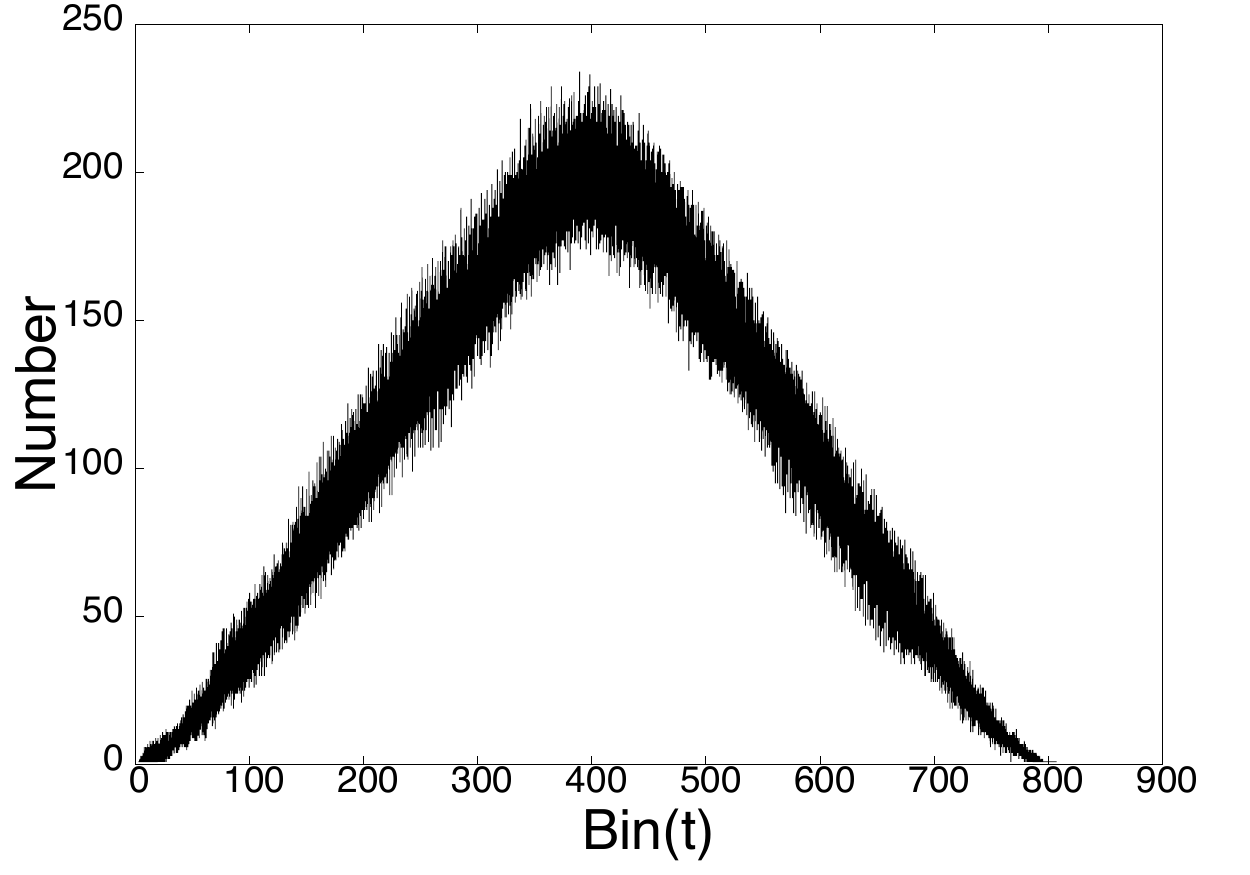}
        }
        \subfigure[\randomnormalfive]{
            \includegraphics[width=0.40\textwidth]{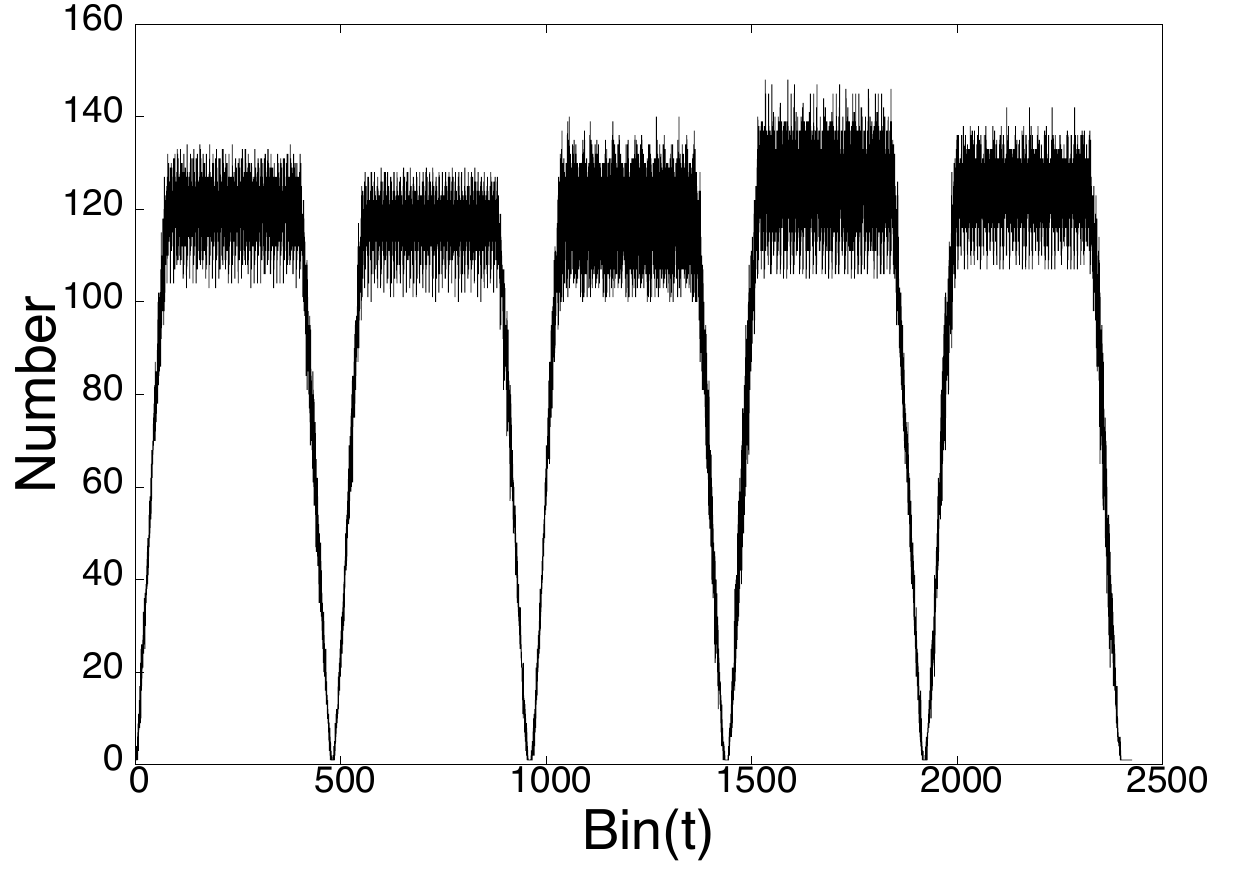}
        }
         \subfigure[\randomexp]{
            \includegraphics[width=0.40\textwidth]{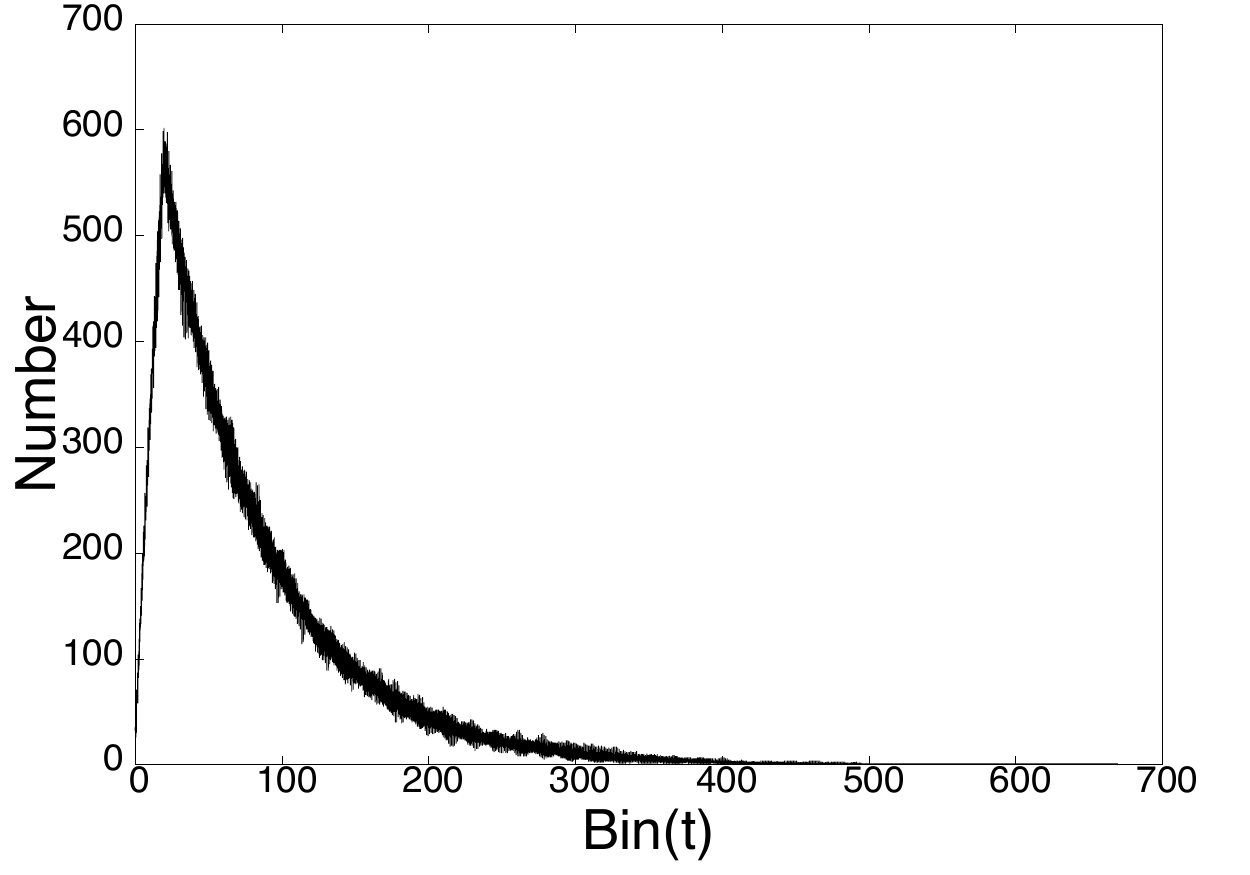}
        }
                \subfigure[\galaxy]{
            \includegraphics[width=0.40\textwidth]{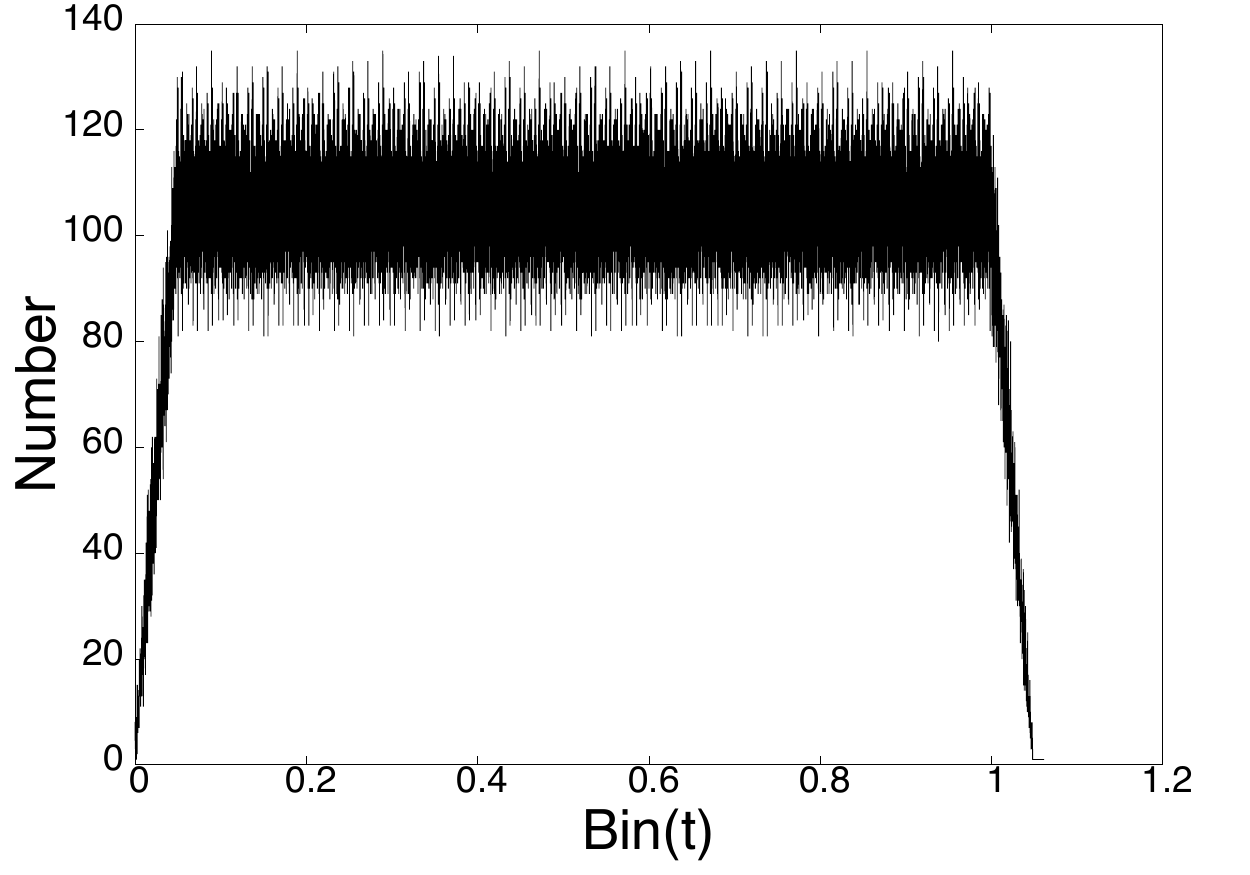}
  		}
          \subfigure[Sample trajectories from \galaxy.]{
            \includegraphics[width=0.40\textwidth]{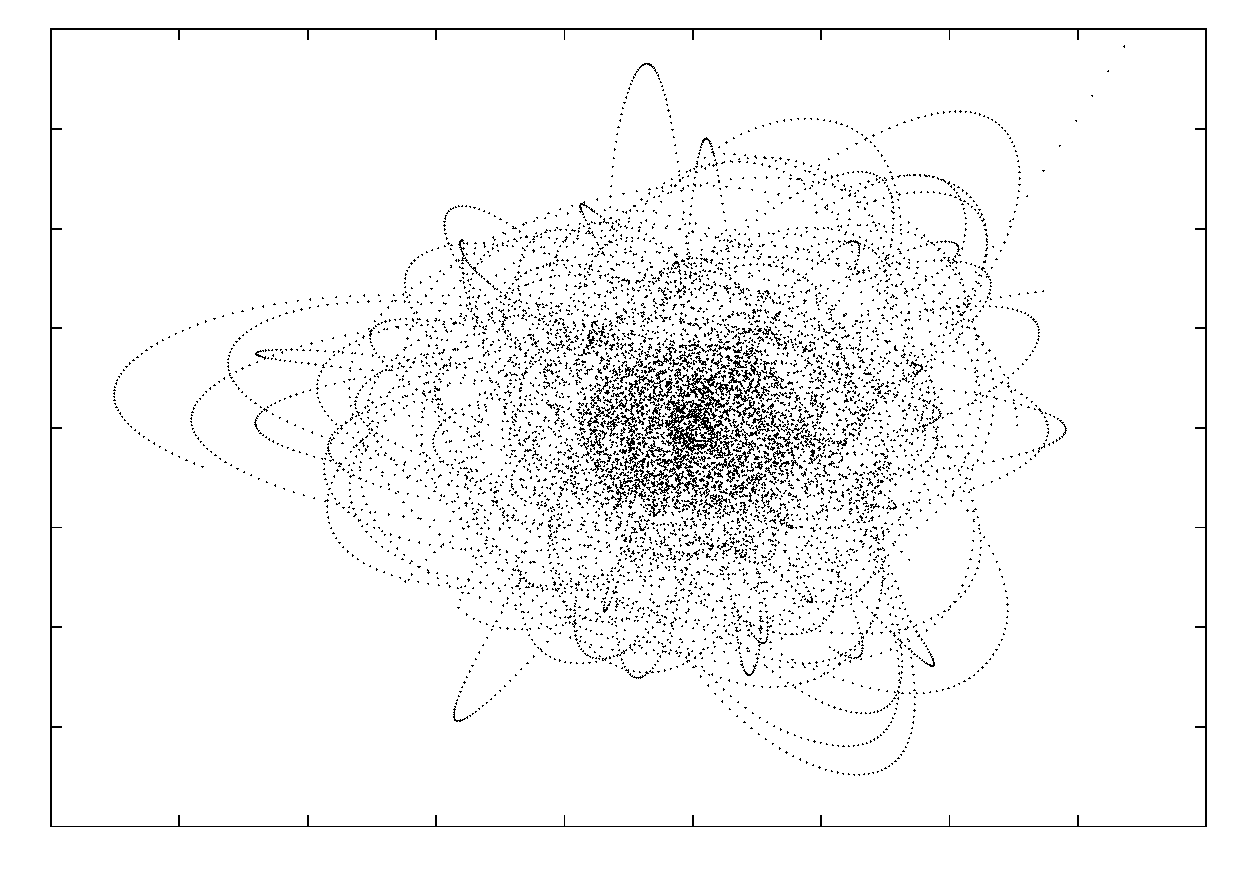}
  		}
    \caption{Temporal distributions of active entry trajectory line segments in the datasets are shown in panels (a) through (e).  The time corresponding to the midpoint of the bin is plotted on the horizontal axis, and the number of segments in the bin is shown on the vertical axis. Panel (f) shows a sample of trajectories from the \galaxy dataset as projected on the x-y plane.}
   \label{fig:dataset_histograms}
\end{figure}

Our random datasets are based on trajectories of bodies subjected to
Brownian motion.  The \randomuniform dataset consists of 400-timestep
trajectories whose start times are sampled from a uniform distribution over
the [0,100] interval. The \randomnormal dataset is similar but uses a
normal distribution to generate start times, with a mean of 200 and
standard deviation of 200, truncated to the [0,400] interval.  The \randomexp
dataset consists of trajectories with numbers of timesteps that are sampled
from an exponential distribution with lambda=1/70, truncated to the
[2,1000] interval, with start times sampled from a uniform distribution
over a the [0,20] timestep interval. The \randomnormalfive dataset is
generated but one of 5 different normal distributions is randomly selected
when generating trajectories. This dataset thus exhibits distinct active
and inactive phases,  as occurs in datasets such as the vehicular traffic
example above. The various parameter values for generating these datasets
were picked so as to produce distinct patterns of numbers of entry segments
assigned to entry bins. These patterns are shown in
Figure~\ref{fig:dataset_histograms}~(a)-(e) for each dataset. In addition,
Figure~\ref{fig:dataset_histograms}~(f) shows a sample of trajectories for
the \galaxy dataset.  Table~\ref{tab:datasets} lists the
number of trajectories and of entry segments in the datasets. The datasets are made publicly available \cite{datasets}.

\begin{table}
\centering
\caption{Characteristics of Datasets}
\begin{tabular}{|c|c|c|} \hline
Dataset&Trajec.&Entries\\ \hline
\hline
\randomuniform&2,500&997,500\\ 
\hline
\randomnormal&2,500&1,000,000\\
\hline
\randomnormalfive&2,500&1,000,000\\
\hline
\randomexp&10,000&684,329\\
\hline
\galaxy&2,500&1,000,000\\ 
\hline\end{tabular}
\label{tab:datasets}
\end{table}

\subsection{Experimental Methodology}\label{sec:exp_method}

The GPU-side implementation is developed in OpenCL, and the host-side
implementation is developed in C++.  The host-side implementation is
executed on one of the 6 cores of a dedicated 3.46 Ghz Intel Xeon W3690
processor with 12 MB L3 cache, while the GPU side runs on an Nvidia Tesla
C2075 card. We measure query response times averaged over 3 trials
(standard deviation over the trials is negligible).  In all experiments the
number of entry bins in our index is set to 10,000.

In our experiments, we utilize the following trajectory searches:
\begin{itemize}
\item S1: From the \galaxy dataset, 100 trajectories are processed with $d=1$, and with a total of 40,000 query line segments.
\item S2: From the \galaxy dataset, 100 trajectories are processed with $d=5$, and with a total of 40,000 query line segments.
\item S3: From the \randomuniform dataset, 100 trajectories are processed with $d=5$, and with a total of 39,900 query line segments.
\item S4: From the \randomuniform dataset, 100 trajectories are processed with $d=25$, and with a total of 39,900 query line segments.
\item S5: From the \randomnormal dataset, 100 trajectories are processed with $d=50$, and with a total of 40,000 query line segments. 
\item S6: From the \randomnormal dataset, 100 trajectories are processed with $d=150$, and with a total of 40,000 query line segments.
\item S7: From the \randomnormalfive dataset, 100 trajectories are processed with $d=50$, and with a total of 40,000 query line segments. 
\item S8: From the \randomnormalfive dataset, 100 trajectories are processed with $d=150$, and with a total of 40,000 query line segments.
\item S9: From the \randomexp dataset, 1000 trajectories are processed with $d=50$, and with a total of 52,044 query line segments.
\item S10: From the \randomexp dataset, 1000 trajectories are processed with $d=100$, and with a total of 69,881 query line segments.
\end{itemize}

For a given entry set the response time depends on the query set. This is
because the spatiotemporal features of the queries determine the number of
interactions to compute.  However, we find that in all of our results,
regardless of the query set, all of our candidate algorithms lead to response
times with a relatively narrow range. For instance, for the \galaxy dataset
and 10 different sample query sets, and for a query distance $d=5$, the
relative response time difference between the fastest and the slowest
algorithm is only 1.99\% on average and at most 3.08\%.  While the ranking
of the particular algorithms may differ from one query set to another,
these variations do not translate to large response time differences.
Consequently, we only present results for a single query set.

\subsection{Sequential Implementation and Multi-core OpenMP}\label{sec:previous_work_exp_eval}

\begin{figure}[t]
\centering
  \includegraphics[width=0.5\textwidth]{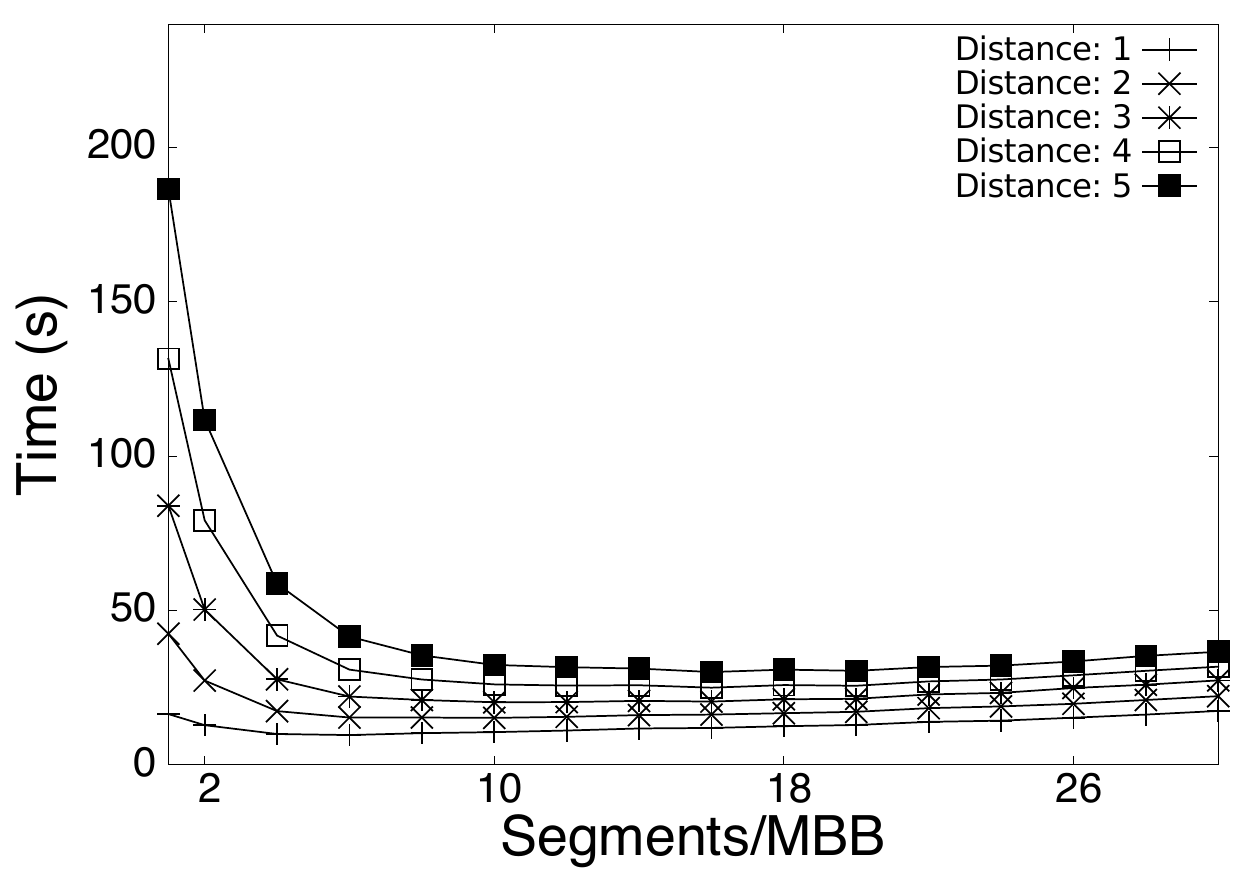}  
    \caption{Response time vs. segments per MBB ($r$) for the \galaxy dataset with the same query set outined in S1, but with $d=1,2,3,4,5$.}
   \label{fig:galaxy_MSB_implementation}
\end{figure}

\begin{figure}[t]
\centering
  \includegraphics[width=0.5\textwidth]{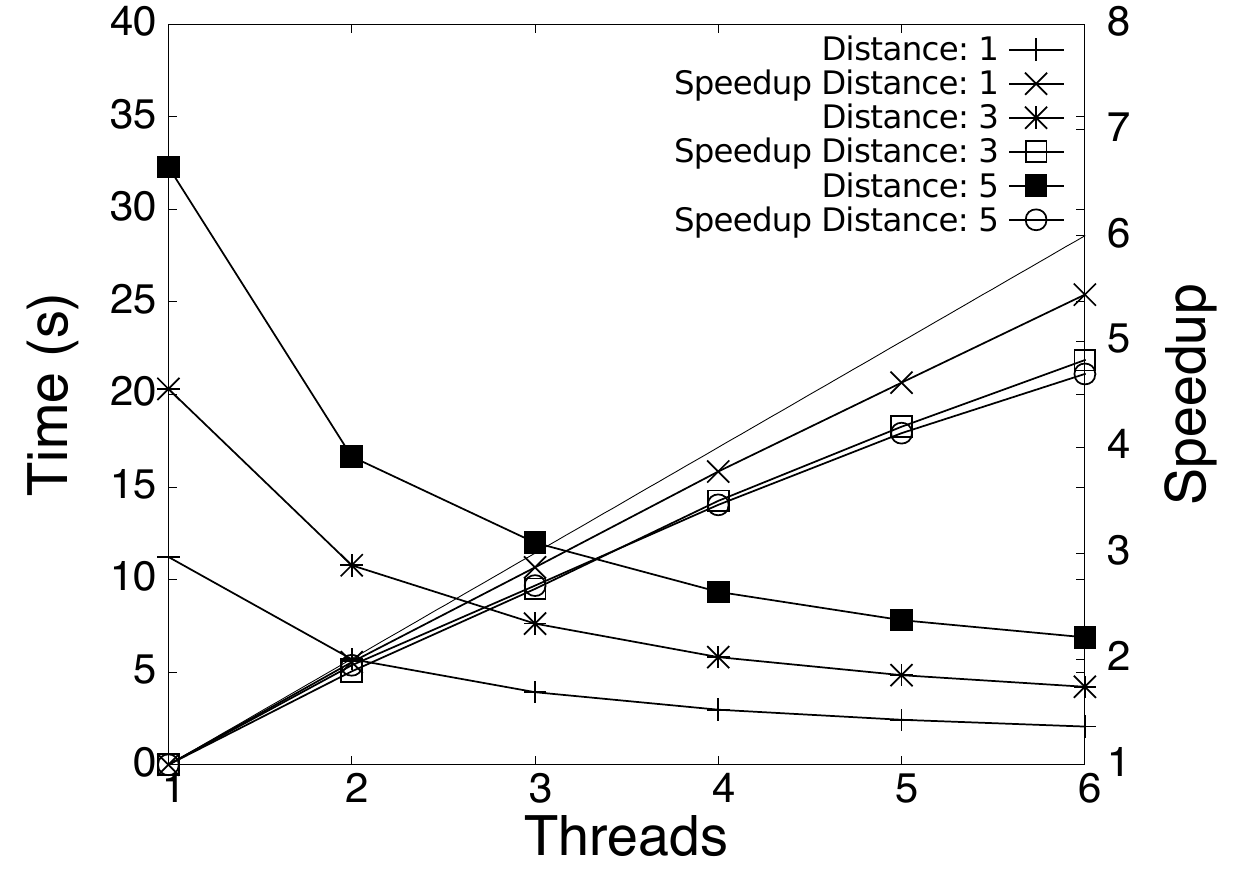}  
    \caption{Response time vs. number of threads for the \galaxy dataset with the same query set outined in S1 with $d=1,3,5$ and $r=12$.}
   \label{fig:galaxy_threads}
\end{figure}

While this work focuses on distance threshold searches on the GPU, in previous
work we have developed sequential and parallel CPU
implementations~\cite{Gowanlock2014, Gowanlock2014b}. The CPU
implementation uses an R-tree index to store trajectory segments inside
MBBs.  One interesting question is how to ``split'' a trajectory, i.e.,
deciding on which (contiguous) segments should be stored in the same MBBs.
In~\cite{Gowanlock2014} we propose a trajectory splitting strategy that achieves a trade-off
between the number of entries in the index, the volume of the space
occupied by the MBBs, and the computational cost of candidate trajectory
segment processing.  Figure~\ref{fig:galaxy_MSB_implementation} shows
average query response time vs. the number of segments indexed per MBB, for
the \galaxy dataset for query distances $d=1,\ldots,5$, when executed
on the host described in Section~\ref{sec:exp_method}. In this case,
indexing $12$ segments per MBB yields the lowest average response time.
See~\cite{Gowanlock2014} for further results and more details.
The sequential CPU implementation can be easily parallelized using OpenMP.
Figure~\ref{fig:galaxy_threads} shows the response time vs. the number of
threads for the \galaxy dataset, with $12$ trajectory segments per MBB.
On our 6-core host parallel efficiency is high (78\%-90\%), with parallel
speedup between 4.69 and 5.44 with 6 threads. In what follows, we 
draw some comparisons between the performance of this CPU-only
implementation and the performance of our GPU implementation.

\subsection{Performance Evaluation}
%\todo[inline]{In comparison to the R-tree implementation, the change in
%distance does not linearly influence response time on the GPU.  So, the GPU
%is probably better when there's lots of clustered interactions vs the
%R-tree type implementations.}

Let us first compare the performance of \algname to that of the sequential
and parallel CPU implementations described in the previous section.
We find that the relative
performance of the GPU and CPU implementation is consistent across
experimental scenarios. Let us consider experimental scenario S2 and only
the \periodic algorithm for creating batches (using a batch size of 120).  Our GPU implementation
achieves average response time as low as 2.08 s, while for the same
experimental scenario our sequential CPU implementation (using the best
number of query segments per MBB for that scenario) leads to an average
response time of 31.62 s.  Our GPU implementation thus gives a speedup of
15.2 over the sequential CPU implementation. When compared to the OpenMP
parallel CPU implementation, the average response time is 6.88 s, so that our
GPU implementation achieves a speedup of 3.3. While these results are tied
to the hardware characteristics of our experimental platform, we conclude
that a GPU implementation of distance threshold query processing is
worthwhile and can yield substantial improvement over a CPU-only version.

%Furthermore, initial results in \cite{Gowanlock2014b}
%shows that the response time of the GPU implementation can be improved by
%overlapping computation with communication between host and device, and
%that we might expect a performance gain of $\sim20\%$, if we utilize 3 work
%queues and kernel instances.

%RW exp d100 S10
%best periodic: 4.59919
%best of the query split algs: 4.52239
%4.59919-4.52239=0.0768

\begin{figure}[!htp]
\centering

        \subfigure[]{
            \includegraphics[width=0.45\textwidth]{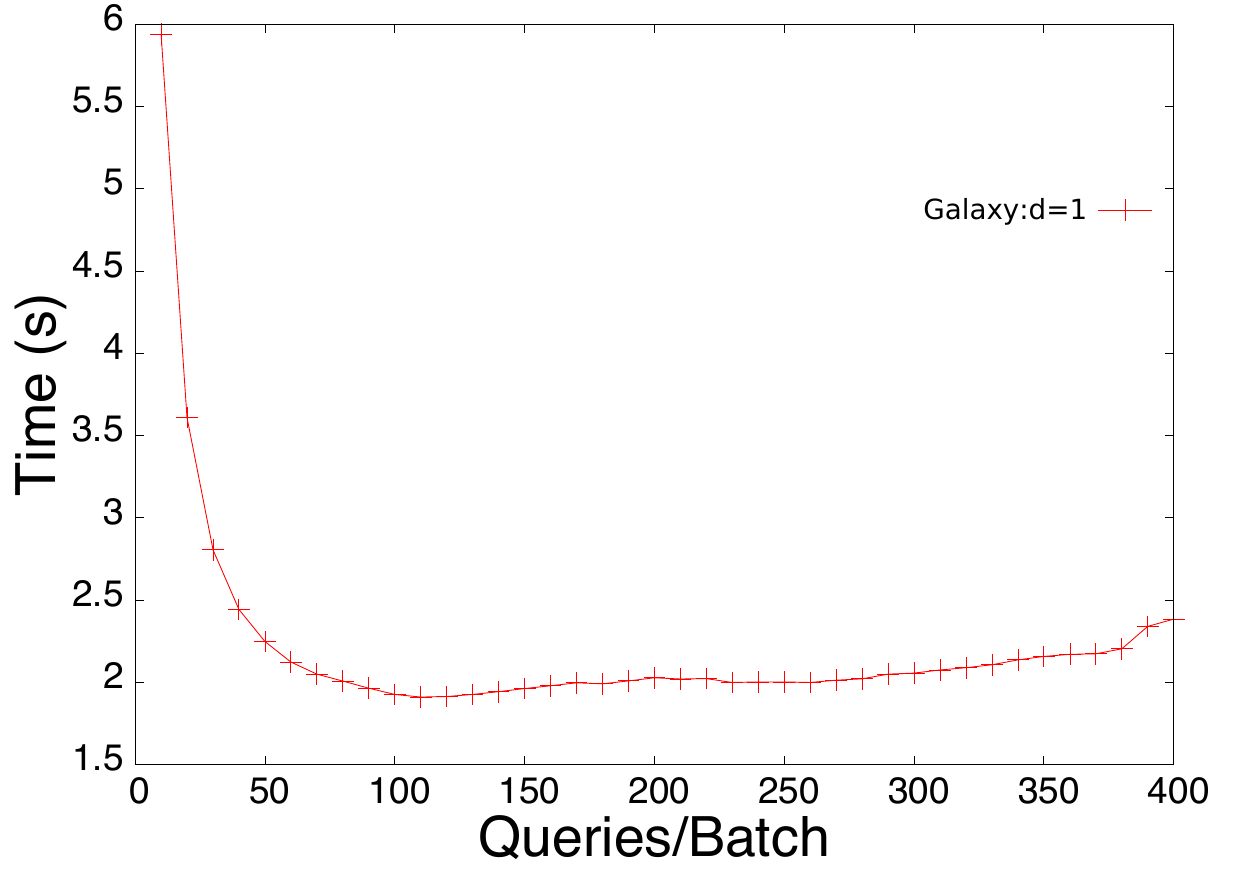}
        }
        \subfigure[]{
            \includegraphics[width=0.45\textwidth]{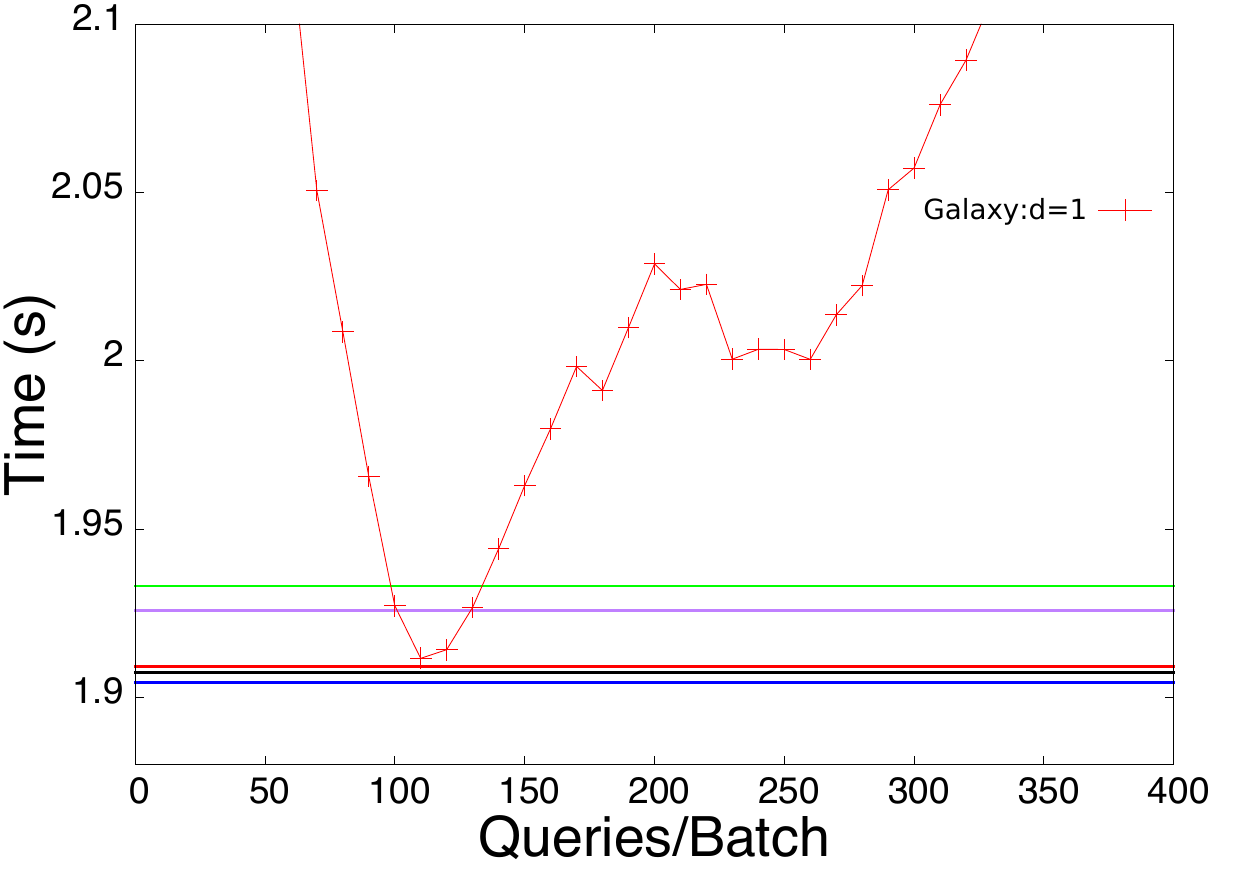}
        }
        \subfigure[]{
            \includegraphics[width=0.45\textwidth]{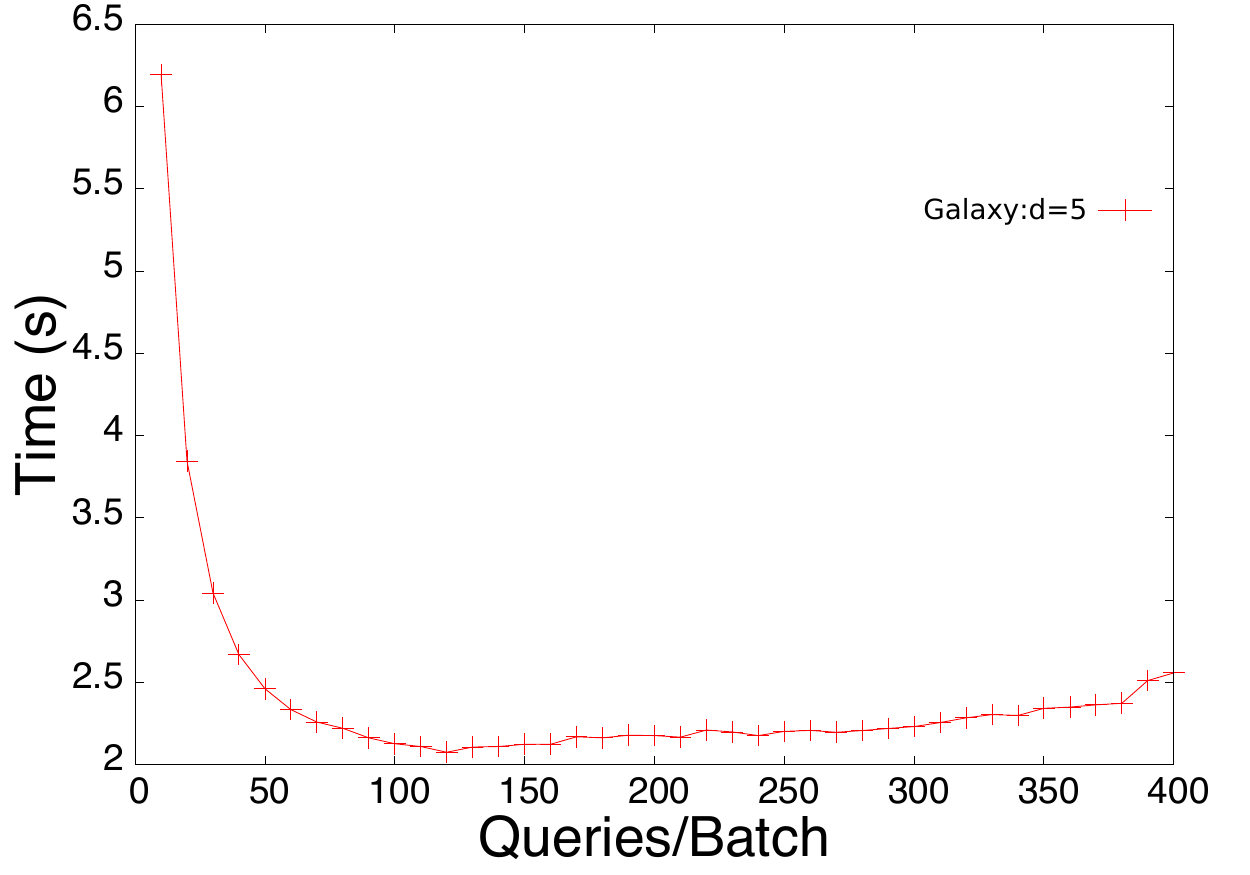}
        }
         \subfigure[]{
            \includegraphics[width=0.45\textwidth]{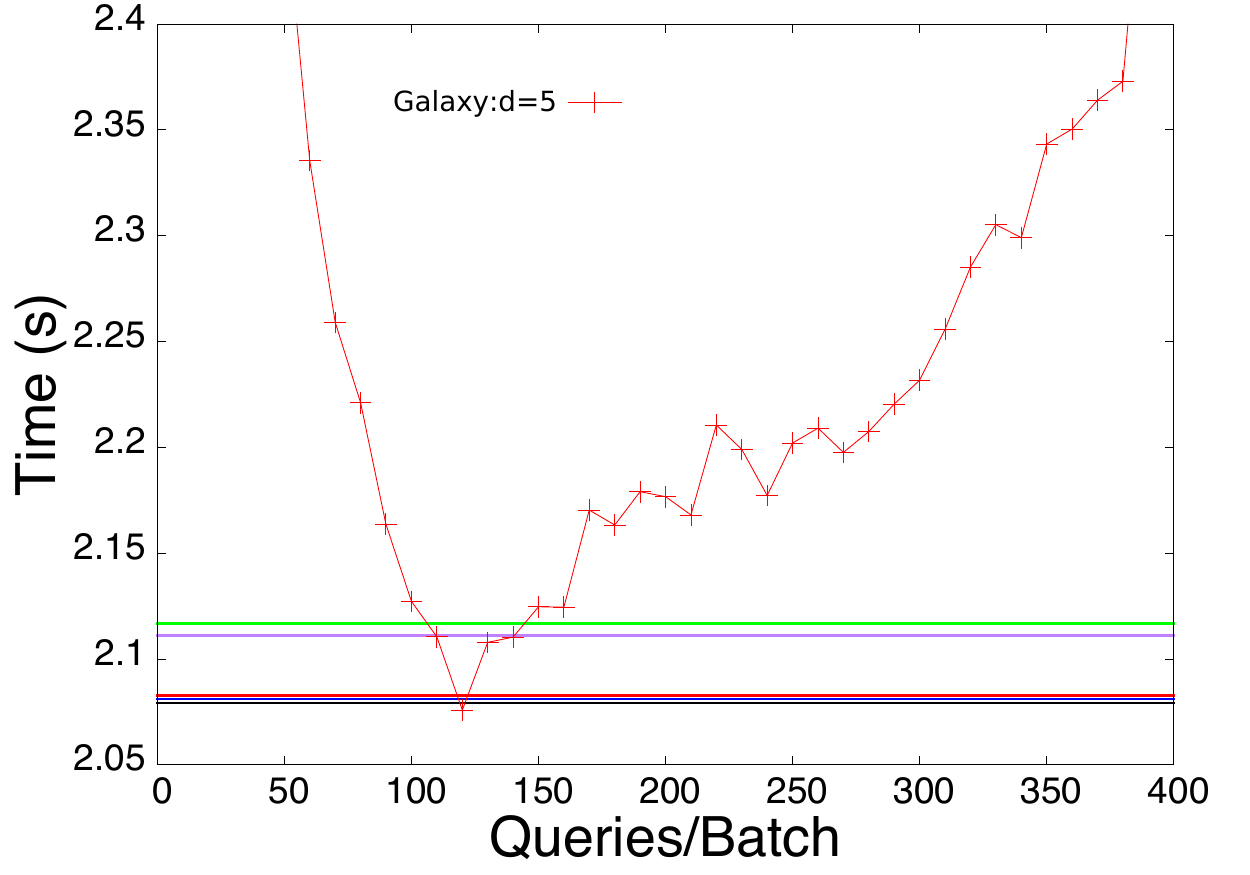}
        }
          
    \caption{Response time vs. queries/batch ($s$) for the periodic query batch method for S1 (a) and S2 (c) (\galaxy dataset). Panels (b) and (d) correspond to zoomed in versions of (a) and (c) respectively, to highlight the minimum response times. The colored lines correspond to the best response time from the query splitting algorithms, where \setsplitF is purple, \setsplitMax is green, \setsplitMM is red, \greedysetsplitMin is blue, and \greedysetsplitMax is black.}
   \label{fig:query_split_galaxy}
\end{figure}

\begin{figure}[!htp]
\centering

        \subfigure[]{
            \includegraphics[width=0.45\textwidth]{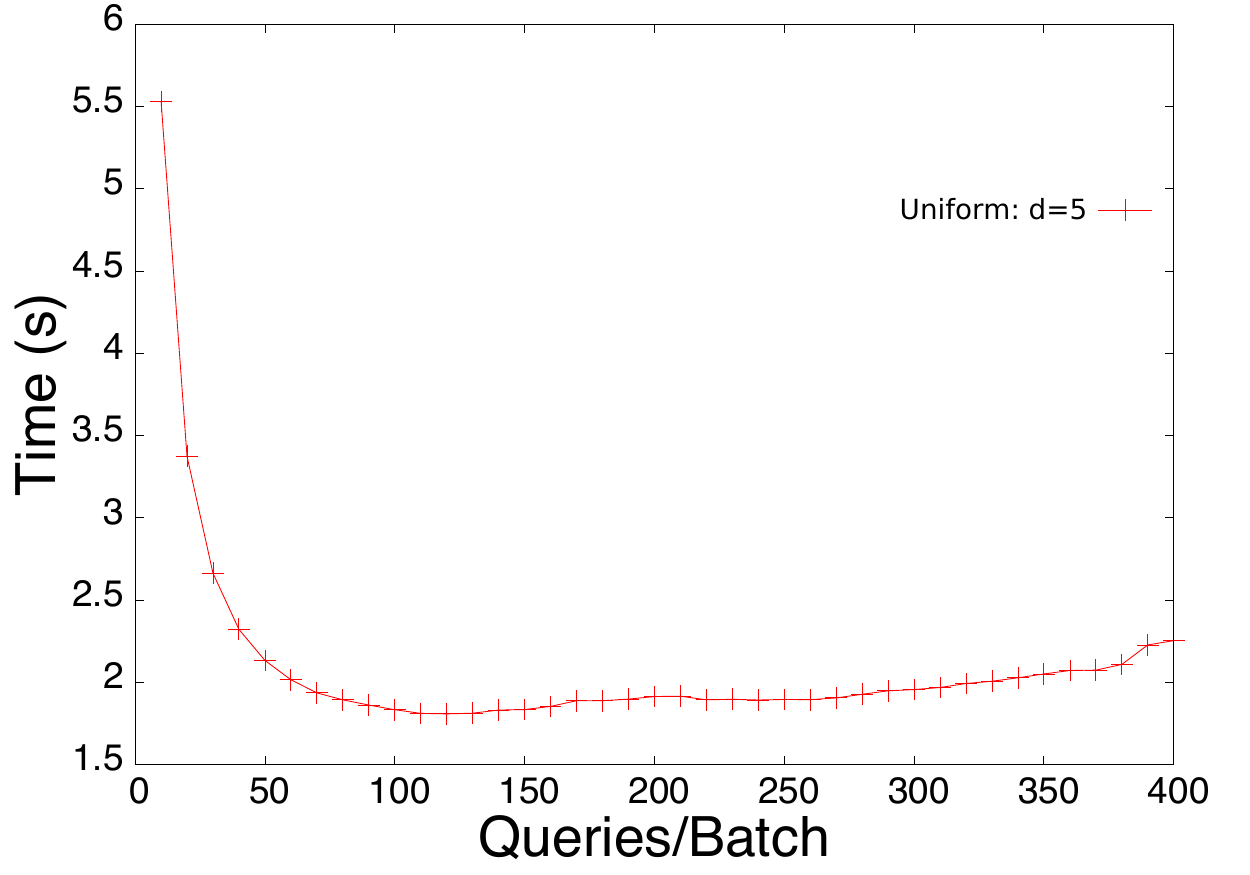}
        }
        \subfigure[]{
            \includegraphics[width=0.45\textwidth]{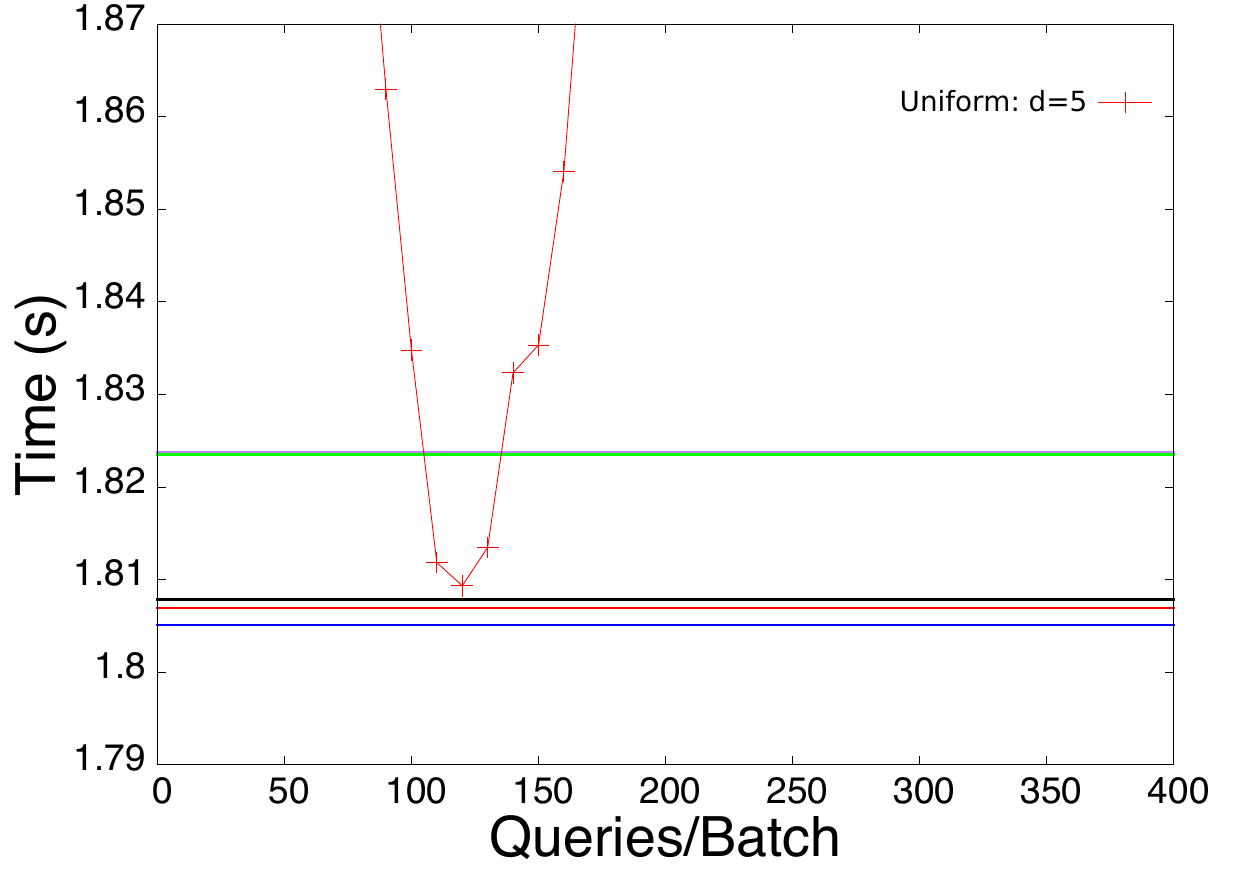}
        }
        \subfigure[]{
            \includegraphics[width=0.45\textwidth]{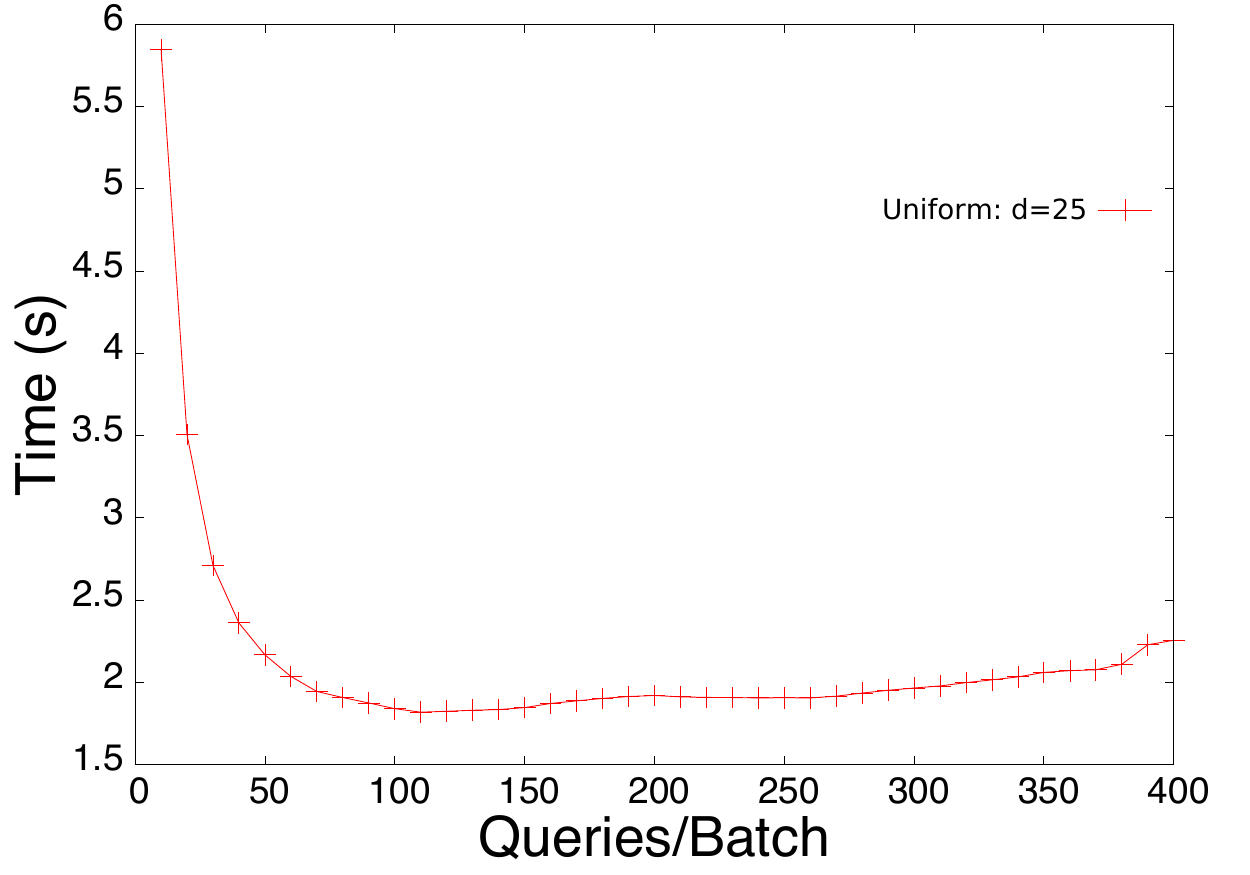}
        }
         \subfigure[]{
            \includegraphics[width=0.45\textwidth]{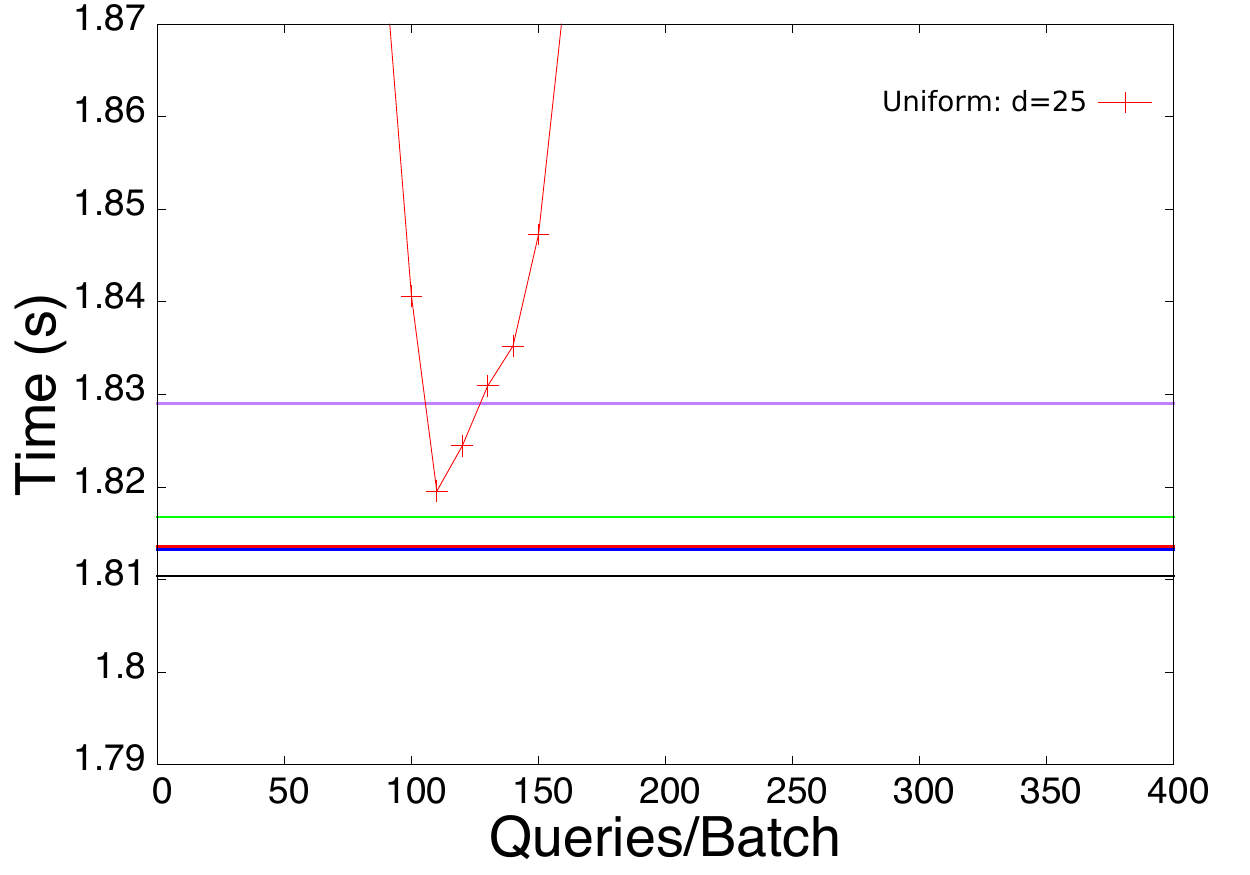}
        }
          
    \caption{Response time vs. queries/batch ($s$) for the periodic query batch method for S3 (a) and S4 (c) (\randomuniform dataset). Panels (b) and (d) correspond to zoomed in versions of (a) and (c) respectively, to highlight the minimum response times. The colored lines correspond to the same algorithms as shown in Figure~\ref{fig:query_split_galaxy}.}
   \label{fig:query_split_rw_uniform}
\end{figure}

\begin{figure}[!htp]
\centering

        \subfigure[]{
            \includegraphics[width=0.45\textwidth]{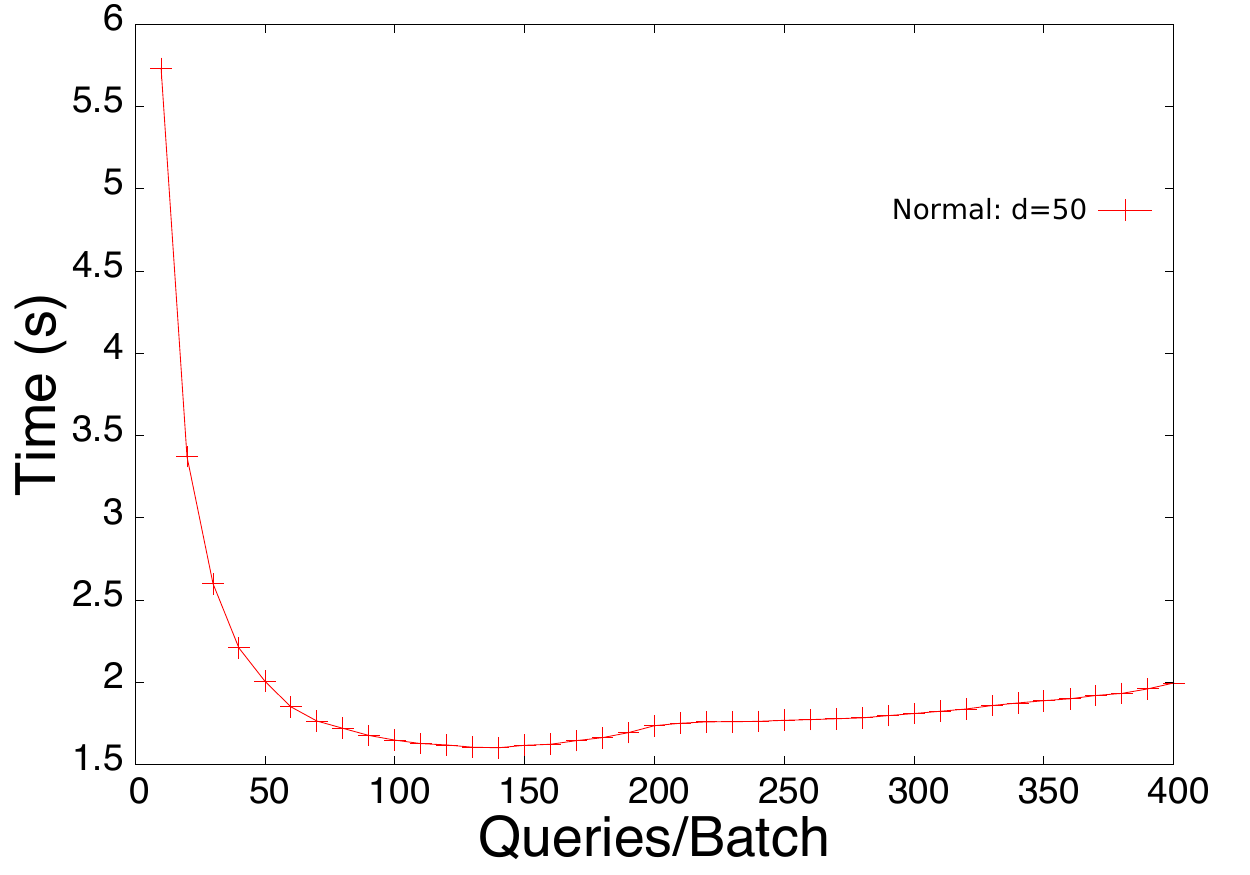}
        }
        \subfigure[]{
            \includegraphics[width=0.45\textwidth]{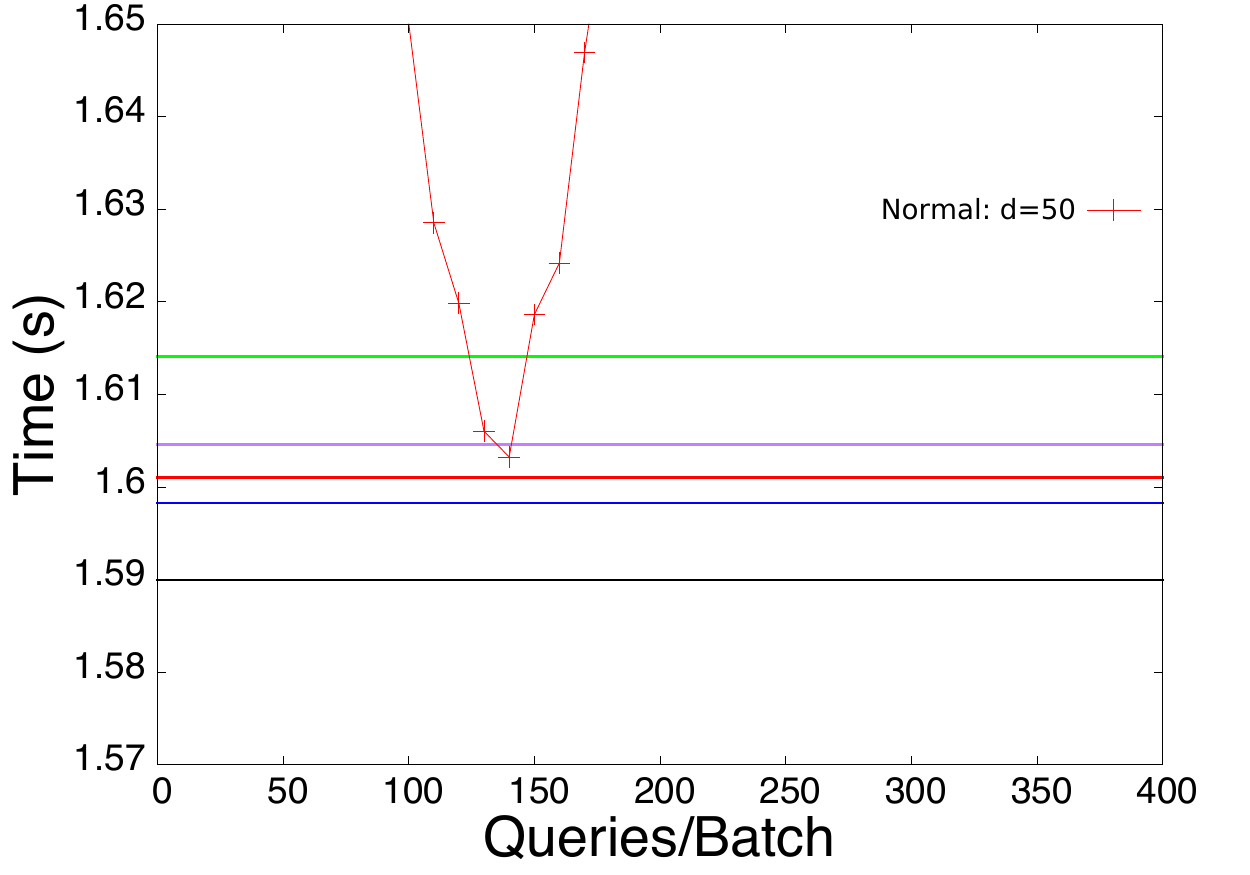}
        }
        \subfigure[]{
            \includegraphics[width=0.45\textwidth]{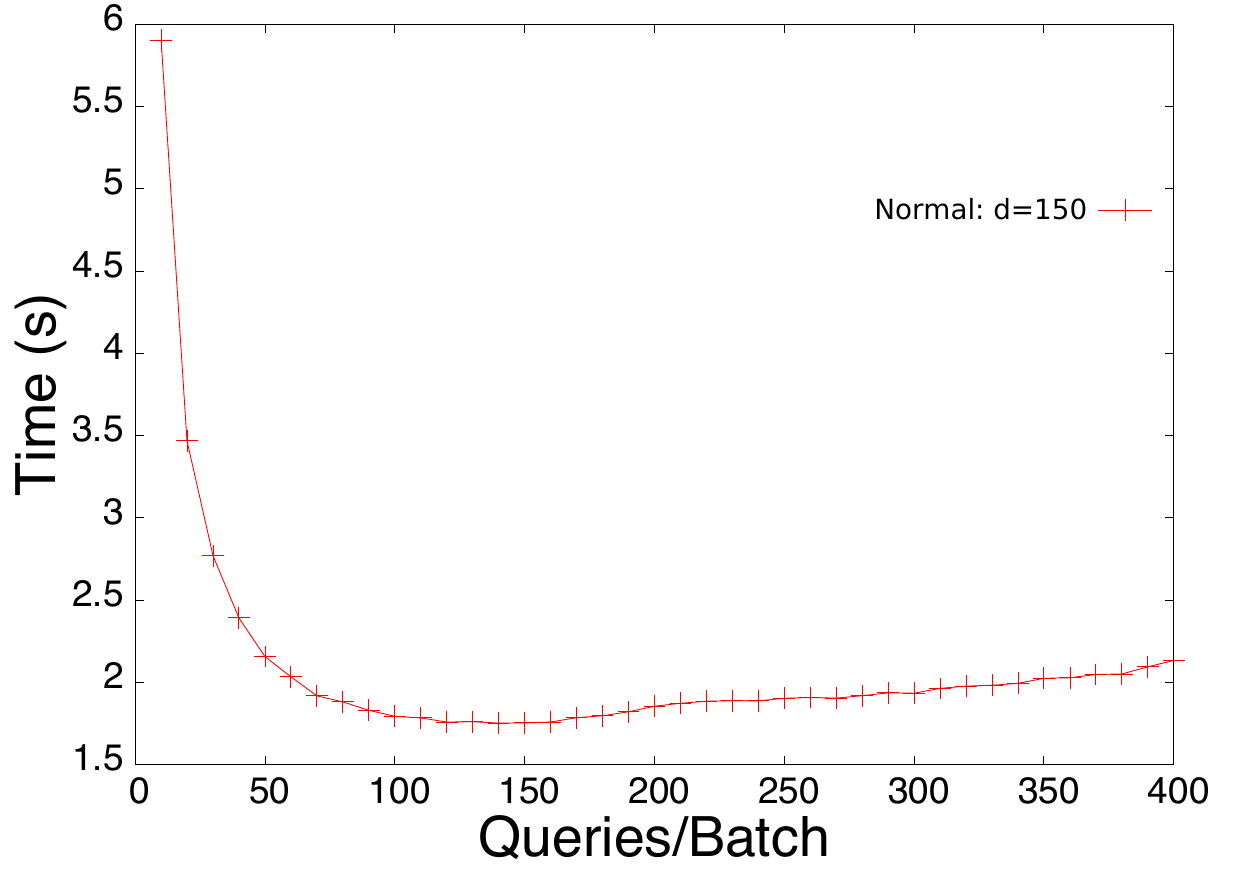}
        }
         \subfigure[]{
            \includegraphics[width=0.45\textwidth]{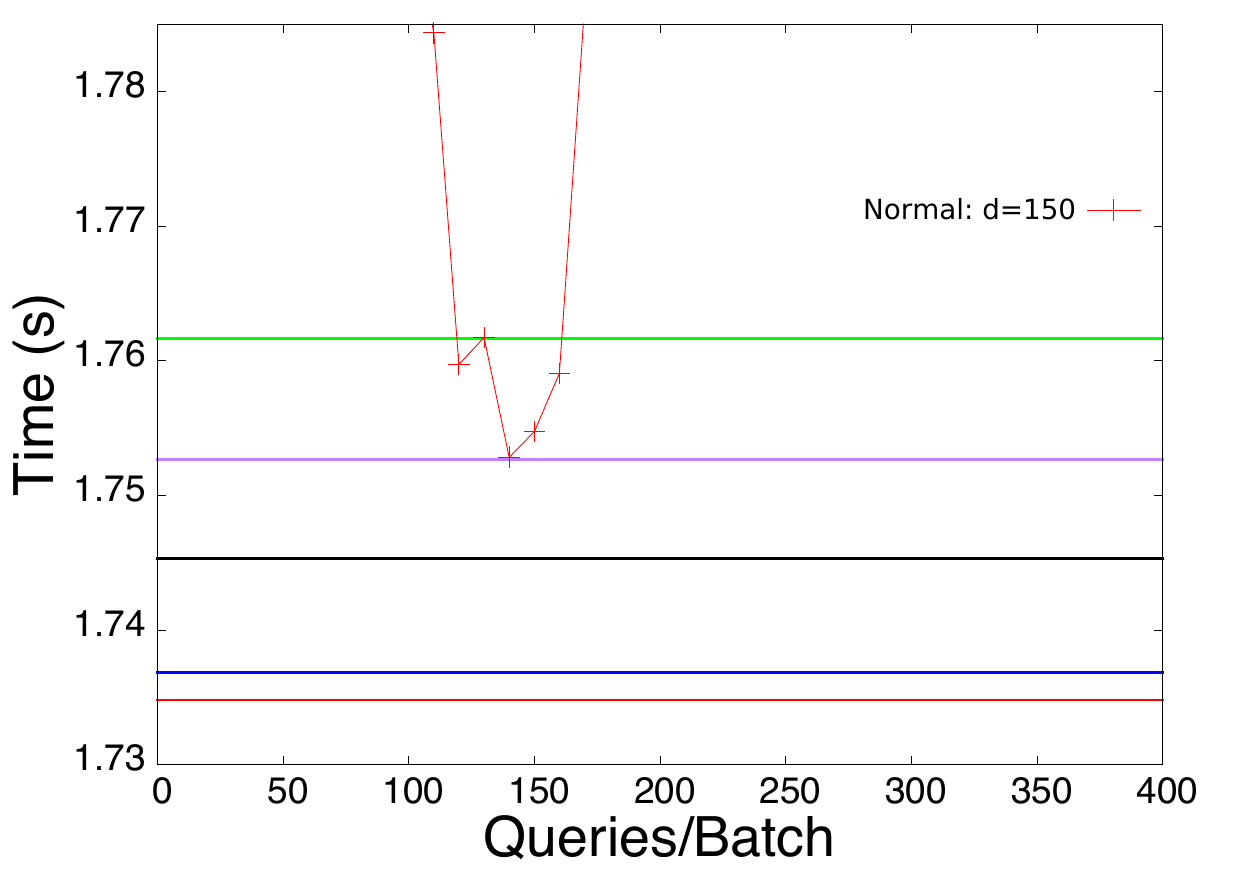}
        }
          
    \caption{Response time vs. queries/batch ($s$) for the periodic query batch method for S5 (a) and S6 (c) (\randomnormal dataset). Panels (b) and (d) correspond to zoomed in versions of (a) and (c) respectively, to highlight the minimum response times. The colored lines correspond to the same algorithms as shown in Figure~\ref{fig:query_split_galaxy}.}
   \label{fig:query_split_normal}
\end{figure}

\begin{figure}[!htp]
\centering

        \subfigure[]{
            \includegraphics[width=0.45\textwidth]{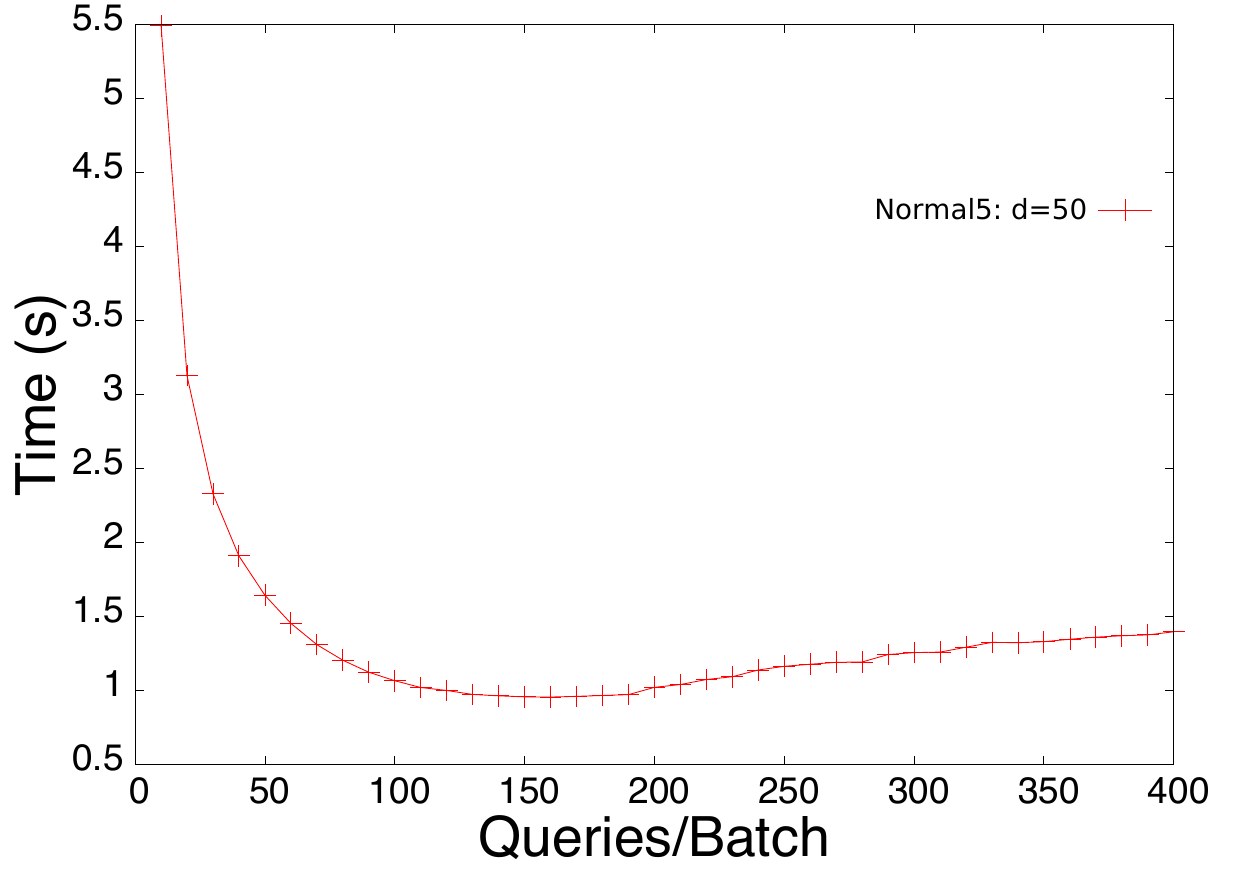}
        }
        \subfigure[]{
            \includegraphics[width=0.45\textwidth]{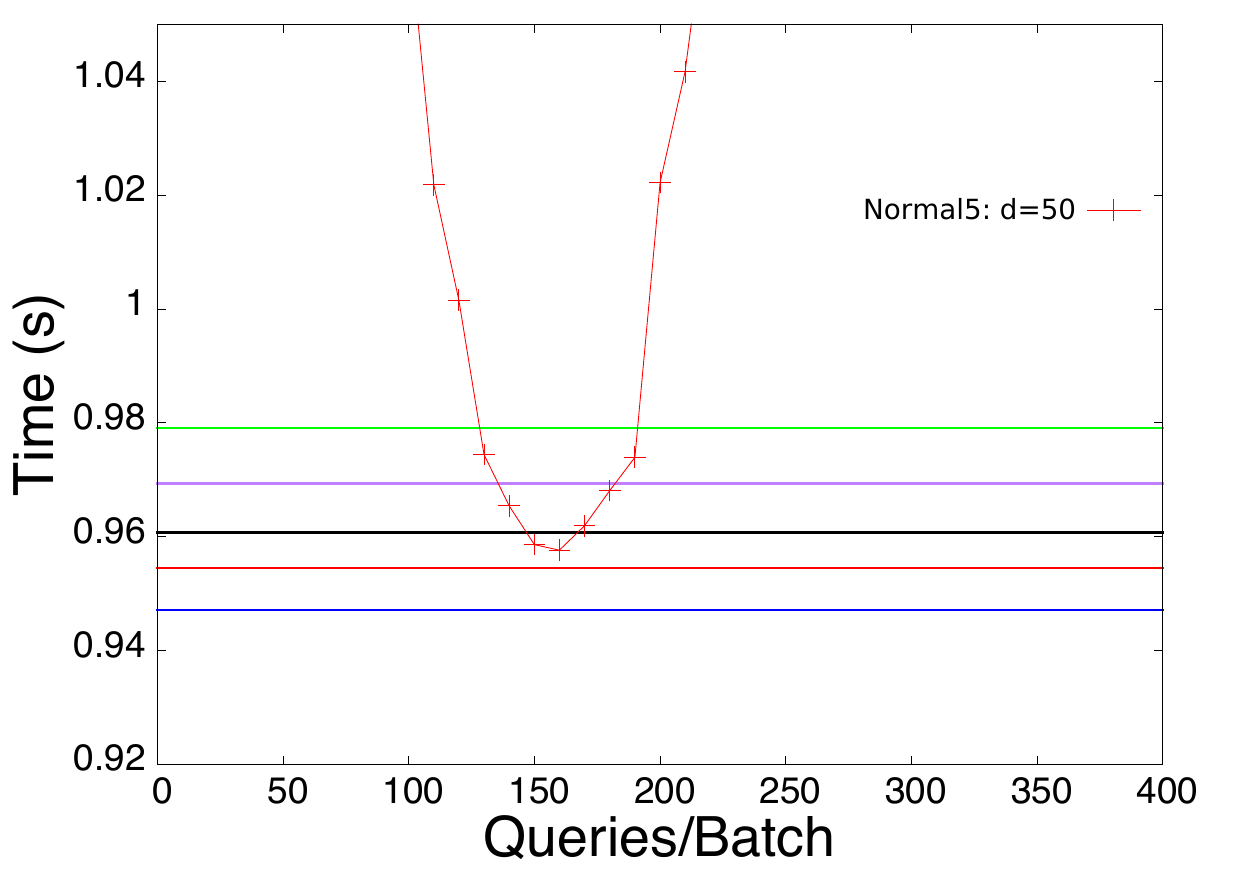}
        }
        \subfigure[]{
            \includegraphics[width=0.45\textwidth]{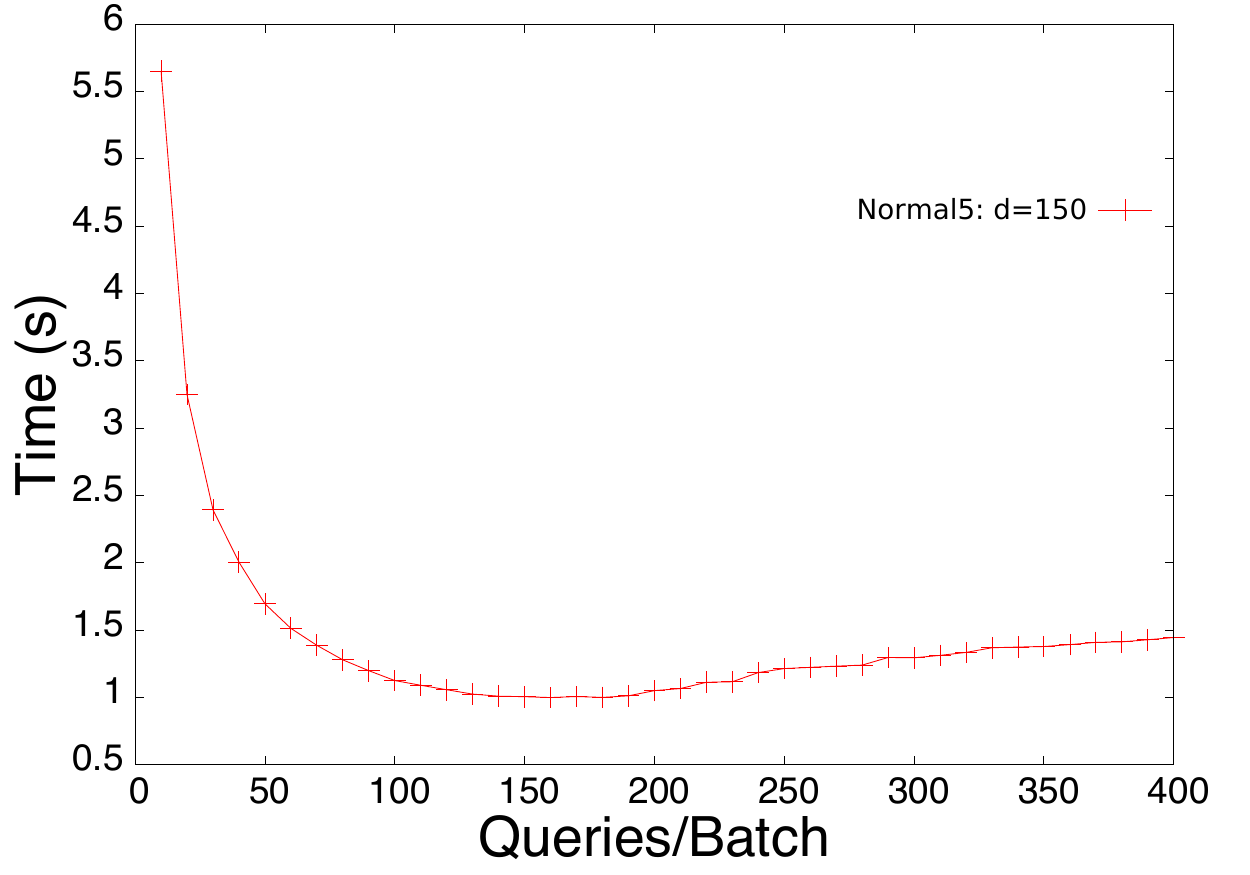}
        }
         \subfigure[]{
            \includegraphics[width=0.45\textwidth]{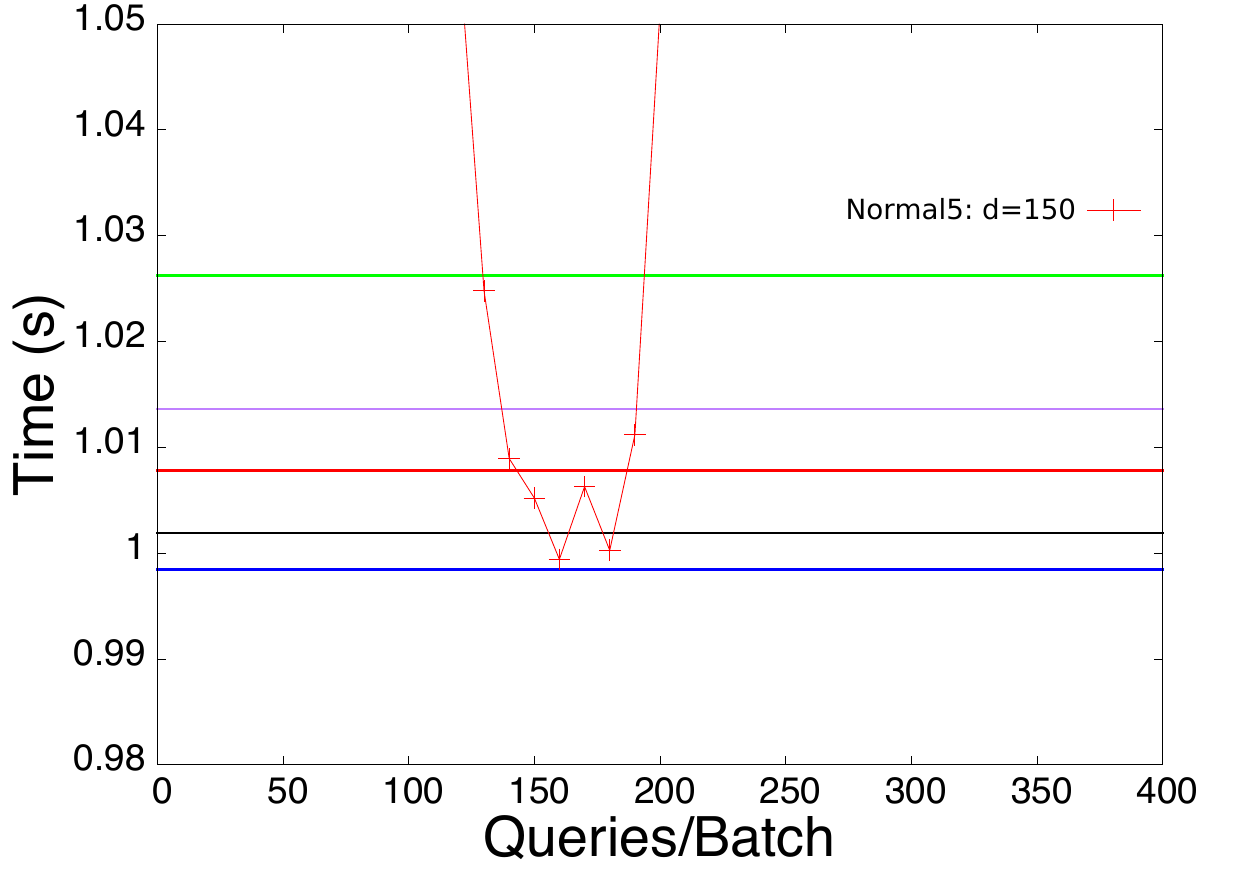}
        }
          
    \caption{Response time vs. queries/batch ($s$) for the periodic query batch method for S7 (a) and S8 (c) (\randomnormalfive dataset). Panels (b) and (d) correspond to zoomed in versions of (a) and (c) respectively, to highlight the minimum response times. The colored lines correspond to the same algorithms as shown in Figure~\ref{fig:query_split_galaxy}.}
   \label{fig:query_split_normal_five}
\end{figure}

\begin{figure}[!htp]
\centering

        \subfigure[]{
            \includegraphics[width=0.45\textwidth]{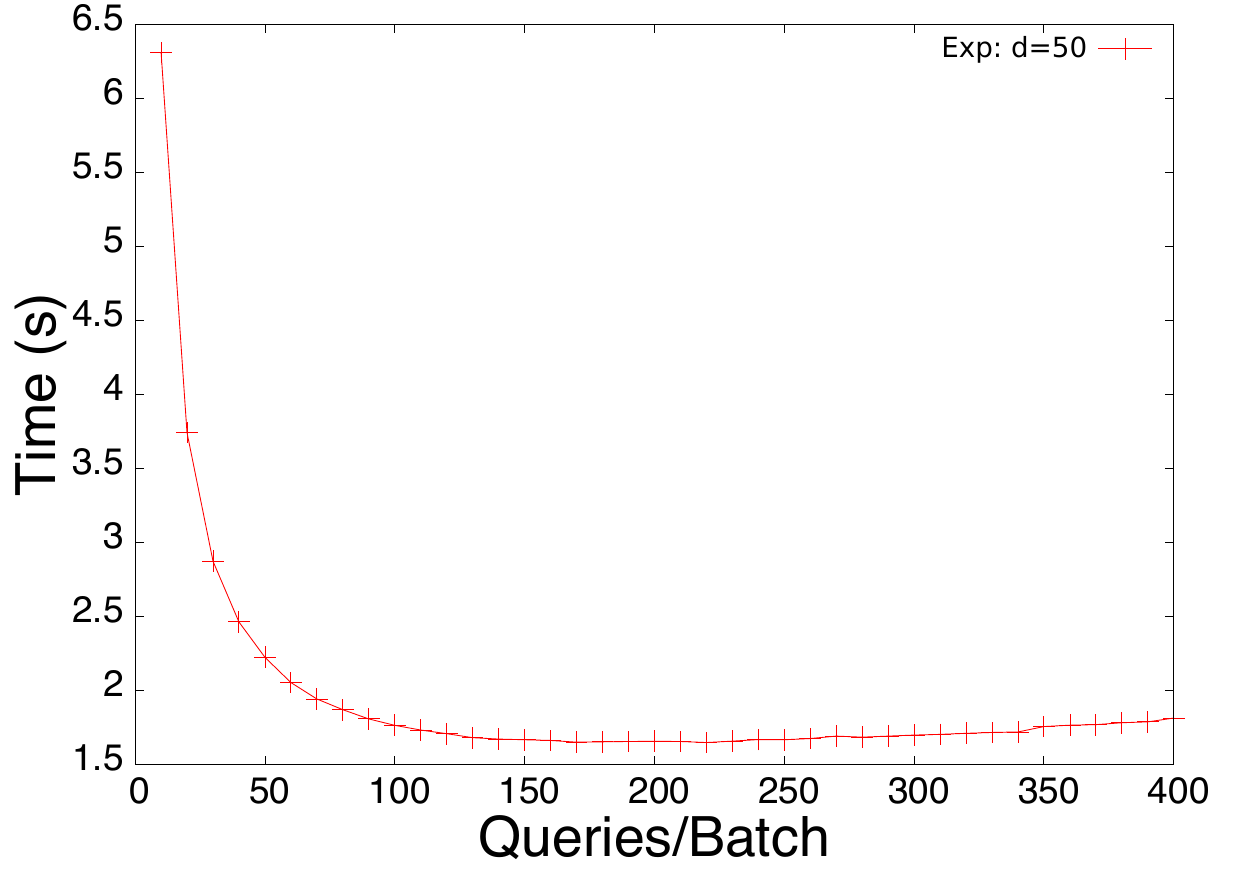}
        }
        \subfigure[]{
            \includegraphics[width=0.45\textwidth]{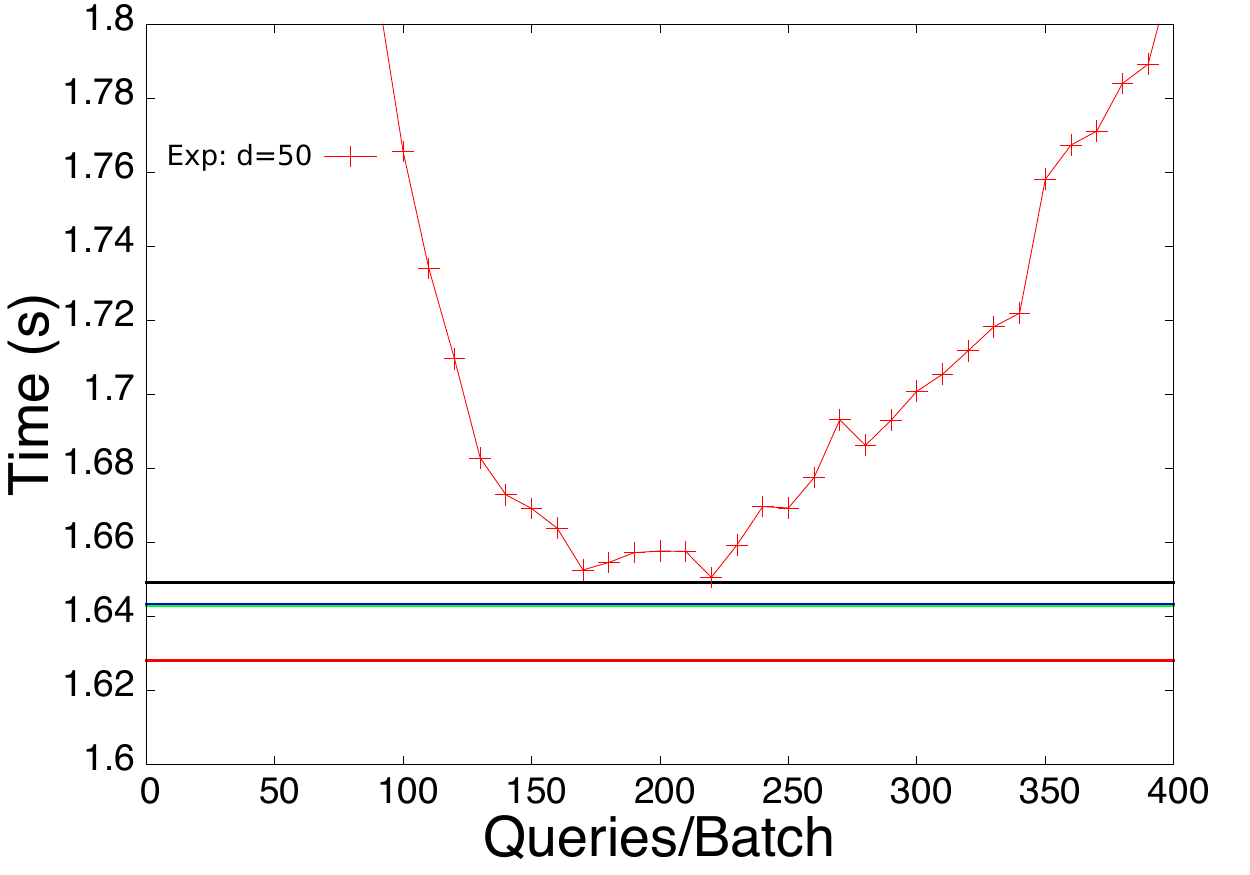}
        }
        \subfigure[]{
            \includegraphics[width=0.45\textwidth]{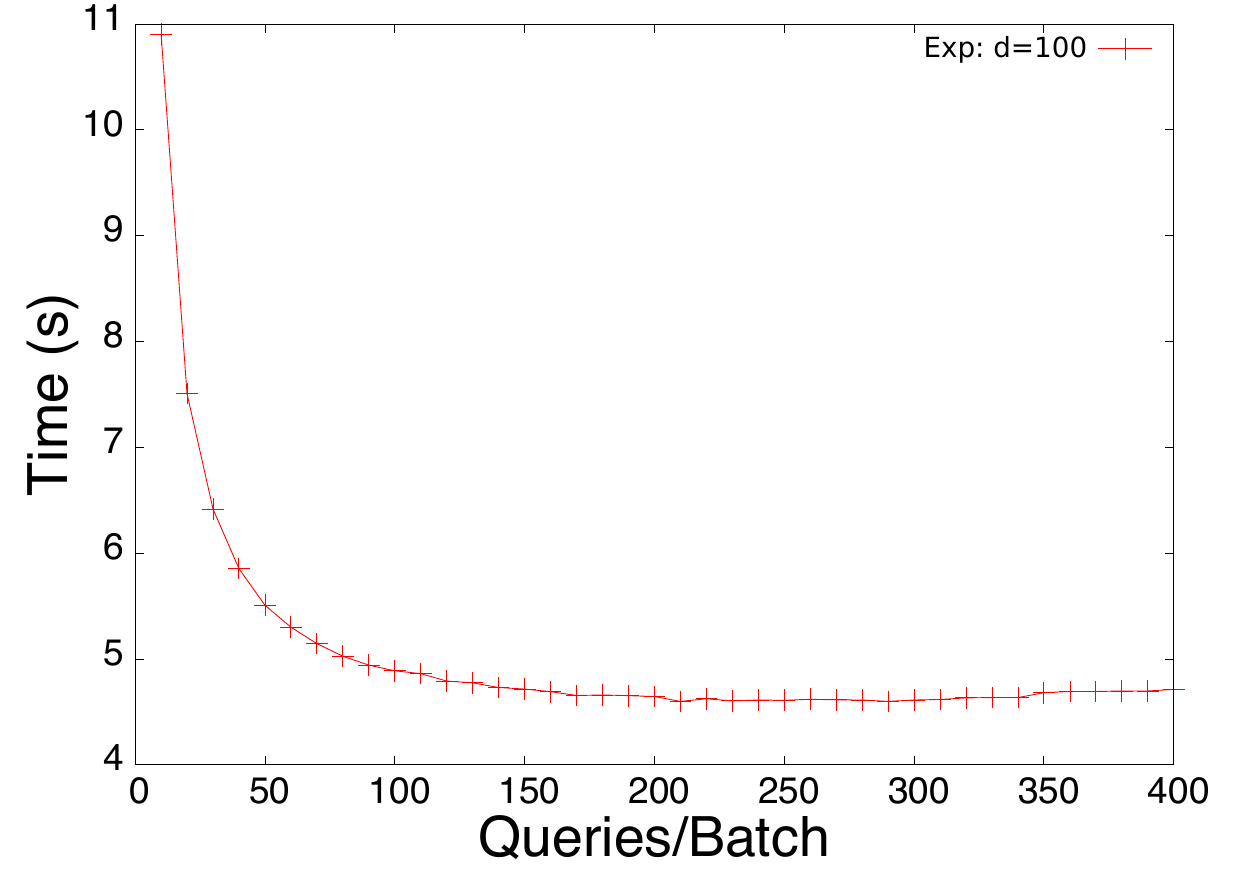}
        }
         \subfigure[]{
            \includegraphics[width=0.45\textwidth]{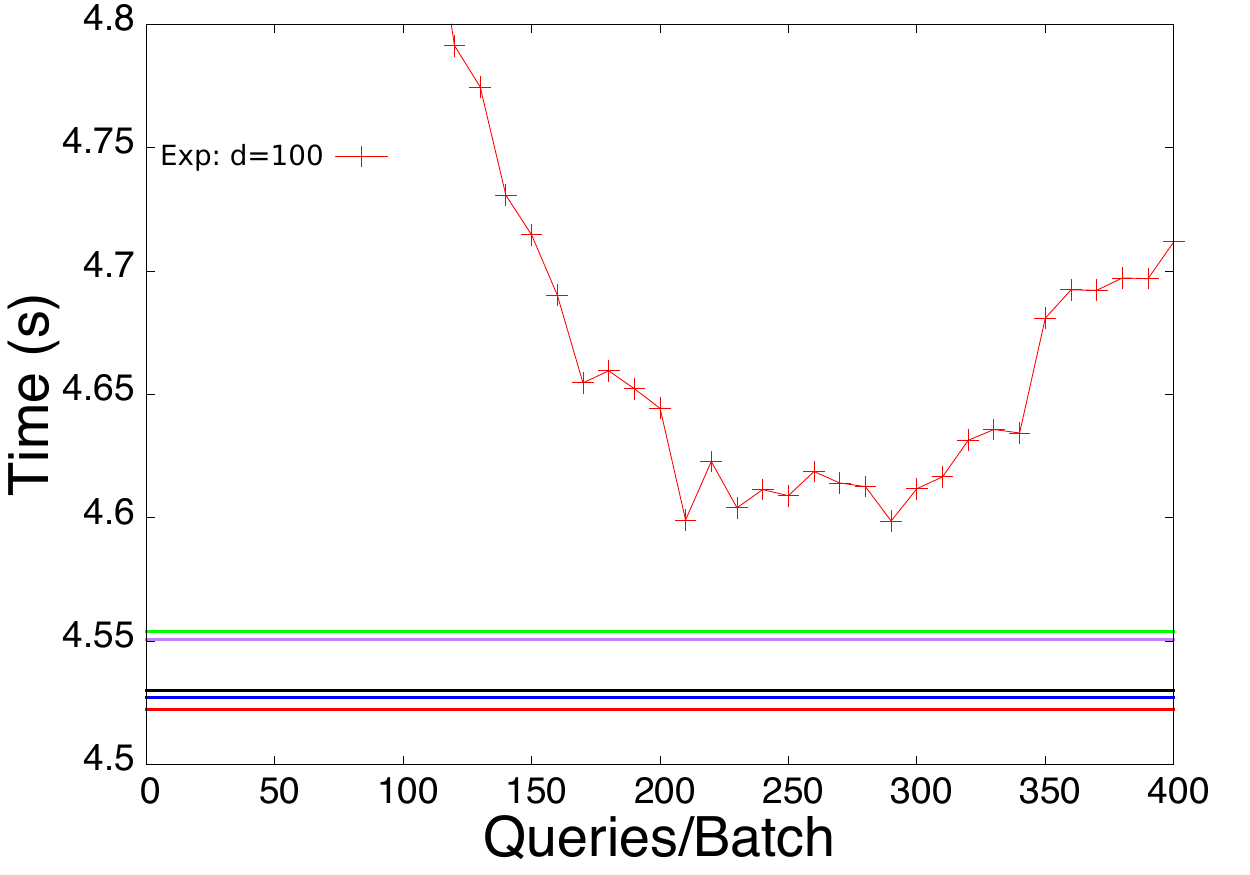}
        }
          
    \caption{Response time vs. queries/batch ($s$) for the periodic query batch method for S9 (a) and S10 (c) (\randomexp dataset). Panels (b) and (d) correspond to zoomed in versions of (a) and (c) respectively, to highlight the minimum response times. The colored lines correspond to the same algorithms as shown in Figure~\ref{fig:query_split_galaxy}.}
   \label{fig:query_split_exp}
\end{figure}

We now evaluate the relative merit of the algorithms for creating query
segment batches (\periodic, \setsplit, \greedysetsplit).  Response time
results for experimental scenarios S1 to S10 are shown in these figures:
Figure~\ref{fig:query_split_galaxy} (for S1 and S2),
Figure~\ref{fig:query_split_rw_uniform} (for S3 and S4),
Figure~\ref{fig:query_split_normal} (for S5 and S6),
Figure~\ref{fig:query_split_normal_five} (for S7 and S8), and
Figure~\ref{fig:query_split_exp} (for S9 and S10).  For each experimental
scenario, the response time of the \periodic algorithm is plotted versus the
batch size on the left-hand side of the figure.  A zoomed-in version of
each plot is shown on the right-hand side, which shows the neighborhood of
the best batch size for \periodic, as well as the response times of the
\setsplit and \greedysetsplit algorithms (which are shown as horizontal
lines). These results correspond to a ``best case'' for the \setsplit and
\greedysetsplit algorithms, for two reasons. First, the response time results do
not include the time necessary to compute the query batches. This time is
negligible for \periodic, but can be significant for \setsplit and even for
\greedysetsplit, as discussed at the end of this section. Second, using an
exhaustive search, for each experimental scenario we have determined the
best parameter configuration for the \setsplit and \greedysetsplit algorithms (i.e., the
best number of batches for \setsplitF, the best maximum batch size
for \setsplitMax, the best minimum and maximum batch size for
\setsplitMM, the best minimum batch size for \greedysetsplitMin, and the best
maximum batch size for \greedysetsplitMax).
The results in Figures~\ref{fig:query_split_galaxy}-\ref{fig:query_split_exp}
are summarized in Table~\ref{tab:result_summary}, which shows for each
algorithm, and each experimental scenario, the percentage response time difference relative
to the response time of the best algorithm for that experimental
scenario. We show two versions of the \periodic algorithm. \periodicbest
corresponds to \periodic when using the batch size that leads to the lowest
response time for the experimental scenario at hand. \periodicgood
corresponds to \periodic but using the worst batch size in a -20/+20
neighborhood of the best batch size (i.e., the batch size in that interval
that leads to the highest response time).

\begin{table}
\centering
\caption{Percentage response time difference relative to the lowest response time for all algorithms and experimental scenarios. Results for the algorithm with the lowest response time shown in boldface.}
\label{tab:result_summary}
{\footnotesize
\begin{tabular}{|l|r|r|r|r|r|r|r|r|r|r|} \hline
Algorithm & S1 & S2 & S3 & S4 & S5 & S6 & S7 & S8 & S9 & S10\\ \hline
\greedysetsplitMax& 0.15&0.15&0.15&{\bf 0.00}&{\bf 0.00}&0.60&1.44&0.34&1.29&0.16\\
\greedysetsplitMin& {\bf 0.00}&0.24&{\bf 0.00}&0.15&0.52&0.11&{\bf 0.00}&{\bf 0.00}&0.93&0.10\\
\hline
\setsplitF& 1.11&1.69&1.03&1.02&0.92&1.03&2.34&1.52&562.90&0.62\\
\setsplitMax& 1.50&1.97&1.02&0.35&1.51&1.54&3.37&2.78&0.90&0.69\\
\setsplitMM& 0.24&0.33&0.10&0.17&0.69&{\bf 0.00}&0.77&0.93&{\bf 0.00}&{\bf 0.00}\\
\hline
\periodicbest& 0.37&{\bf 0.00}&0.23&0.50&0.83&1.03&1.11&0.09&1.50&1.69\\
\periodicgood & 3.21&2.47&1.64&3.56&2.15&1.43&2.21&1.05&2.52&2.69\\
\hline\end{tabular}
}
\end{table}

Some trends are clearly seen in the results. Over the
10 experimental scenarios, the \setsplit and \greedysetsplit algorithms all
lead to response times that are close to each other (within 3.4\%). One
exception is for \setsplitF, which leads to a significantly larger response
time for S9 (about a factor 10 larger than the other algorithms).  Recall
that \setsplitF creates a fixed number of batches without any constraint on
the maximum batch size.  S9 contains entries with exponentially
distributed temporal extents, which causes \setsplitF to create a few very large
batches at the tail of the distribution (which leads to the smallest
\emph{minDiff} value - line~\ref{algsetsplitA:test} in
Algorithm~\ref{alg:setsplitA}).  These large batches are the reason for the
high response time of \setsplitF. This problem does not occur for
experimental scenario S10 due to the larger total number of query segments.
An interesting finding is that the \greedysetsplit algorithms, even though
they use a single pass through the query segments, do well.
\greedysetsplitMax, resp. \greedysetsplitMin, leads to the lowest response time
in 2, resp. 4, of the 10 experimental scenarios.  Overall, the
\greedysetsplit algorithms are among the 3 best algorithms for each
experimental scenario.  This suggests that the quadratic complexity of the
\setsplit algorithm to attempt a less local optimization is in fact
unnecessary.

The key observation from our results is that \periodic leads to good
performance.  As seen in Figure~\ref{fig:query_split_galaxy}, the response
time of \periodic can be high for some batch sizes. However, when using the
best batch size, \periodic can produce response time on par or even better
than that of the \greedysetsplit and \setsplit algorithms. Overall, for each
experimental scenario \periodicbest leads to response times at most 1.69\%
larger than that of the best \greedysetsplit or \setsplit algorithm for that
scenario.  It even leads to the lowest response time for experimental
scenario S2. Even when \periodic does not use the best batch size it leads
to good results. \periodicgood still leads to response times at most 3.56\%
 larger than the best \greedysetsplit or \setsplit algorithm over
the 10 experimental scenarios.

%%%%%%%%% TIME %%%%%%%%%%%
As explained above, our results do not include the time to compute the
batches. Due to quadratic complexity, for the \setsplit algorithms this time
is large, factors larger than the query response time for our experimental
scenarios. Overall, when adding the time to compute the batches (on the
CPU), we find that the \setsplit algorithms lead to average response time
more than 4.69 times larger than \periodicbest and up to 8.84 times larger
(discounting \setsplitF for experimental scenario S9, which leads to
response time 12.76 times larger).  The \greedysetsplit algorithms fare better
when compared to \periodicbest, with response times only up to 2.9\% larger
over all experimental scenarios. This is because these algorithms have
linear complexity. 

We conclude that although computing batches that reduce wasteful
interactions, as in the \setsplit and \greedysetsplit algorithms, is an
appealing idea, in practice it does not outperform a simple periodic
approach.  This is because the small response time benefit due to the use
of better batches is offset by the CPU time overhead of computing these
batches.  One drawback of \periodic is that one must specify a good batch
size, i.e., a batch size in a neighborhood of the best batch size.  In the
next section, we propose performance modeling techniques that can be used
to determine such a good batch size.

\section{Performance Modeling}\label{sec:performance_model}

Most previous works on spatiotemporal database querying, and in particular
works that consider distance threshold
queries~\cite{Arumugam2006,Gowanlock2014,Gowanlock2014b}, rely on
index-trees, such as R-trees. These index-trees have complex performance
behavior as the traversal time depends on the set of pointers followed on a
path toward a leaf node, which is highly data dependent.  As a result,
predicting query response time is challenging.  An
added difficulty in the case of distance threshold queries is that one
query may lead to a large result set while another may lead to an empty
result set.

Because designed for GPU execution, the indexing scheme proposed in
this work (Sections~\ref{sec:traj_indexing} and~\ref{sec:search_alg}) does
not rely on index-trees. While not completely free of data-dependent
behavior, the more deterministic behavior of this scheme makes it possible
to predict query response time. And in particular, such prediction is
sufficiently accurate to determine a good batch size for the \periodic
algorithm.

\begin{figure}[t]
\centering
  \includegraphics[width=0.5\textwidth]{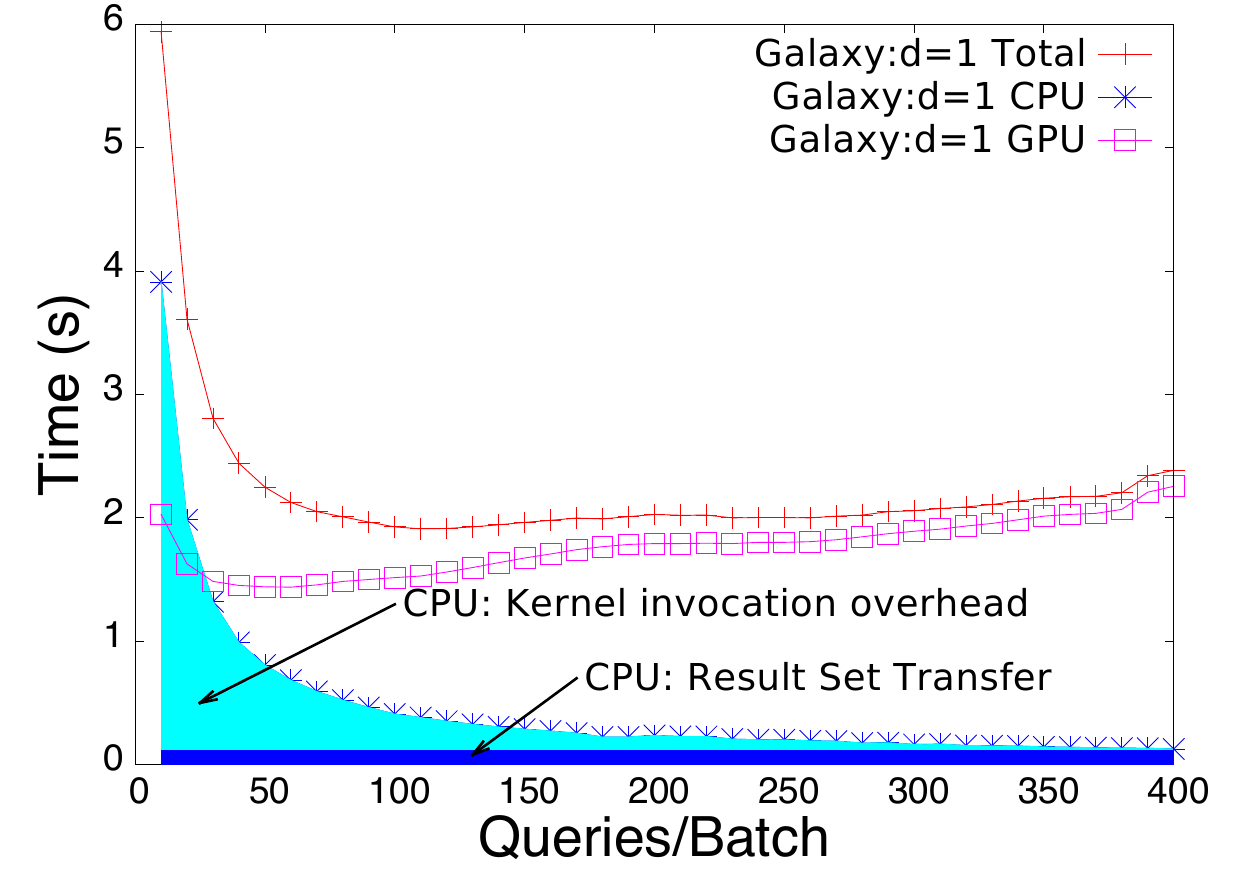}  
    \caption{Response time vs. queries/batch ($s$) for S1 (\galaxy dataset with $d=1$).  The individual CPU and GPU components are shown.}
   \label{fig:Galaxy_d1_cpu_vs_gpu_times}
\end{figure}

The model consists of a GPU component and CPU component.  The GPU component
accounts for the invocation overhead and execution time of each individual
kernel invocation, so that summing over all invocations gives the estimated
GPU time for processing the entire set of query segments.  The CPU
component accounts for the time to perform memory allocations, set kernel
parameters, send query data to the GPU, receive result sets from the GPU,
marshal data, and perform other CPU-side computations (e.g., counter and
pointer updates).  Figure~\ref{fig:Galaxy_d1_cpu_vs_gpu_times} shows
response time results for the S1 experimental scenario, showing both the
CPU and GPU components.  The GPU curve shows an initial decrease as the
batch size, $s$,  increases in the interval $10\leq s \leq 40$. This
decrease is because for low $s$ values the GPU device is underutilized and
the kernel invocation overhead is large due to many such invocations. For
$s\geq 50$, the GPU time increases due to the increasing number of
interactions that must be computed (as explained in
Section~\ref{sec:set_algs}).  The CPU time curve shows a steady decrease as
$s$ increases. This is because the smaller the value of $s$ the more kernel invocations
and thus the more work done on the CPU. We show two portions of the CPU
time. The time necessary to perform kernel invocations, including the
transfer of query segments from the CPU to the GPU, is shown as a shaded
cyan portion below the CPU curve.  The shaded blue portion corresponds to
the time necessary for transferring result sets from the GPU back to the
CPU.

%Although small values of $S$ will have more data transfers, we assume that
%the time to transfer the results is roughly independent of batch size.
%as we would not observe the asymptotic behaviour close to a CPU response time of 0 s if there was a dependency.   

\subsection{GPU Component}

\subsubsection{Model}

In this section, we derive an empirical model for the GPU component of our
performance model.  Let us use $T^{GPU}(i,c)$ to denote the GPU time for a
kernel invocation that computes the $i$ interactions necessary for
comparing a batch of $i/c$ query segments against $c$ candidate segments (using $c$ GPU threads).  Let us
use $\Theta^{GPU}(i,c)$ to denote the overhead of launching a no-op kernel
for $q$ query segments and $i$ interactions (the overhead depends both on
the number of queries and on the number of GPU threads).  Given the $i$
interactions to be computed, we denote by $\alpha$ the fraction of these
interactions that lead to an item being added to the result set (i.e., both
a temporal hit and a spatial hit), by $\beta$ the fraction of these
interactions for which the entry segment does not overlap the query segment
temporally (i.e., a temporal miss), and by $\gamma$ the fraction of these
interactions for which the entry segment overlaps the query segment
temporally but not spatially (i.e., a temporal hit but a spatial miss).  We
have $\alpha + \beta + \gamma = 1$.  We distinguish these three cases
because the computational cost is different in each. Candidate
segments that are temporal misses can be determined with only a few
instructions (i.e., comparing temporal extremities of query and candidate
segments). Candidate segments that are temporal hits but spatial misses
require more instructions (i.e., spatial extremities comparisons).
Candidate segments that should be added to the result set require even more
instructions to be performed (i.e., determining the actual overlapping
temporal interval). One can view the computation of an interaction as a set
of comparisons and moving distance calculations, but these comparisons and computations
are short-circuited whenever a segment is found to be a
temporal or spatial miss.

We denote by $T_1(i,c)$, $T_2(i,c)$, and $T_3(i,c)$ the time for a kernel
invocation with $i$ interactions so that all $c$ candidate segments 
are temporal and spatial hits, temporal misses, and
temporal hits and spatial misses, respectively. This leads us to the
following model:
\begin{equation*}
T^{GPU}(i,c) = T^{GPU}_1(\alpha i,c) + T^{GPU}_2(\beta i,c) + T^{GPU}_3(\gamma i,c) - 2 \Theta^{GPU}(i,c).
\label{eqn:performance_model}
\end{equation*}
The first three terms above each include a $\Theta^{GPU}(i,c)$ component,
hence the subtracted fourth term.  $T_{GPU}(i,c)$ is computed for each
batch, and the sum gives the total GPU time assuming the batch size is $s$:
\begin{equation*}
T^{GPU}(s) = \sum_{j=0}^{|Q|/s} T^{GPU}(i_j,i_j/s).
\end{equation*}
$Q$ is the total set of query segments (for simplicity this equation
assumes that $s$ divides $|Q|$). $i_j$ is the number of interactions that
must be computed for the $j$-th query batch, which is determined based on
the entry segment bins (see Section~\ref{sec:traj_indexing}). Therefore
$i_j/s$ is the number of candidate entry segments for the $s$ query segments
in the batch.

In this model, parameters $\alpha$, $\beta$ and $\gamma$ depend on the
dataset and the query. They must thus be determined empirically for typical
scenarios.  By contrast, the functions $\Theta^{GPU}$, $T^{GPU}_1$,
$T^{GPU}_2$, $T^{GPU}_3$ depend only the hardware characteristics of the
platform. In what follows we describe how we estimate these parameters.
Note that this estimation is done for each batch of $s$ queries.

\subsubsection{Estimating $\alpha$, $\beta$ and $\gamma$}
\label{sec:alphabetagamma}

Recall that $\alpha$ is, for a kernel invocation on a query batch, the fraction of
interactions that lead to a new item being added to the result set. Given
that it is dataset dependent, we use a pragmatic approach to estimate
$\alpha$ for a particular dataset once and for all, i.e., before the
dataset is being queried in ``production'' use.  Depending on the temporal
distribution of the entry segments, there may be time periods with few
active trajectories and some with many, resulting in a non-uniform
distribution of query hits throughout time. To estimate $\alpha$, we divide
the dataset into $numEpochs$ temporal epochs.  For each epoch we select a
batch of $s$ sample queries that fall within the epoch.  We do this by
randomly selecting $s$ consecutive query segments from a representative
query dataset.  We then execute our kernel and calculate the fraction of
interactions that produced result items. We perform this over enough trials
such that the predicted total number of result set items is within 5\% of
the total true number of result set items.  This procedure yields an
$\alpha$ estimate for each epoch.  This estimate may be inaccurate if the
sample queries are not representative of queries that will be processed in
production. Also, if too low a value of $numEpochs$ is used, then the
$\alpha$ estimates are more likely to be inaccurate since transient
temporal patterns are then averaged over larger epochs.  Using $numEpochs=1$
is a degenerate case in which our model would assume that for any kernel
invocation the query hit probability is the same. This may be accurate for a
temporally uniform dataset, but vastly inaccurate for datasets that exhibit
temporal transience. In all the experiments presented hereafter we use
$numEpochs=50$.

Unlike $\alpha$, $\beta$ can be computed precisely.
For a given set of $s$ query segments, one can 
determine which entry segments they may temporally overlap using the bins
in our indexing scheme (see Section~\ref{sec:traj_indexing}). Then, with
two nested loops one can simply compare the temporal extremities of each
query segment to that of each entry segment, yielding an exact value for
$\beta$. Parameter $\gamma$ is computed as $1-\alpha-\beta$.

To summarize, for a given dataset we compute once and for all a set of
$\alpha$ estimates for each epoch and for the full range of (reasonable)
$s$ values.  Then, for each batch of $s$ queries we compute an $\alpha$
estimate, $\beta$ and $\gamma$.  Therefore, for each candidate $s$ value we
can plug appropriate values of these three parameters into the $T^{GPU}$
performance model.

\subsubsection{Estimating $T^{GPU}_1$, $T^{GPU}_2$, $T^{GPU}_3$ and $\Theta^{GPU}$}

The $T^{GPU}_1$, $T^{GPU}_2$, $T^{GPU}_3$ and $\Theta^{GPU}$ functions
depend only on the implementation of the kernel and the hardware
characteristics of the platform. As a result, we can empirically estimate these time components
based on benchmark results. Let us consider $T^{GPU}_1$, i.e., the kernel
response time when all interactions are both temporal hits and spatial
hits.  The same approach is used to estimate $T^{GPU}_2(i,c)$,
$T^{GPU}_3(i,c)$ and $\Theta^{GPU}(i,c)$.

\begin{figure}[t]
\centering
  \includegraphics[width=0.5\textwidth]{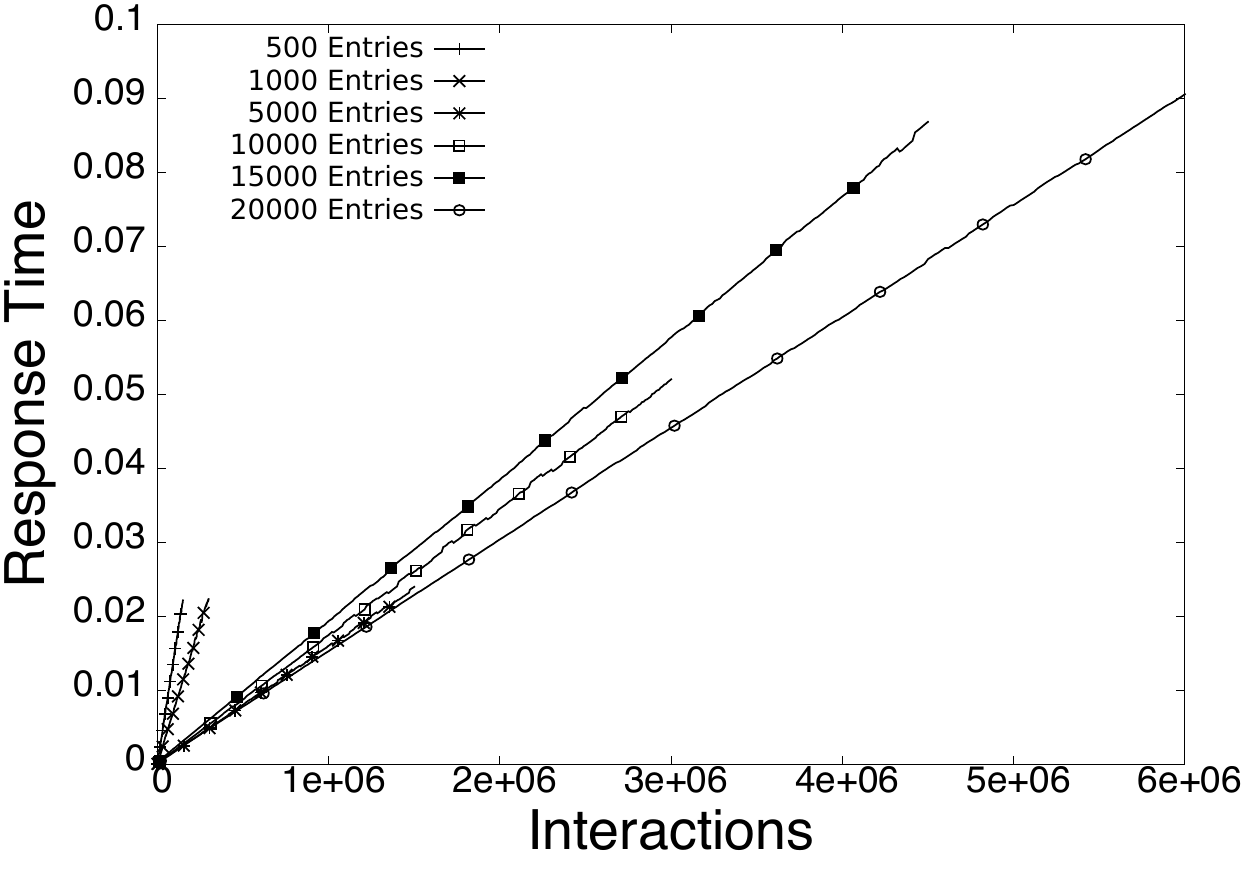}  
    \caption{Interactions vs. response time for a selection of entries.  The data shown corresponds to a range of 1-300 benchmarked queries.}
   \label{fig:all_hits_benchmark_interactions_entries_response_time}
\end{figure}

We generate a synthetic dataset and query set in which all
interactions are guaranteed to be both temporal and spatial hits.
Figure~\ref{fig:all_hits_benchmark_interactions_entries_response_time}
shows a subset of our benchmark results as response time vs. number of
interactions for various numbers of candidate entry segments, as measured
on our target platform. Given a number of interactions, $i$, and a number
of candidate entries, $c$, we simply use linear interpolation to determine
a response time prediction $T^{GPU}_1(i,c)$ from the benchmark results.

%To calculate
%response time, consider the scenario where $\alpha=\beta=\gamma=1/3$, which
%means that $1/3$ of the interactions are hits, temporal misses, and
%temporal hits, but spatial misses, respectively, $E=5031$, $Q=99$, and
%$I=5031\times99=498069$. $\alpha I=498069/3=166023$, then, $A(\alpha
%I,E)=A(166023,5031)$.  From
%Figure~\ref{fig:all_hits_benchmark_interactions_entries_response_time}, the
%response time is calculated by interpolating between the 5000 entry and
%10000 entry curves at x=166023 interactions.  Note that this is just an
%illustration of the data, and only a select portion of it is shown.

%From Figure~\ref{fig:all_hits_benchmark}~(a), and using Equation~\ref{eqn:interactions}, this would correspond to looking up the response time values of $Q_A=32967/999=33$, where $Q_A$ is the number of queries associated with the all hits benchmark, and $E=999$.  

\begin{figure}[t]
\centering
        \subfigure[1-300 queries]{
            \includegraphics[width=0.35\textwidth]{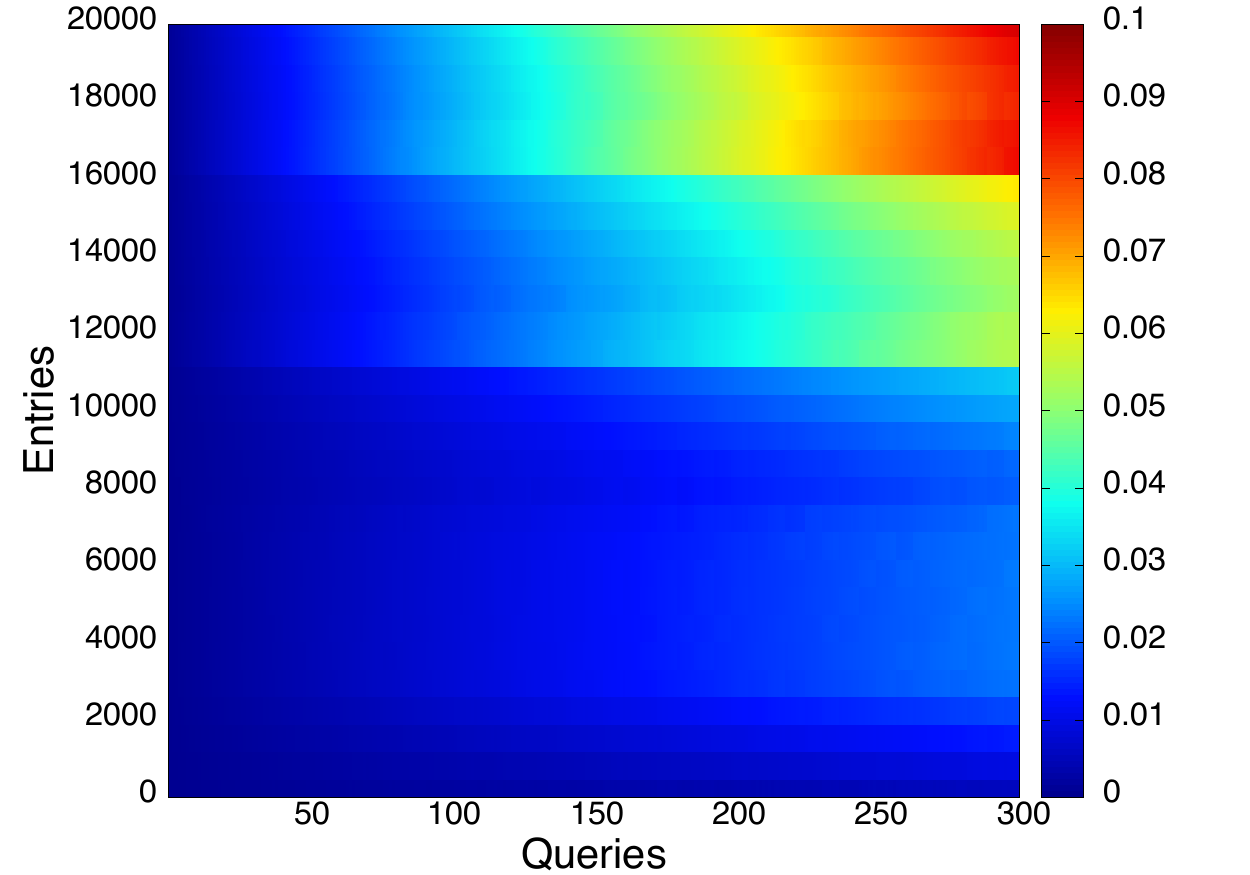}
  }
        \subfigure[1-20 queries]{
            \includegraphics[width=0.35\textwidth]{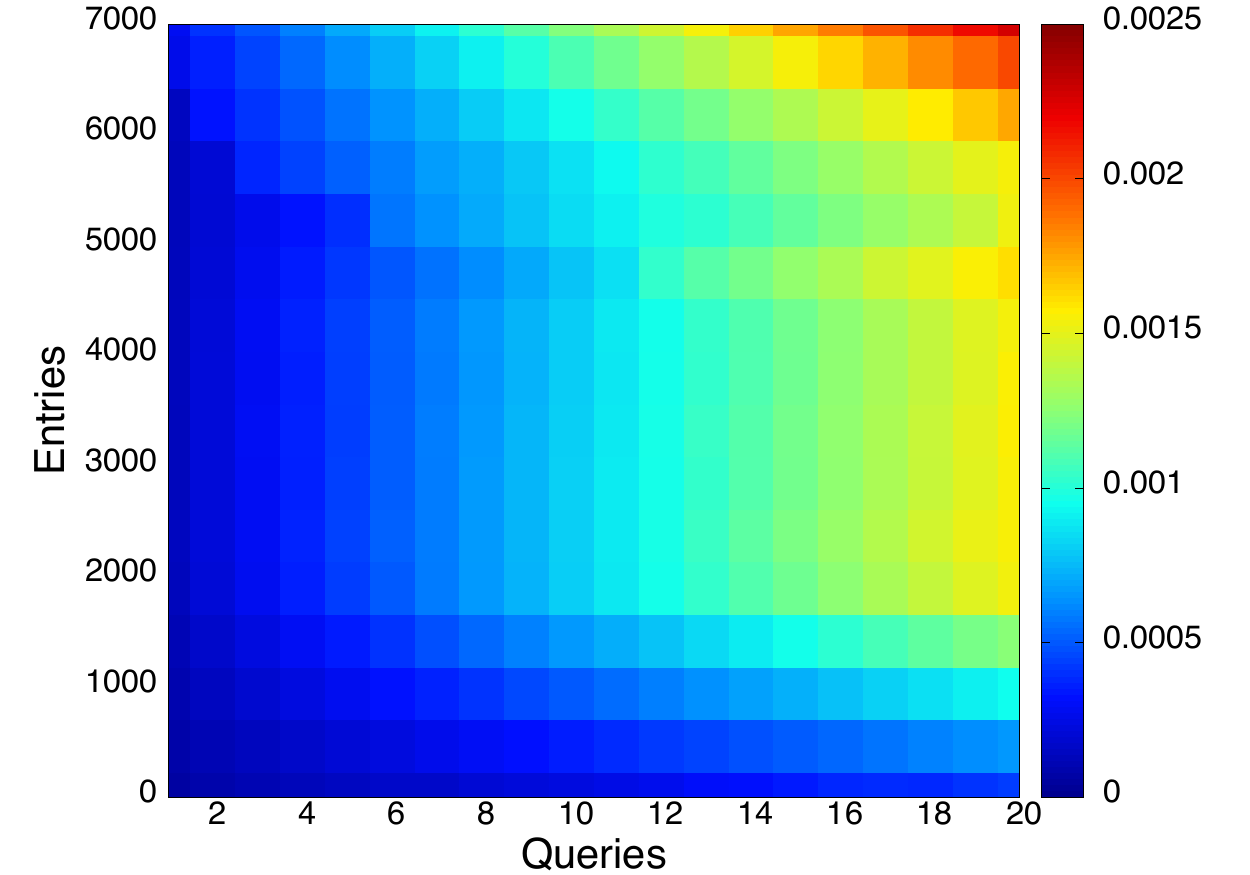}
        }
        \subfigure[21-40 queries]{
            \includegraphics[width=0.35\textwidth]{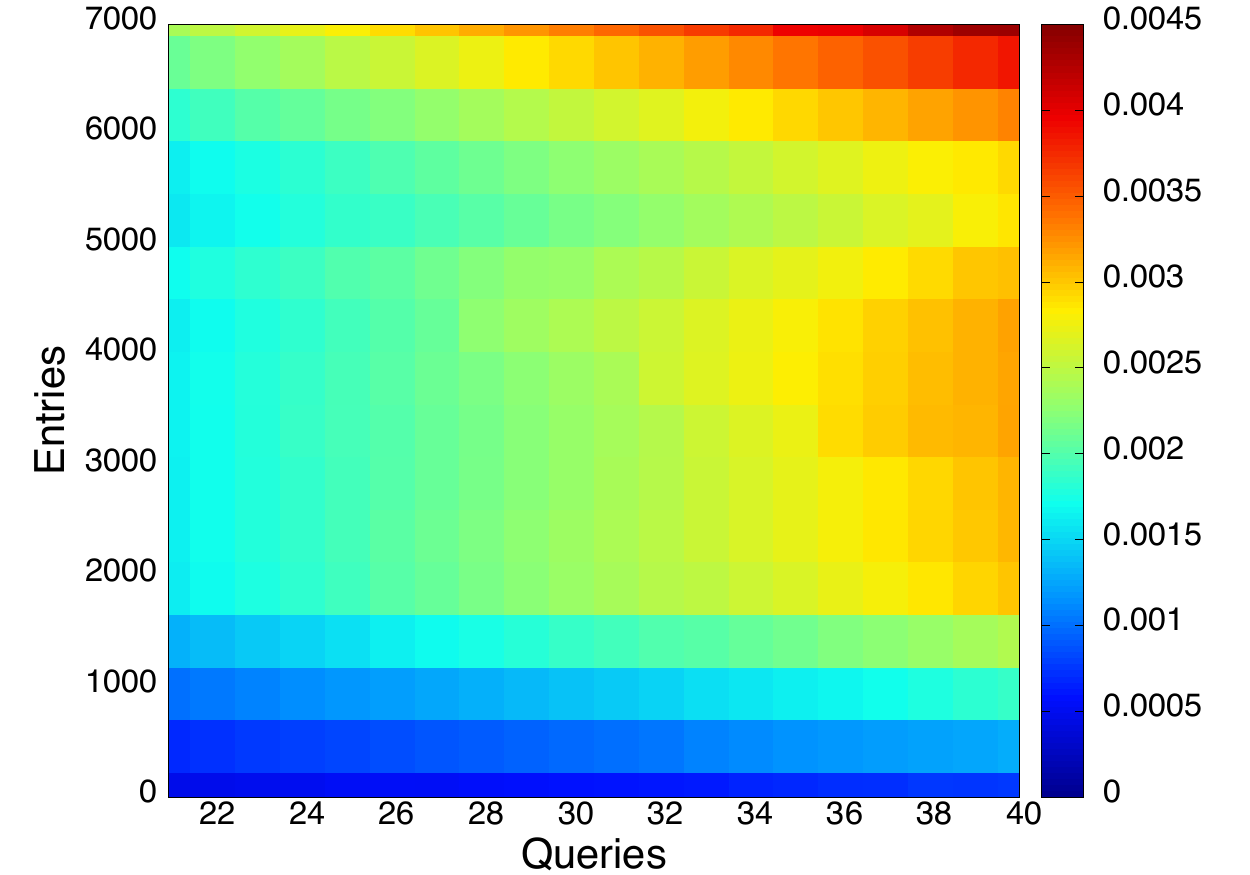}
        }
        \subfigure[41-60 queries]{
            \includegraphics[width=0.35\textwidth]{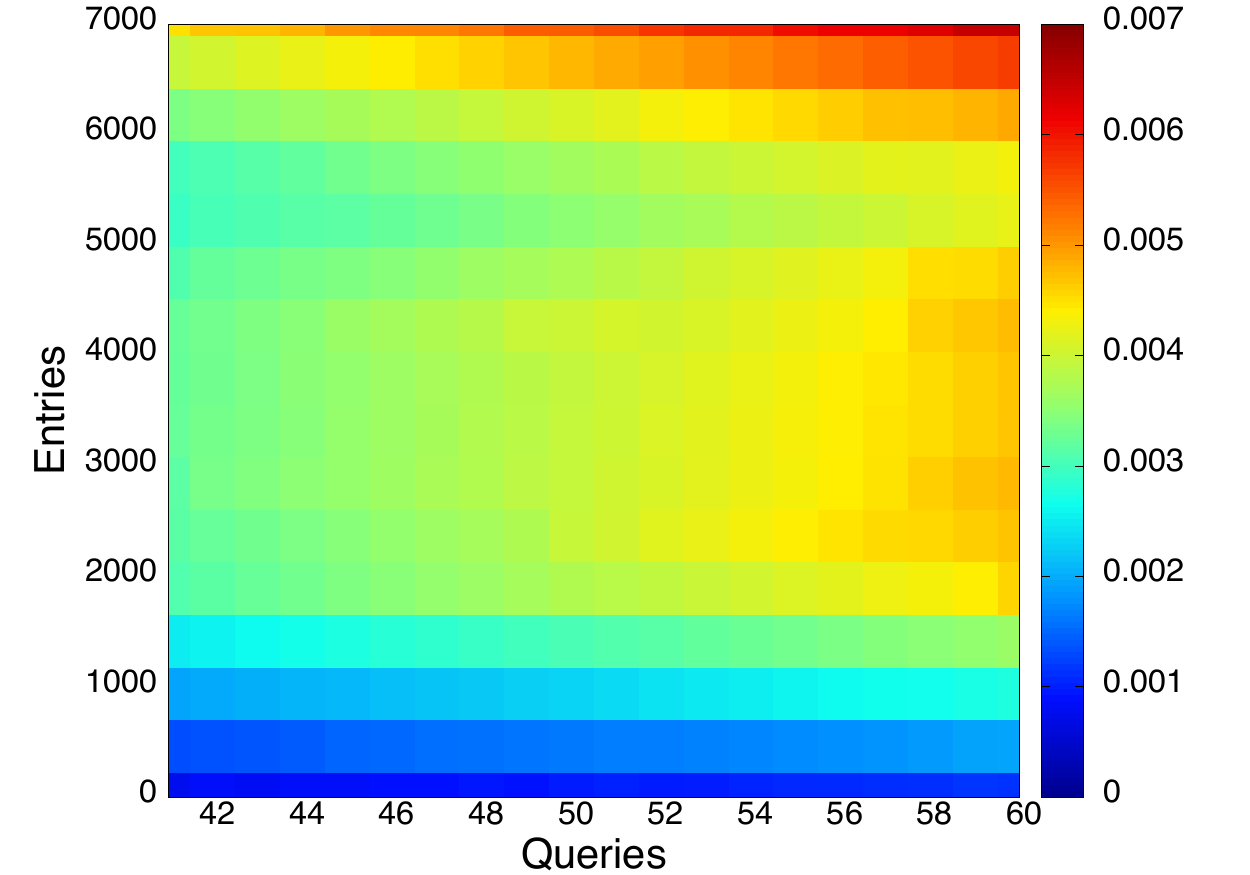}
        }
    \caption{Benchmark of interactions that all are within the query distance. Panel~(a) shows a large selection of the data, across a range of 300 queries, and (b), (c) and (d) show detailed versions of the data between 1-60 queries.}
   \label{fig:all_hits_benchmark}
\end{figure}

Figure~\ref{fig:all_hits_benchmark}~(a) shows a broader range of benchmark
results, shown as heat maps of the response time vs. the number of
candidate entries, $c$ and the number of queries $q = i/c$.  In
Figure~\ref{fig:all_hits_benchmark}~(a) we observe that there are
discontinuities in response time. We attribute these discontinuities mainly
to thread scheduling factors on the GPU.  Regardless, they will be a source
of modeling error due to our use of linear interpolation.
Figure~\ref{fig:all_hits_benchmark}~(b),~(c),~and~(d) show
plots with queries in the range of 1 to 60. We see somewhat
smoother response time trends, particularly in Figure~\ref{fig:all_hits_benchmark}~(c)~and~(d) suggesting that in that range the use of
linear interpolation should lead to less error. Since we know that the
batch size, and thus the number of candidate segments, should be relatively
small, then one may expect that modeling error due to linear interpolation
could also be small.

%From Figure~\ref{fig:all_hits_benchmark}~(b), we see that the distribution of
%response times is not smooth, suggesting that the benchmarks will be a
%source of model error, especially for low period values.  However,
%Figure~\ref{fig:all_hits_benchmark}~(c)~and~(d) show smoother transitions,
%which suggest that estimating response times with a moderate number of
%interactions, where the number of queries in the figure is $\gtrsim30$ will
%have less error, than at lower interaction values.  This is not
%problematic, as we know that low periods do not lead to the best response
%times overall, due to increased kernel overhead, in comparison to a higher
%period, which has fewer kernel invocations over the entire execution.  The
%other values, $B(\beta I,E), C(\gamma I,E)$ and $\theta(I,E)$ required to
%compute the total response time using Equation~\ref{eqn:performance_model},
%are calculated in the same manner, with their own sets of benchmarks.  We
%omit showing these benchmarks, as the benchmark values for $A(\alpha I,E)$
%have the highest response times.

The above modeling approach assumes that the kernel response time can be
estimated from the benchmark-based models of $T^{GPU}_1$, $T^{GPU}_2$,
$T^{GPU}_3$, executed separately.  However, an actual kernel execution
consists of a mix of temporal and spatial hits, temporal misses, and
temporal hits but spatial misses. One may thus wonder whether the notion of
separating the model into three components can lead to reasonable response
time predictions. To answer this question we compared executions of the
three benchmark kernels to a mixed execution, using various synthetic
datasets and query sets with $\alpha=\beta=\gamma=1/3$.

\begin{figure}[t]
\centering
  \includegraphics[width=1.0\textwidth]{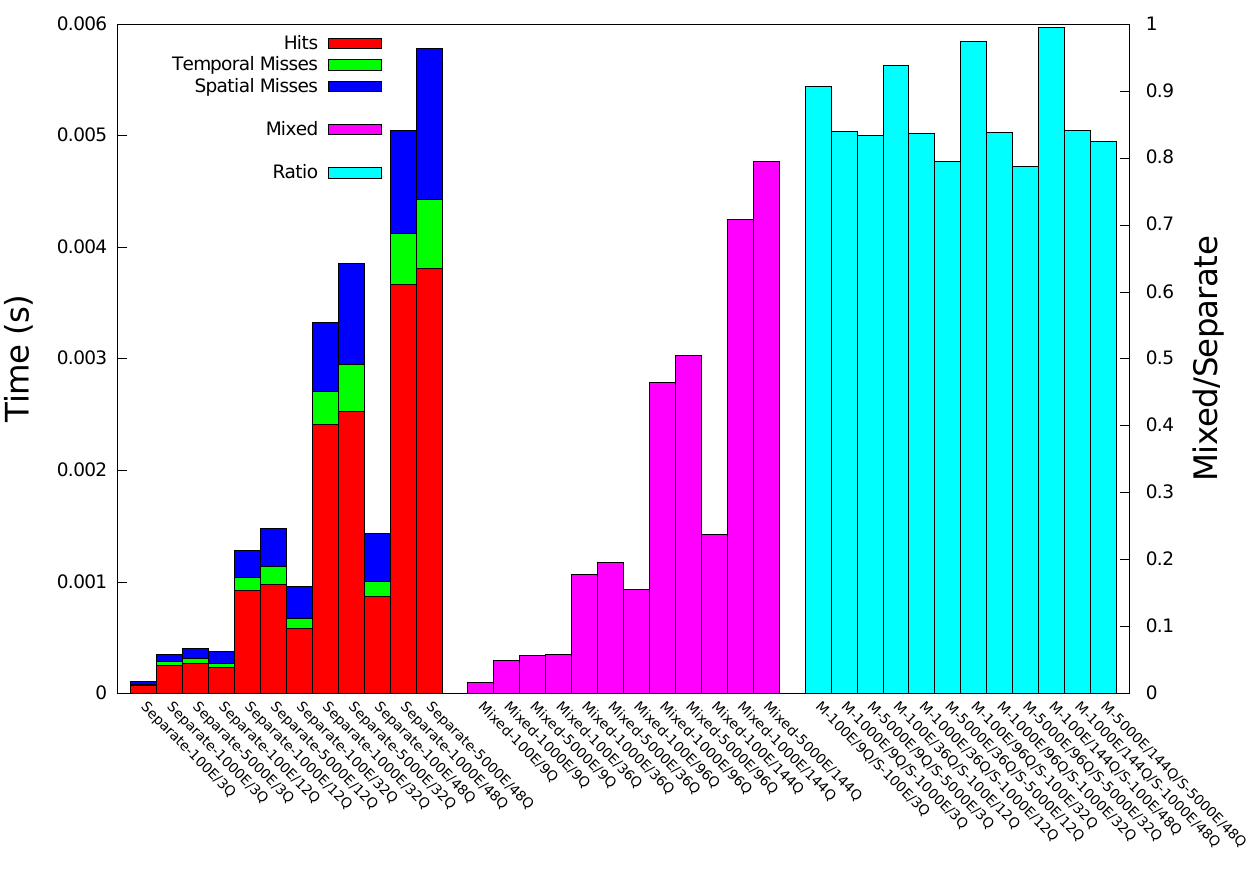}  
    \caption{GPU response time vs. test cases of mixed and separated kernel invocations.}
   \label{fig:mixed_vs_separate}
\end{figure}

Figure~\ref{fig:mixed_vs_separate} shows the response time vs. test case,
where the first two histograms correspond to the separate and mixed tests,
respectively, and the rightmost histogram shows the ratio of mixed to
separate response times.  The test cases are identified as ``Separate'' or
``Mixed'', followed by the number of entry segments and the batch size.
Since $\alpha=\beta=\gamma=1/3$, then the number of queries of each type
($\alpha$, $\beta$, $\gamma$) is equal to the total number of queries in an
experimental scenario divided by 3.  For example, consider the scenario
where there are 9 queries in total, Separate-100E/3Q (3 queries of each
type) is compared to Mixed-100E/9Q, where we compare three kernel
executions with queries of type $\alpha$, $\beta$, and $\gamma$ each with a
query batch size of 3, to one kernel execution with 9 queries, which are a
mixture of query types $\alpha$, $\beta$, and $\gamma$.  To allow for fair
comparisons we have discounted the kernel invocation overhead from all
results (this overhead occurs three times in the Separate results but only once
in the Mixed results).

The first observation is that the Mixed executions have lower response
times than the Separate counterparts.  This may seem counter-intuitive
because the Mixed executions, unlike the Separate executions, should have a
high degree of branch divergence, thus causing partially serialized thread
executions on the GPU~\cite{Han:2011:RBD:1964179.1964184}.  However, in
Mixed executions the entry segments are retrieved from the GPU's global
memory and stored into private memory once, and are then reused.  This does
not occur in the Separate executions as entry segments have to be reloaded
from global memory into private memory at each kernel invocation.
Regardless, the rightmost histogram in Figure~\ref{fig:mixed_vs_separate}
shows that the error due to using Separate executions is relatively consistent and in
the 1\%-20\% range.  As a result, we expect that our modeling approach
should lead to reasonable response time predictions. Furthermore, since the
error is consistent, the same bias should apply when comparing the
estimated response time for various candidate batch sizes.

\subsection{CPU Component}

To model the CPU time, we propose an empirical model for each of the two
portions of the CPU time shown in
Figure~\ref{fig:Galaxy_d1_cpu_vs_gpu_times} as shaded cyan and blue
areas. These models consist of simple curve fitting based on benchmark
results ($R^2$ values for the fits are above 0.9999).

%r2=0.9999087373653627 
%r2adj=0.9999011321458096 
%StdErr=0.006119618051903412 
%Fstat=202693.1582106867

To estimate $T^{CPU}_1(s)$, the portion of the CPU time that corresponds to the kernel
invocation overhead for a batch size $s$  (the cyan area in
Figure~\ref{fig:Galaxy_d1_cpu_vs_gpu_times}), we generate a synthetic
dataset with $\alpha \approx 0$. With a very low value of $\alpha$, the result set has negligible
size.  As a result, kernel response time is approximately equal to the
aggregate kernel invocation overhead.  We thus obtain a kernel invocation
overhead curve for the full range of possible batch sizes.  This benchmark
must be executed for various total numbers of query segments so that for a
query set $Q$, our model uses benchmark results obtained for approximately
$|Q|$ query segments.  In practice, one would thus run the benchmark for
various numbers of query segments, obtaining a family of CPU response time
curves.  In all experiments hereafter, $|Q|$=40,000, and we thus use 40,000
queries as well in our benchmark. We obtain the following response time
fitted curve:
%In actuality this value was computed by taking the Galaxy dataset d=5 with something like 8 query batches in total (Chunksize/queries per batch=5000?), which is where the CPU curve is nearly 0.  We know with so few kenrel invocations, the kernel overhead time is negligible, but the transfer time still exists.  So, the constant below is the CPU time divided by the number within the query distance.
 \begin{equation}T^{CPU}_1(s)=-0.0017+32.2946\times s^{-0.9528}.\end{equation}

To estimate $T^{CPU}_2$, the portion of the CPU time that corresponds to
the transfer of the result set from the GPU to the CPU (the blue area in
Figure~\ref{fig:Galaxy_d1_cpu_vs_gpu_times}), we rely on the $\alpha$
parameter defined in the GPU component of our performance model and
estimated as described in Section~\ref{sec:alphabetagamma}. Using $\alpha$, we
can determine the number of result set items generated by each kernel invocation.
Summing over all kernel invocations and multiplying by the size in bytes
of a result set item yields the total size of the result set, which we denote
by $\sigma$.  Assuming that $T^{CPU}_2$ does not depend on $s$, we can then
estimate it by dividing $\sigma$  by the GPU-CPU bandwidth
measured on the platform. On our target platform the model is as follows:
\begin{equation}T^{CPU}_2(\sigma)=1.54\times10^{-8} \times \sigma.\end{equation}

In the end, the total CPU time is modeled as:
$T^{CPU}(s,\sigma)=T^{CPU}_1(s)+T^{CPU}_2(\sigma)$,
and the total response time is modeled as $T^{CPU}(s) + T^{GPU}(s,\sigma)$. 

\subsection{Model Evaluation}

\begin{figure}[t]
\centering
  		\subfigure[S1: \galaxy.]{
            \includegraphics[width=0.3\textwidth]{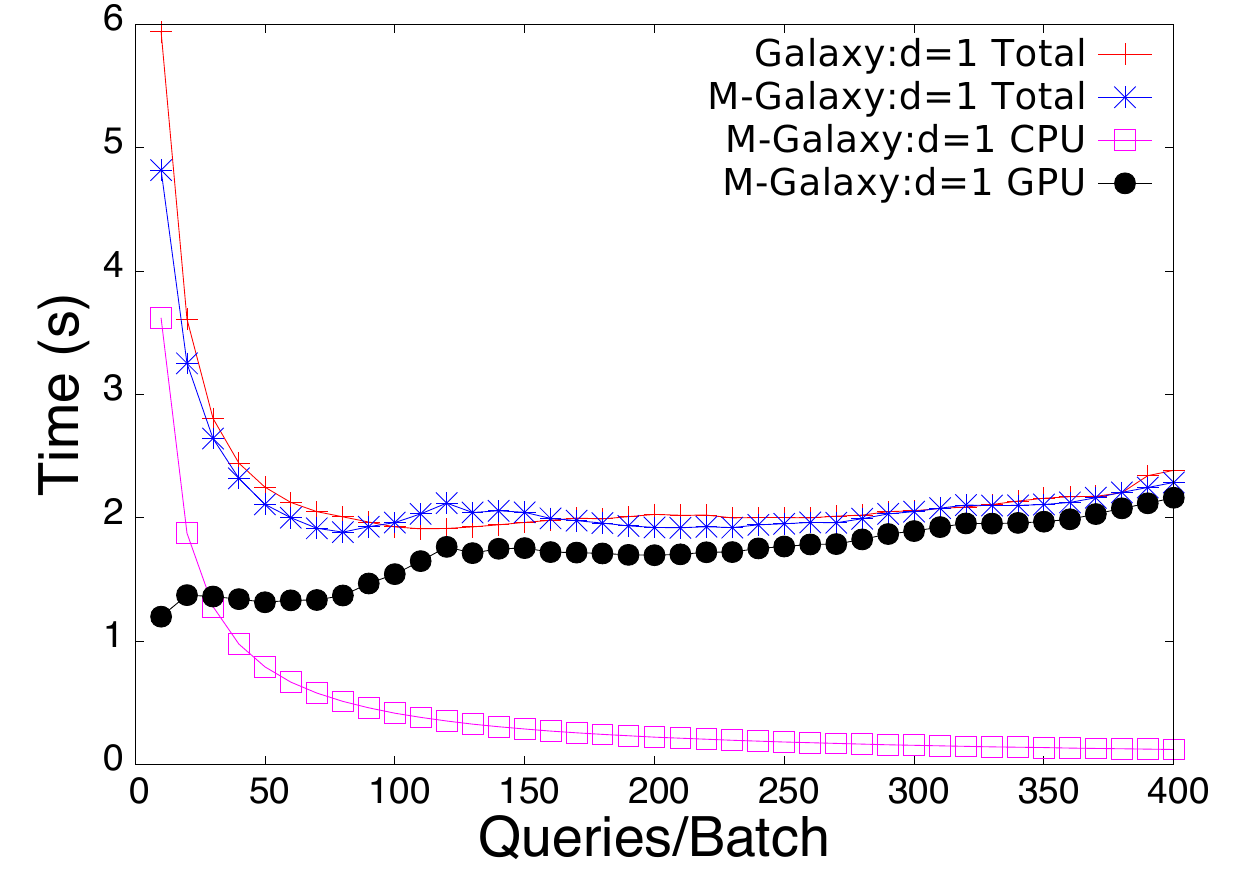}
  		}  		
        \subfigure[S2: \galaxy.]{
            \includegraphics[width=0.3\textwidth]{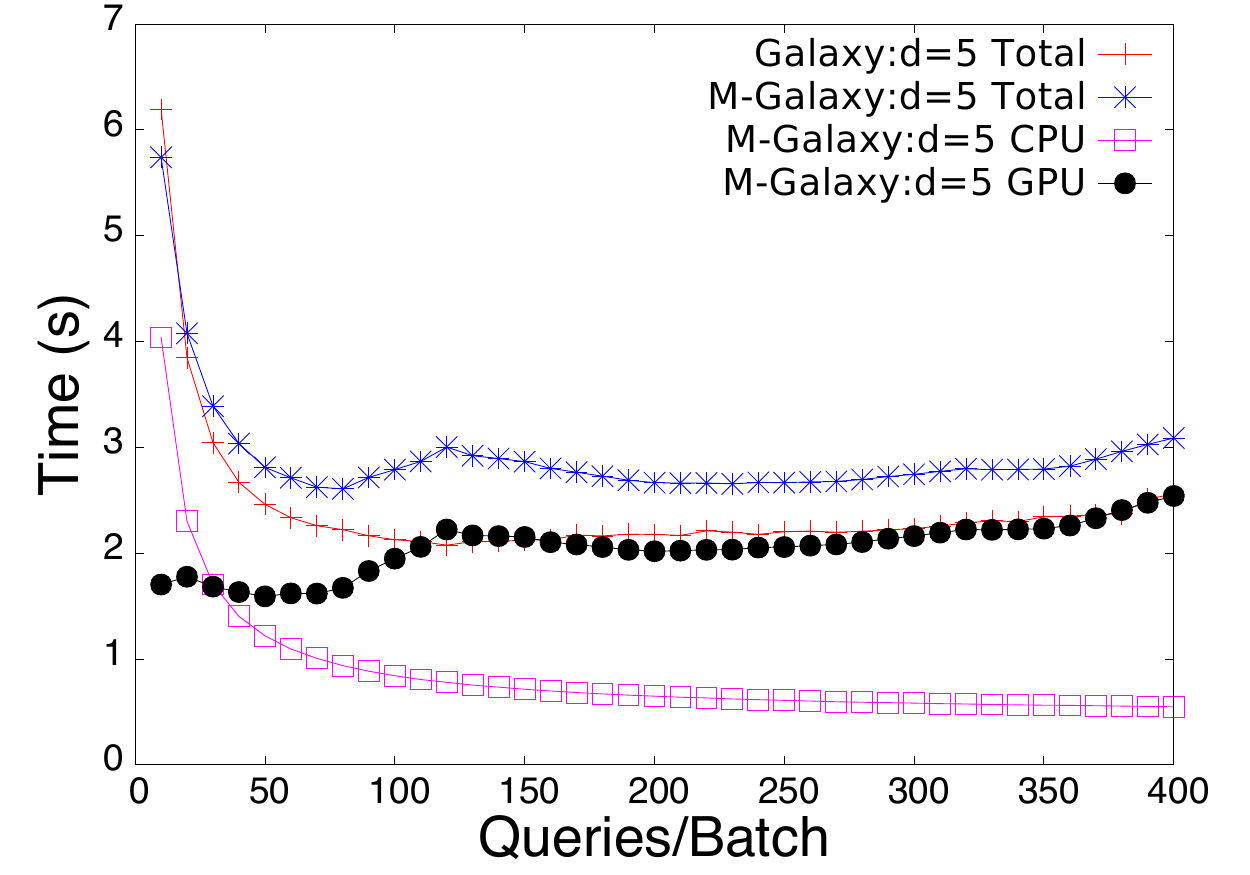}
  		}  
  		 \subfigure[S3: \randomuniform.]{
            \includegraphics[width=0.3\textwidth]{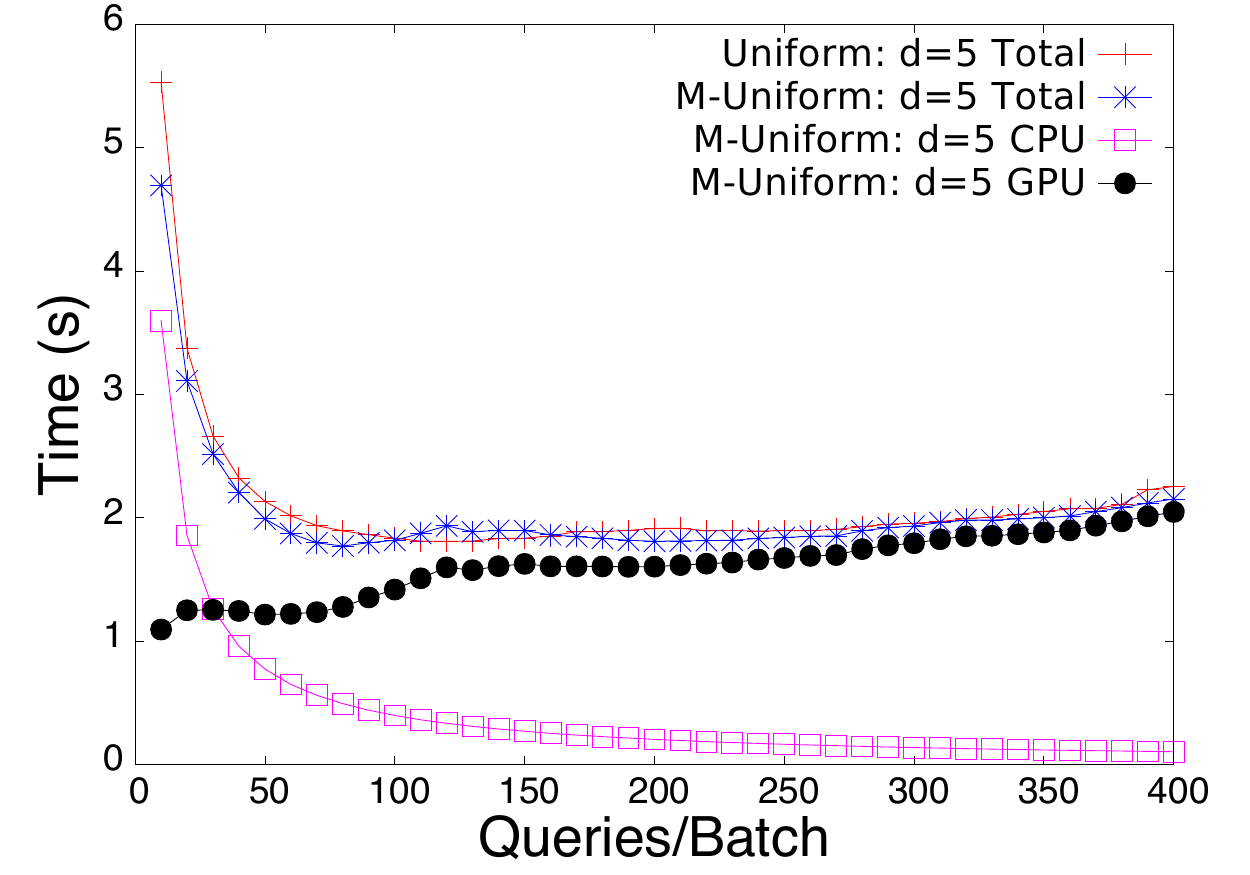}
        }		
        \subfigure[S4: \randomuniform.]{
            \includegraphics[width=0.3\textwidth]{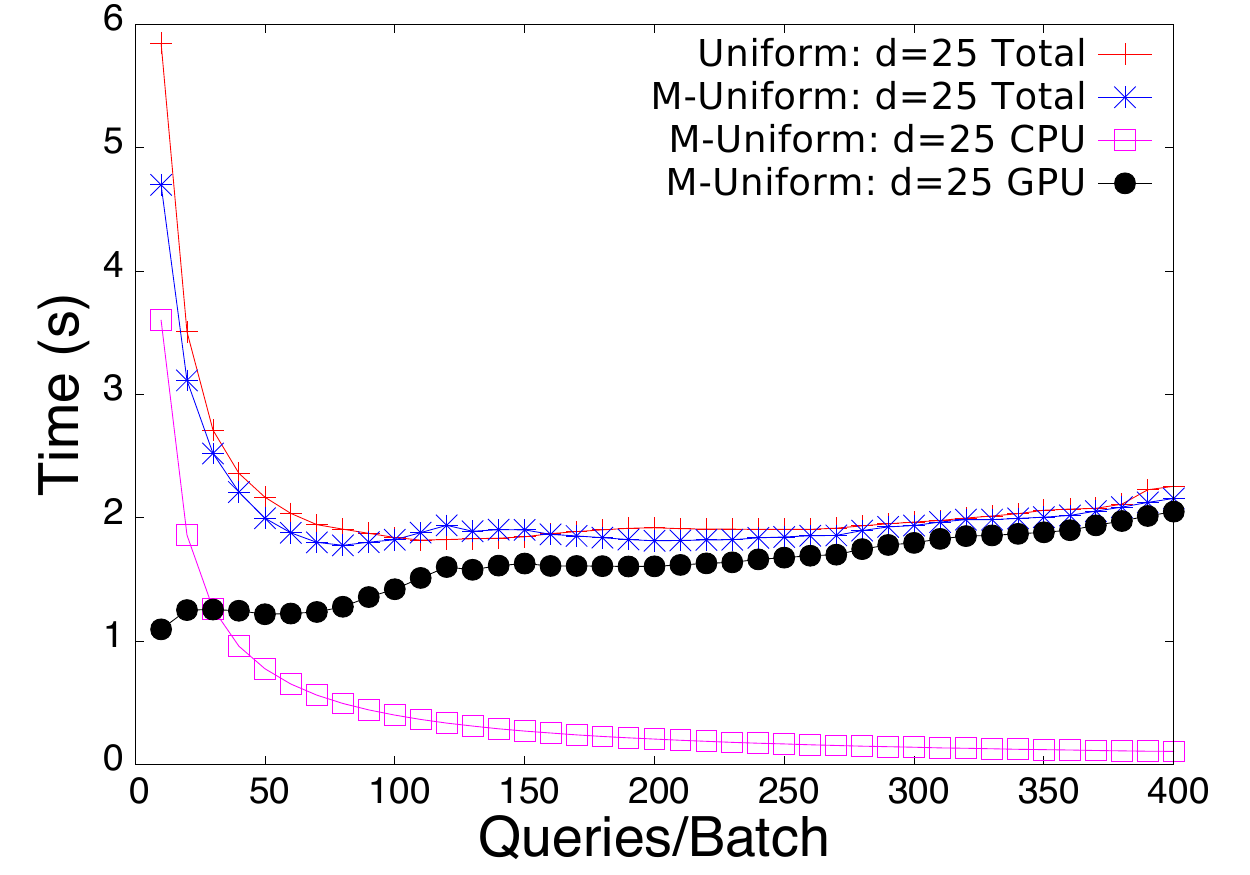}
        }
         \subfigure[S5: \randomnormal.]{
            \includegraphics[width=0.3\textwidth]{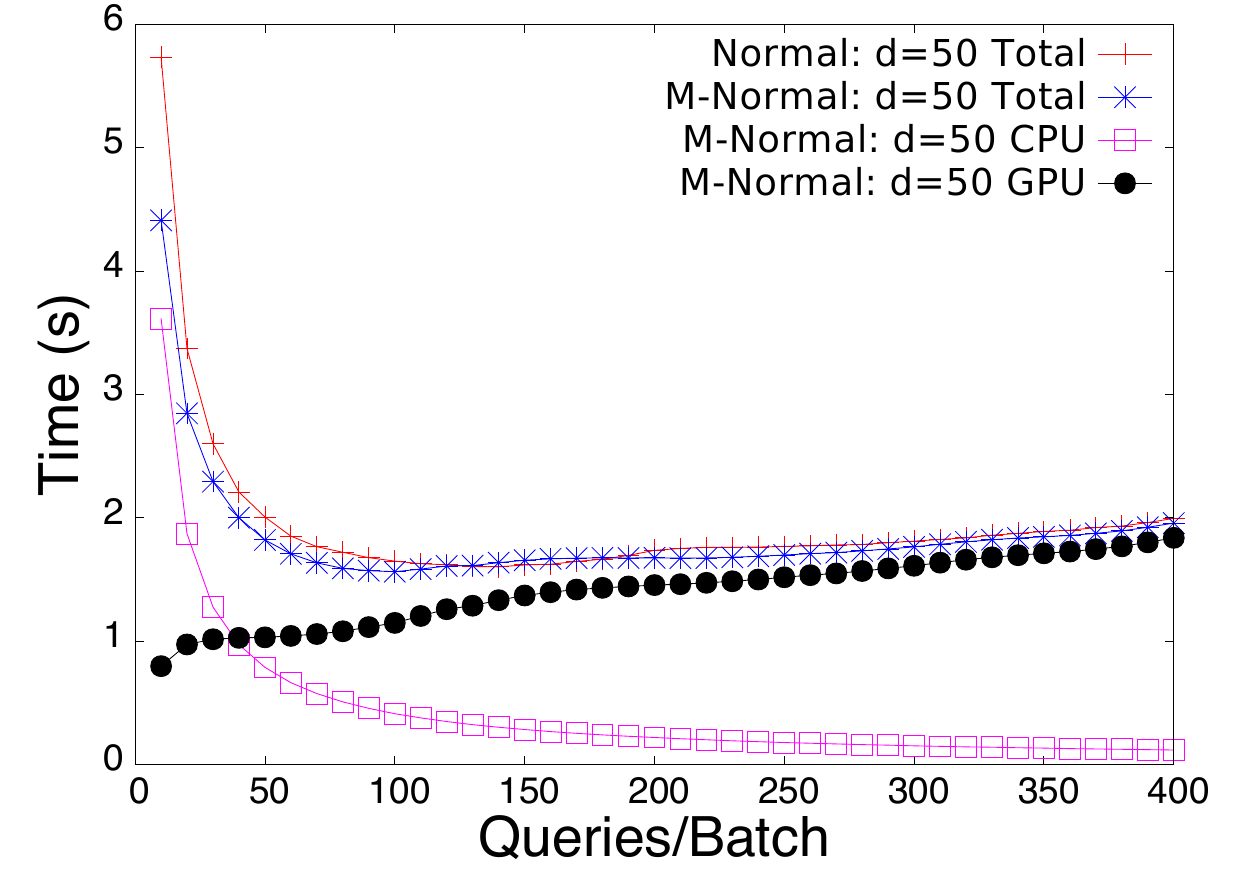}
        }
         \subfigure[S6: \randomnormal.]{
            \includegraphics[width=0.3\textwidth]{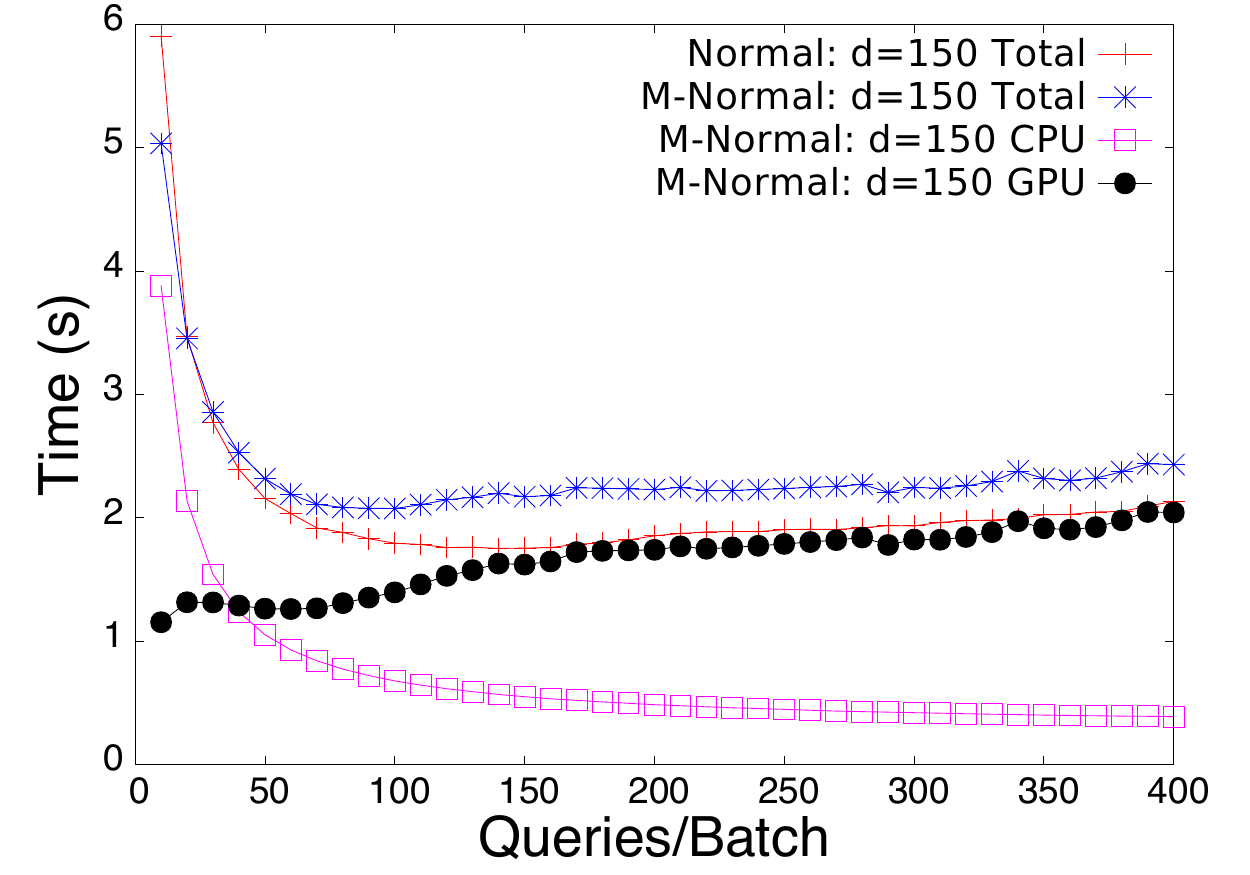}
        }
         \subfigure[S7: \randomnormalfive.]{
            \includegraphics[width=0.3\textwidth]{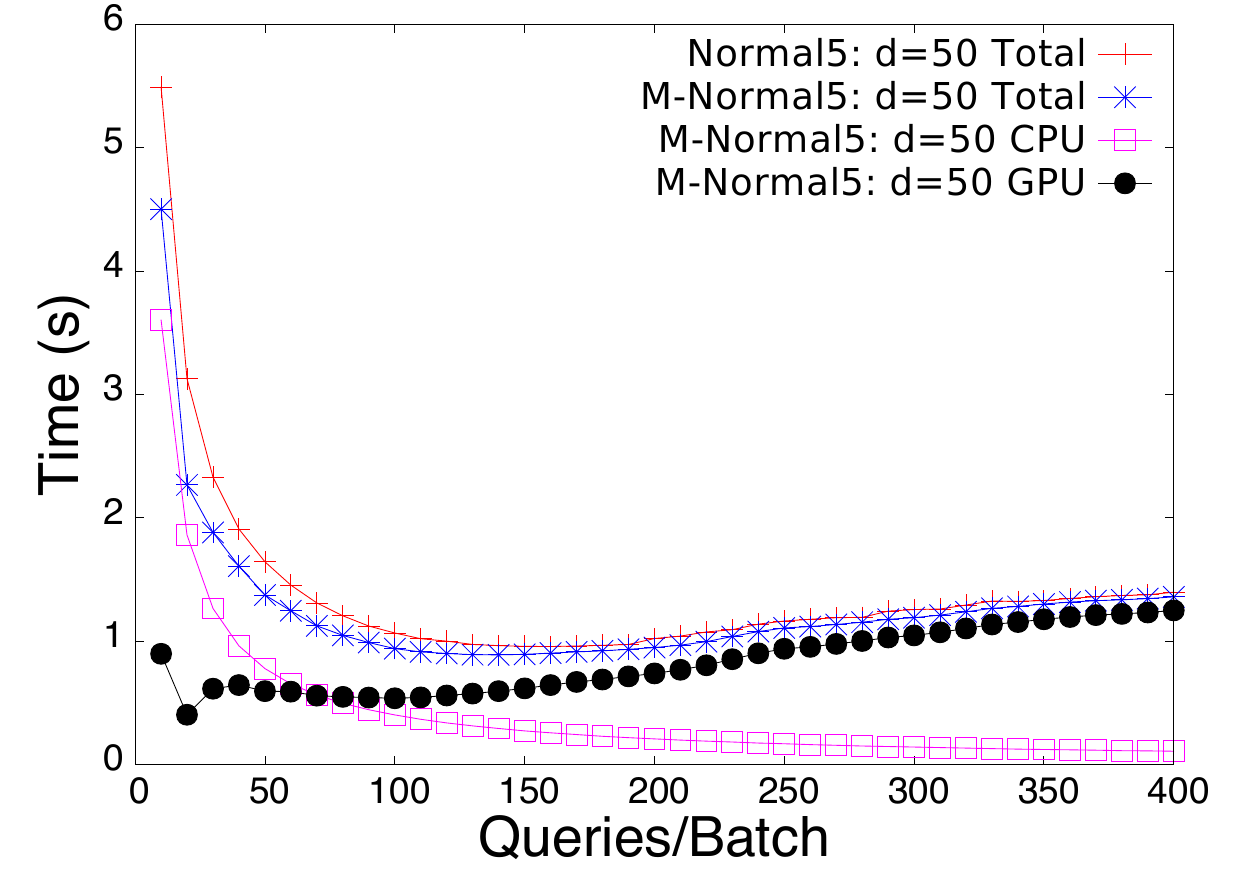}
        }
         \subfigure[S8: \randomnormalfive.]{
            \includegraphics[width=0.3\textwidth]{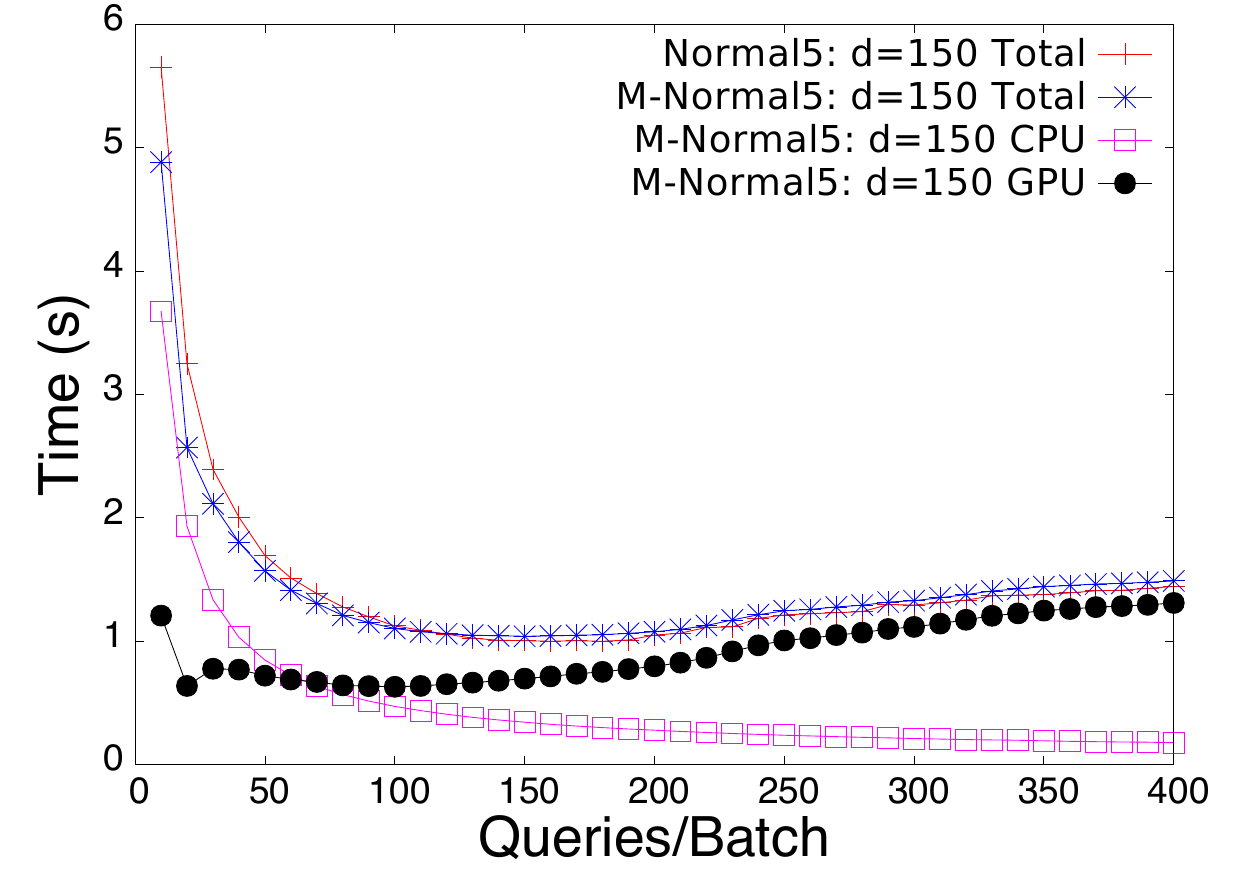}
        }
          \subfigure[S9: \randomexp.]{
            \includegraphics[width=0.3\textwidth]{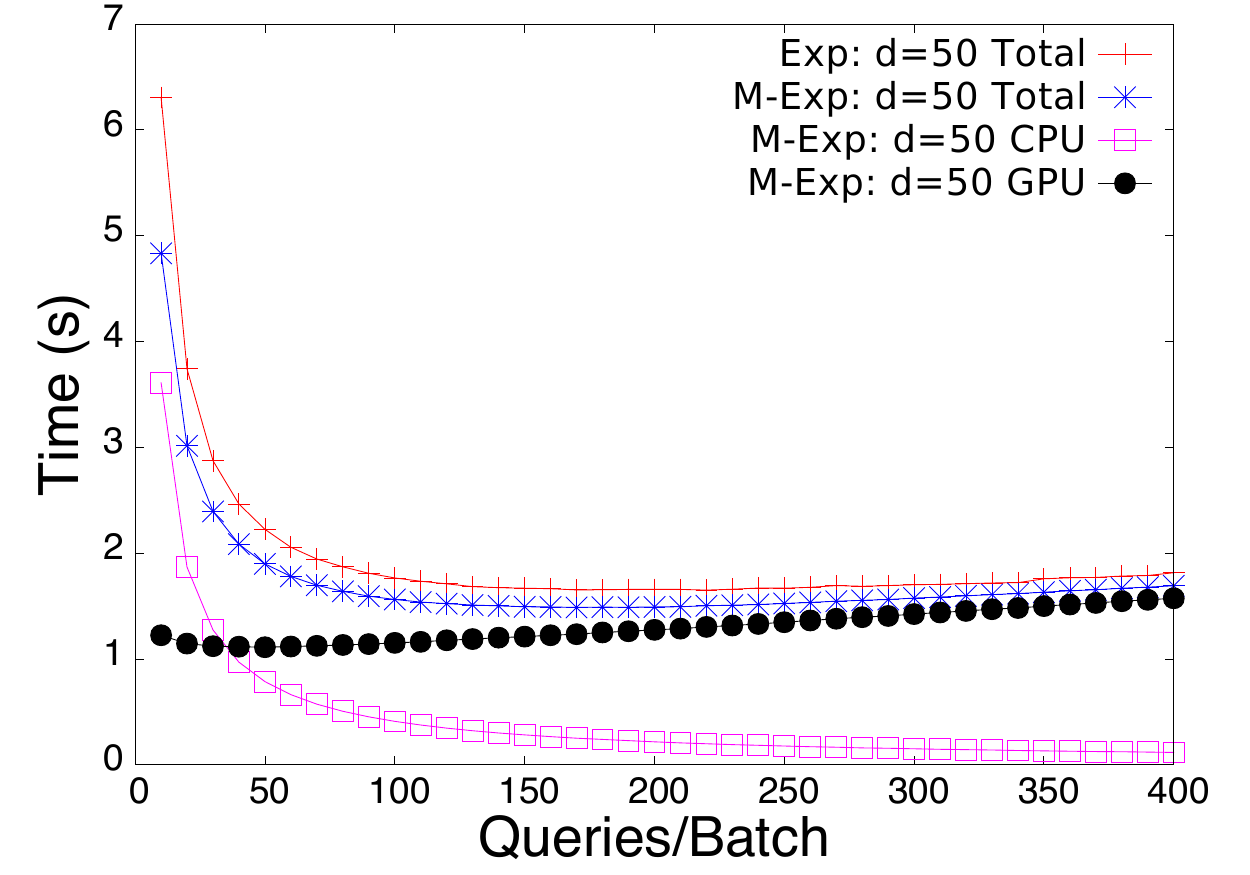}
        }
         \subfigure[S10: \randomexp.]{
            \includegraphics[width=0.3\textwidth]{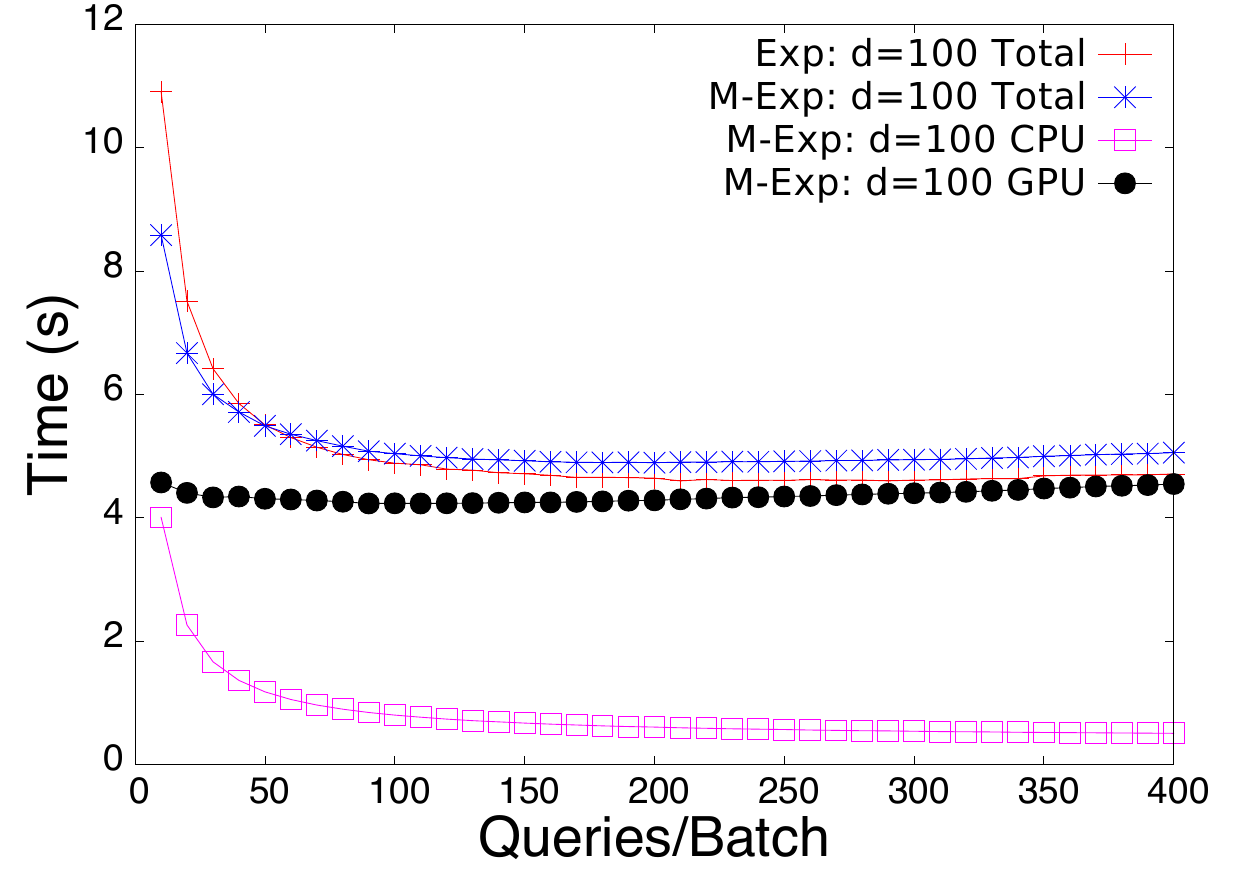}
        }
        
    \caption{Modeled response times vs. queries per batch ($s$) for searches on each dataset.  The red curve shows the actual response time, the blue curve shows the modeled total response time, where the CPU (magenta) and GPU (black) model components added together equal the modeled total (blue) curve.}
   \label{fig:model_results}
\end{figure}

Figure~\ref{fig:model_results} shows actual and modeled response times vs.
the batch size for a selection of our experimental scenarios. The CPU and
GPU model components are shown with separate curves. The general trends of
the actual response time are respected by the model. In some instances, the
model tracks the actual response time well, while on other it exhibits some
deviations. However, the main purpose of our model is not to predict
response time perfectly, but to produce a sufficiently coherent prediction
so that a good batch size can be selected.
Figure~\ref{fig:model_results}~(b) presents the model for the \galaxy
dataset with $d=5$.  The model suggests that $s=80$ yields the best
response time; however, the actual best response time occurs when $s=120$.
Had $s=80$ been chosen, $s=120$ would be 6.3\% faster.  Such results are
summarized in Table~\ref{tab:model} for our 10 experimental scenarios,
where the Model column gives the batch size based on the model, the Actual
column gives the empirically best batch size, and the Slowdown column gives
the response time slowdown due to using the model-driven batch size, as a
percentage.  Note that S3 has few elements in its result set, and thus has a low value of $\alpha$ across all epochs.  For this scenario, it was not possible to assure that the total estimated number of result set items is within 5\% of the actual number across all values of $s$. 

%slowdown: 
%Galaxy d1
%model: 80
%actual: 110
% (2.00862-1.91159)/2.00862=0.0483067

%Galaxy d5
%(2.22-2.08)/2.22=0.06306

%RW uniform d5
%model: 80
%actual: 120
% (1.89469-1.80934)/1.89469=0.045046

%RW uniform d25
%(1.9083-1.8195)/1.9083=0.04653

%RW normal d50
%model: 100
%actual: 140
% (1.65016-1.60324)/1.65016= 0.0284336

%RW normal d150
%(1.79469-1.75283)/1.79469=0.0233243

%RW normal five d50
%model:140
%actual:160
%(0.965424-0.957676)/0.965424= 0.008025489

%RW normal five d150
%(1.00521-0.999409)/1.00521=0.00577

%RW EXP d50
%model: 170
%actual: 220
%(1.65260-1.65058)/1.65260=0.0012223

%RW EXP d100
%(4.64440-4.59919)/4.64440=0.0097343

\begin{table}
\centering
\caption{Model Results.}
\begin{tabular}{|c|c|c|c|} \hline
Search&Model &Actual & Slowdown\\ \hline
S1&80&110&4.8\%\\ \hline
S2&80&120&6.3\%\\ \hline
S3&80&120&4.5\%\\ \hline
S4&80&110&4.5\%\\ \hline
S5&100&140&2.8\%\\ \hline
S6&100&140&2.3\%\\ \hline
S7&140&160&0.8\%\\ \hline
S8&150&160&0.58\%\\ \hline
S9&170&220&0.1\%\\ \hline
S10&200&210&0.97\%\\ 
\hline\end{tabular}
\label{tab:model}
\end{table}

In these results we find that the worst slowdown is less than 7\% and that
in many cases the slowdown is negligible.  We conclude that, in spite of
data dependency challenges, our model is useful for determining a good
batch size to use with the \periodic algorithm and for predicting the total query response time.  An interesting question
is whether our modeling approach can be used for response time prediction
purposes for other spatiotemporal queries.

\section{Conclusions}\label{sec:conclusions}
%intro paragraph
In this work, we have studied the efficient execution of an in-memory
trajectory similarity search, the distance threshold search, on the GPU.
The objective is to minimize response time in an online setting in which a
series of kernel invocations are performed to process a potentially large
query set.  We have shown that the parallelism afforded by the GPU,
provided a GPU-friendly indexing method is used, can outperform
multithreaded CPU implementations that use an in-memory R-tree index.  We
have proposed such a GPU-friendly indexing method.  While conceptually
simple, this method may be suitable for indexing spatial and spatiotemporal
objects for parallel architectures in general, as described in
\cite{Zhang2014}.  We have proposed several algorithms for partitioning a
query set into individual batches, so as to reduce memory pressure and computational cost on the
GPU. We have found that, when considering the cost to compute the
batches, a simple algorithm that partitions the query set into fixed-size
batches leads to competitive response times.

Modeling the performance of algorithms that process moving objects is a
challenge due to the spatiotemporal nature of data.
Furthermore, in the context of spatiotemporal databases, where index-trees
are paramount, the non-deterministic nature of tree traversals adds an
additional source of performance uncertainty.  The indexing method proposed
in this work obviates some of this data-dependent uncertainty.  As a
result, we are able to derive a reasonably accurate response time model.
This model, which considers both CPU and GPU time, is sufficient for
predicting a good batch size for a given dataset.  Furthermore, in some
instances, the model is adequate to estimate the actual query response time
across a range of query batch sizes.  This result is encouraging, as it
suggests that predicting query response time on the GPU, at least with some
indexing techniques, is feasible, making it possible to assess the
tractability of spatiotemporal queries across a range of application
domains.  In particular, such query response time prediction will be
crucial for for estimating the compute time for the astrophysical
application that is the initial motivation for this work.  Future work directions include utilizing multiple work queues to overlap computation with communication between host and device, investigating other GPU-friendly indexes and applying our performance model to other spatiotemporal queries.

\section*{Acknowledgments}
This material is based upon work supported by the National Aeronautics and Space Administration through the NASA Astrobiology Institute under Cooperative Agreement No. NNA08DA77A issued through the Office of Space Science.

\bibliographystyle{plain}

\end{document}